\documentclass[11pt]{article}
\usepackage{epsfig}
\usepackage{amssymb}
\usepackage[]{times}

\makeatletter \@addtoreset{equation}{section}

\makeatother
\newcommand{\be}{\begin{equation}}
\newcommand{\ee}{\end{equation}}
\newcommand{\bea}{\begin{eqnarray}}
\newcommand{\eea}{\end{eqnarray}}

\newtheorem{remark}{Remark}
\begin{document}
\title{Theory of Pendular Rings Revisited}
\author{Boris Y. Rubinstein$^1$ and Leonid G. Fel$^2$
\footnote{Corresponding author: lfel@technion.ac.il}
\\ \\
$^1$Stowers Institute for Medical Research,
1000 E 50th St, Kansas City, MO 64110, USA\\
$^2$Department of Civil and Environmental Engineering,\\
Technion -- Israel Institute of Technology, Haifa, 32000, Israel}
\date{\today}
\vspace{-.5cm}
\maketitle
\begin{abstract}
We present the theory of liquid bridges between two axisymmetric solids, sphere
and plane, with prescribed contact angles in a general setup, when the solids
are non-touching, touching or intersecting, We give a detailed derivation of
expressions for curvature, volume and surface area of pendular ring as
functions of the filling angle $\psi$ for all available types of menisci:
catenoid ${\sf Cat}$, sphere ${\sf Sph}$, cylinder ${\sf Cyl}$, nodoid ${\sf
Nod}$ and unduloid ${\sf Und}$ (the meridional profile of the latter may have
inflection points).

The Young-Laplace equation with boundary conditions can be viewed as a
nonlinear eigenvalue problem. Its unduloid solutions, menisci shapes $z_n^s(r)$
and their curvatures $H_n^s(\psi)$, exhibit a discrete spectrum and are
enumerated by two indices: the number $n$ of inflection points on the meniscus
meridional profile ${\cal M}$ and the convexity index $s=\pm 1$ determined by
the shape of a segment of ${\cal M}$ contacting the solid sphere: the shape is
either convex, $s=1$, or concave, $s=-1$.

For the fixed contact angles the set of the functions $H_n^s(\psi)$ behaves in
such a way that in the plane $\{\psi,H\}$ there exists a bounded domain where
$H_n^s(\psi)$ do not exist for any distance between solids. The curves $H_n^s(
\psi)$ may be tangent to the boundary of domain which is a smooth closed curve.
This topological representation allows to classify possible curves and introduce
a saddle point notion. We observe several types of saddle points, and give their
classification.
\end{abstract}

{\em This paper is dedicated to the memory of our friend and bright scientist
A. Golovin (1962--2008)}
\tableofcontents
\section{Introduction}\label{r1}
The problem of pendular ring (PR) arises when a small amount of fluid forms an
axisymmetric liquid bridge with interface (meniscus) between two axisymmetric
solids. This problem includes a computation of liquid volume $V$, surface area
$S$ and surface curvature $H$ and was one of gems in mathematical physics
of the 19th century. In the last decade the PR problem became again an area of
active research due to investigations on stability of the PR shapes, and its
importance has grown for various applications in soil engineering, physics of
porous media, etc.

The history of the problem dates back to 1841 when Delaunay \cite{Delaunay1841}
classified all non-trivial surfaces of revolution with constant mean curvature
in ${\mathbb R}^3$ by solving the Young-Laplace (YL) equation and showed that
they are obtained by tracing a focus of a conic section when rolled on a line,
and revolving the resulting curve around the axis of symmetry. These are {\em
cylinder} (${\sf Cyl}$), {\em sphere} (${\sf Sph}$), {\em catenoid} (${\sf
Cat}$), {\em nodoid} (${\sf Nod}$) and {\em unduloid} (${\sf Und}$). The two
last of them are defined through the elliptic integrals and may appear of two
kinds, {\em concave} (-) and {\em convex} (+), depending on constant sign of the
meridional profile ${\cal M}$ curvature. One more type of meniscus, an {\em
inflectional unduloid}, appears when meridional section ${\cal M}$ curvature
changes its sign along the meniscus.

In 1864 Plateau \cite{Pl1864} applied this classification to analyze the
figures of equilibrium of a liquid mass, and was the first who discovered
\cite{Pl1873} a standard sequence of meniscus evolution observed with increase
of the liquid volume in absence of gravity. According to \cite{Orr1975} the
Plateau sequence reads:
\bea
{\sf Nod^-}\rightarrow {\sf Cat}\rightarrow {\sf Und_0^-}\rightarrow
{\sf Und_1^-}\rightarrow {\sf Und_0^+}\rightarrow {\sf Sph}\rightarrow
{\sf Nod^+}\;,\label{a1}
\eea
where subscripts denote the number of inflection points on the meniscus
meridional section ${\cal M}$. Even so, the actual algorithm for solution of the
PR problem leads to the eigenvalue problem for mean curvature $H$ that requires
extensive and accurate computation of the elliptic integrals and was not
available before the computer era has been started. A complete review on
different methods used to find actual solutions of the YL equation or the
equivalent variational problem (Howe \cite{Ho1887} in 1887 and Fisher
\cite{Fisher1926} in 1926) throughout the last century can be found in
\cite{Orr1975}.

In 1966 Melrose \cite{Mel1966} gave a detailed analysis of the ${\sf Nod}^-$
meniscus and derived the formulas for $V$, $S$ and $H$ in the case of two
touching spheres of equal radii. Orr, Scriven and Rivas in 1975 extended this
result in seminal article \cite{Orr1975} for the menisci of various profiles in
the case of solid sphere of radius $R$
above the solid plane for $d=D/R \geq 0$ with prescribed contact
angles $\theta_1$ and $\theta_2$ on sphere and
plane, respectively; here $D$ denotes a distance between the sphere and the
plane. In the case of touching solids, $d=0$, they \cite{Orr1975} have performed
numerical computations and verified the Plateau sequence (\ref{a1}) when the
liquid volume is increasing.

\begin{figure}[h!]\begin{center}\begin{tabular}{cc}
\psfig{figure=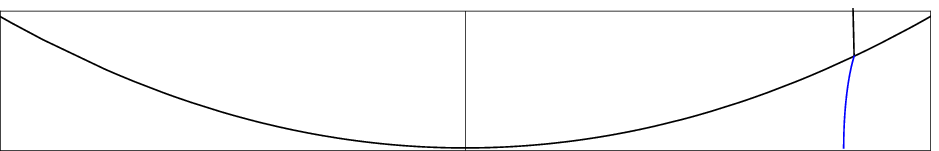,height=1.5cm,width=8cm} &
\psfig{figure=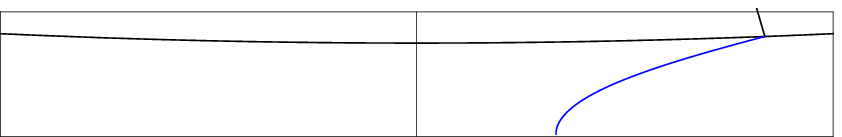,height=1.5cm,width=8cm}\\(a) & (b)
\end{tabular}\end{center}
\vspace{-.2cm}
\caption{(a) The shape of meniscus for $d=0$, $\theta_1=\pi/6$, $\theta_2=\pi/
2$, $\psi=\pi/30$, $H_aR=-69.57$. (b) The shape of meniscus for $d=0.076$,
$\theta_1=\pi/6$, $\theta_2=\pi/2$, $\psi=\pi/30$, $H_bR=2.14$.}\label{a4a}
\end{figure}

During the past decades the work \cite{Orr1975} became classical, albeit
throughout a vast number of references no attempt was made to extend this
explicit analysis using modern computer algebra technique. What is more
important, formulas in \cite{Orr1975} for the non-touching solids were left
without detailed analysis. The reason why we have noticed this fact is based on
substantial difference in the behavior of the function $H(\psi,d)$ in two
different setups: $d=0$ and $d>0$. This can be seen easily in Figure \ref{a4a}
where we consider for small filling angles $\psi$ two types of menisci between
touching (a) and non-touching (b) sphere and plane having the same set of
contact angles $\theta_1<\pi/2$ and $\theta_2=\pi/2$.

\vspace{.4cm}
\noindent

In the case (a) the sphere-plane geometry approaches its {\em wedge} limit while
in the case (b) it approaches the {\em slab} geometry. Estimate two principal
radii, meridional $R_v$ and horizontal $R_h$ in both cases for $\psi\ll 1$.
In the first case (a) they are of different signs, i.e., $R_v<0$, $R_h>0$.
Keeping in mind $H=1/2(R_v^{-1}+R_h^{-1})$, a simple trigonometry gives
\bea
\frac{R_v}{R}\simeq-\frac{\psi^2}{2\cos\theta_1}\;,\quad\frac{R_h}{R}\simeq\psi
\;,\quad RH\simeq -\frac{\cos\theta_1}{\psi^2}\;.\label{b1}
\eea
The dependence $H\simeq-\psi^{-2}$, $\psi\ll 1$ describes the asymptotics $H(
\psi)$ of the ${\sf Nod^-}$ meniscus for two touching solids found in
\cite{Orr1975}. In section \ref{nod_as0} we justify formula (\ref{b1}) by
rigorous derivation of nodoidal asymptotics.
In the second case (b) we have another estimate,
\bea
\frac{R_v}{D}\simeq-\frac1{\cos\theta_1}\;,\quad\frac{R_h}{R}\simeq\psi\;,\quad
RH\simeq \frac1{2\psi}\;.\label{b2}
\eea
We have found meniscus with $H>0$ that according to the Plateau
classification has the concave unduloid type ${\sf Und}_0^-$ which is absent in
the range $\psi\ll 1$ in the sequence (\ref{a1}). Here another asymptotics
holds, $H\simeq\psi^{-1}$. This leads to dramatic changes for the whole sequence
(\ref{a1}) giving rise to the ${\sf Cat}$ menisci for two different filling
angles $\psi$, or to a single degenerated ${\sf Cat}$ meniscus, or to
disappearance of both of them.

There exists one more case which cannot be reduced to the previous ones. This is
a meniscus between intersecting sphere and plane ($d<0$) having the same set of
contact angles $\theta_1$ and $\theta_2=\pi/2$ and wedge geometry for small
filling angles $\psi$. The calculation gives two principal radii, $R_v$ and
$R_h$, and its curvature as follows (we refer to the Figure
\ref{intersect_menis}(b) in section \ref{r13}),
\bea
\frac{R_v}{R}\simeq-\frac{\psi-\psi_*}{\cos\left(\theta_1+\psi_*\right)}\;,
\quad\frac{R_h}{R}\simeq\sin\psi_*\;,\quad RH\simeq -\frac{\cos\left(\theta_1+
\psi_*\right)}{2(\psi-\psi_*)}\;,\quad \psi_*=\arccos(1+d)\;.\label{b1a}
\eea
The dependence $H\simeq-\left(\psi-\psi_*\right)^{-1}$, $\psi-\psi_*\ll 1$
describes the asymptotics $H(\psi)$ of the ${\sf Nod^-}$ meniscus for two
intersecting solids and is intermediate between (\ref{b1}) and (\ref{b2}). Thus,
even a change of a single governing parameter $d$ only, when other two
$\theta_1,\theta_2$ are fixed, changes drastically the evolution of menisci.

Our analysis of solutions of the YL equation shows that the changes become more
essential (non-uniqueness of solutions, excluded domain $\overline{{\cal B}}$ in
the $\{\psi,H\}$ plane where $H(\psi)$ does not exist,
etc.) when we deal with the whole 3-parametric space
${\mathbb P}^3=\{\theta_1,\theta_2,d\}$. From this point of view the article
\cite{Orr1975} has dealt with 2-parametric subspace $\{\theta_1,\theta_2,0\}$.
This creates an additional challenge to describe the menisci in different areas
of ${\mathbb P}^3$, i.e., to give a complete theory.

In this paper we present the theory of pendular rings located between two
axisymmetric solids, sphere and plane, in a general setup, when the solids are
non-touching, and touching or intersecting. We give a detailed derivation of
expressions for curvature $H(\psi)$, volume $V(\psi)$ and surface area $S(\psi)$
as the functions of the filling angle for all available types of menisci
including those omitted in \cite{Orr1975}. We give also an asymptotic analysis
of these functions in the vicinity of singular points where they diverge.

The paper is organized in eight sections and four appendices. In section
\ref{r2} we give a setup of the problem and derive the YL equation and its
solution through the elliptic integrals. The integrals introduced in this
section are evaluated in section \ref{r3} and the explicit expressions for the
meniscus curvature, shape, volume and surface area are found. We introduce also
a new function $\alpha(\psi)$ which is intimately related to the curvature $H(
\psi)$ and becomes a main tool in analysis of an evolution of pendular rings. In
section \ref{s4} we discuss the general curvature behavior for different types
of menisci. Existence of catenoids in the menisci sequence for the cases of
non-touching solids is considered in section \ref{r41}.

In contrast to the case of touching solids discussed in \cite{Orr1975},
when for the small $\psi$
the ${\sf Nod^-}$ menisci exist, in the case of non-touching solids the ${\sf Und}_1^-$
menisci come first. This allows existence of two catenoids in the
menisci sequences, while in some cases catenoids do not appear at all. We find
the critical value of the distance between solids at which the catenoids merge
and estimate the curvatures for different types of menisci. We show that the
values of $\alpha$ can be bounded for some types of the menisci. When these bounds
are crossed, the corresponding menisci transform one into another. Analysis of
these transitions, their sequence and smoothness, is given in section
\ref{r45}.

In section \ref{rr10} we elaborate a topological approach to study different
curves $\alpha_n(\psi)$ enumerated by the number $n$ of inflection points at
menisci. A behavior of these curves in the plane $\left\{\psi,\alpha\right\}$ is
confined within domain $\Delta^{\prime}=\left\{0\leq\psi\leq\pi-\theta_1,\;0\leq
\alpha\leq1\right\}$ with embedded subdomain ${\cal B}\subset\Delta^{\prime}$
which is prohibited for $\alpha_n(\psi)$ to pass through; this makes $\Delta^{
\prime}$ not simply connected. The curves $\alpha_n(\psi)$ may be tangent to the
subdomain's boundary which is a smooth closed curve described by symmetric
transcendental function. This global representation allows to classify possible
curves and introduce a saddle point notion in the PR problem. We observe several
types of saddle points, their classification is presented in section \ref{r511}.

In section \ref{r13} we give a brief analysis of menisci evolution in the cases
of touching and intersecting solid bodies which is essentially different from a
general setup of non-touching bodies. Concluding remarks and open problems are
listed in section \ref{conclus}.

Four appendices are inseparable parts of the paper. Appendices \ref{appendix1}
and \ref{ar5} contain a list of formulas with technical details for $H$, $V$,
$S$ and the shape of meniscus of each type separately. They give an exhaustive
description of menisci and build a basis for further investigation of basic
properties of PRs like stability, rupture, hysteresis etc. In appendix
\ref{appendix5} we show that the nodoid meniscus ${\sf Nod^+}$ always has a
local minimum of curvature and local maxima of the surface area and volume. We
also find the asymptotic behavior of the nodoidal curvature in vicinity of
singular point when $-2<d\leq 0$. Appendix \ref{appendix3} is completely devoted
to elliptic integrals and their applications to computation various expressions
throughout the paper.
\section{Young-Laplace Equation and its Solutions}\label{r2}
The problem of PR can be posed as a search of the surface of revolution
characterized by a constant mean curvature $\tilde H$ satisfying the YL
equation valid in case of negligibly small gravity effect
\be
2\tilde H=\frac{z''}{\left(1+z'^2\right)^{3/2}}+\frac{z'}{r\left(1+z'^2\right)
^{1/2}},\label{YLeq}
\ee
where $z(r)$ and $r$ are cylindrical coordinates of the meniscus. Introducing
new variables $x=r/R$ and $y=z/R$ and a parameter $u=\sin t$ (where $t$ is an
angle of the normal to meniscus with the vertical axis), we transform this
equation into the problem for nondimensional curvature $H=R\tilde H$
\be
2H=du/dx +u/x.\label{main_eq}
\ee
The contact angles with the solid bodies are $\theta_1$ (with the sphere) and
$\theta_2$ (with the plane). The boundary conditions read
\bea
t_1=\theta_1+\psi, & y_1=1+d-\cos\psi, &\ \ x_1=\sin\psi,\nonumber\\
t_2=\pi-\theta_2, & y_2=0.\label{BC}
\eea
Here $\psi$ is the filling angle, and $d=D/R$ is the scaled distance between
the sphere and the plane. It is easy to show that
\be
dy/dx=\tan t\;.\label{y(u)}
\ee
The solution in parametric form reads
\bea
x &=&\frac{1}{2H}\left[\sin t+s\sqrt{\sin^2 t+c}\;\right],\label{solx}\\
y &=&\frac{1}{2H}\int_{t_2}^{t}\left[\sin t +\frac{s\;\sin^2 t}
{\sqrt{\sin^2 t+c}}\;\right]\;dt\;,\label{soly} \eea
where we used
the relation $dx/dt=xs\cos t/\sqrt{\sin^2 t+c}$ and the parameter
$c$ depends on curvature
\be
c=4H\sin\psi\;(H\sin\psi-\sin t_1)\;.
\label{cdef}
\ee
Here and below $s=\pm 1$; its computation
will be described in section \ref{r45}, formula (\ref{curv_und_pm}).

Introducing a parameter $\alpha$, such that
\be
H\sin\psi=\alpha\sin t_1,\label{alpha_def}
\ee
we rewrite relation (\ref{cdef}) as
\be
c=4\alpha(\alpha-1)\sin^2 t_1.\label{cdef_alpha}
\ee
Making use of the boundary conditions we find for the curvature
\be
2H\Psi=I_s,\quad I_s=\int_{t_2}^{t_1}\left[\sin t +\frac{s\;\sin^2 t}{\sqrt{
\sin^2 t+c}}\right]\;dt\;,\quad\Psi=d+1-\cos\psi\;.\label{curv}
\ee
The meniscus surface area $S$ is computed as $S=2\pi\int x\sqrt{1+(dx/dy)^2}\;dy$
and is given by the integral
\be
S=\frac{\pi}{2H^2}K_s\;,\quad K_s=s\int_{t_2}^{t_1}\frac{\sin t\left(\sin t+s
\sqrt{\sin^2 t+c}\right)^2}{|\sin t|\sqrt{\sin^2 t+c}}\;dt\;.\label{surf}
\ee
The volume $V_r=\pi\int x^2 dy$ of the solid of rotation reads
\be
V_r=\frac{\pi}{8H^3}J_s\;,\quad J_s=s\int_{t_2}^{t_1}\frac{\sin
t\left(\sin t+s\sqrt{\sin^2 t+c}\right)^3}{\sqrt{\sin^2 t+c}}\;dt\;.
\label{volume}
\ee
Then the volume $V$ of liquid inside the PR can be computed by subtracting
from the above expression the volume $V_{ss}=\pi(2-3\cos\psi+\cos^3\psi)/3$ of
the spherical segment corresponding to the filling angle $\psi$.
\subsection{Integral Evaluation}\label{r3}
Consider evaluation of three integrals $I_s$, $J_s$ and $K_s$ and give
expressions for $H$, $V$, $S$ in terms of $c$, filling angle $\psi$, contact
angles $\theta_1$, $\theta_2$ and distance $d$ from the sphere to the plane.
The integral $I_s$ can be written as $I_s=I_1+s I_2$, where
\bea
I_1(t_1,t_2)&=&\int_{t_2}^{t_1}\sin t\;dt=-\cos t_1+\cos t_2,\label{I_1}\\
I_2(t_1,t_2)&=&\int_{t_2}^{t_1}\frac{\sin^2 t\; dt}{\sqrt{\sin^2 t+c}}\;.
\label{I_2_def}
\eea
Making use of the elliptic integrals of the first $F(t,k)$ and the second
$E(t,k)$ kind, respectively, we find (see Appendix \ref{app31})
\be
I_2(t_1,t_2)=\sqrt{c}\;\left[{\overline E}(t_1,k)-{\overline F}(t_1,k)-
{\overline E}(t_2,k)+{\overline F}(t_2,k)\right]\;,\quad k^2=-1/c,\label{I_2}
\ee
where by ${\overline A}(z)$ we denote a complex conjugation of complex
function $A(z)$.

Consider the integral $K_s$ required for computation of the surface area in
(\ref{surf}). The value of $\sin t_1$ at the upper limit $t_1$ can take both
positive (for $t_1<\pi$) and negative (for $t_1>\pi$) values. In the first case
the integral reads $K_s=s(2I_2+I_3)+2I_1$, where
\be
I_3(t_1,t_2)=\int_{t_2}^{t_1}\frac{c\; dt}{\sqrt{\sin^2 t+c}}=\sqrt{c}\;
\left[{\overline F}(t_1,k)-{\overline F}(t_2,k)\right].\label{I_3}
\ee
In the last case the integral is broken into two parts as follows
\be
K_s(t_1,t_2)=K_s(\pi,t_2)+K_s(\pi,t_1)\;.\label{K_sum}
\ee
The integrals $I_2$ and $I_3$ follow another relation
\be
I_2(t_1,t_2)=I_2(\pi,t_2)-I_2(\pi,t_1),\quad I_3(t_1,t_2)=I_3(\pi,t_2)-I_3(
\pi,t_1)\;.\label{I2_3_sum}
\ee
Using (\ref{I_2} -- \ref{I2_3_sum}) we find for $t_1>\pi$
\bea
K_s(t_1,t_2)&=&s(2I_2(\pi,t_2)+2I_2(\pi,t_1)+I_3(\pi,t_2)+I_3(\pi,t_1))
+2I_1(\pi,t_2)+2I_1(\pi,t_1)\nonumber\\
&=&s(2I_2(t_1,t_2)+I_3(t_1,t_2))+2I_1(t_1,t_2)+4(\cos t_1+1)\nonumber\\
&+&2s\sqrt{c}\;\left[4{\overline E}(k)-2{\overline K}(k)-2{\overline
E}(t_1,k)+{\overline F}(t_1,k)\right].\nonumber
\eea
Finally we have
\be
K_s(t_1,t_2)=\left\{\begin{tabular}{lc}
$s(2I_2+I_3)+2I_1,$&$t_1\le\pi,$\\$s(2I_2+I_3)+2I_1+4(\cos t_1+1)$+ &\\
$\;\;\;2s\sqrt{c}\;\left[4{\overline E}(k)-2{\overline K}(k)-2{\overline E}
(t_1,k)+{\overline F}(t_1,k)\right],$ &$t_1>\pi.$
\end{tabular}\right.
\label{K}
\ee
Both surface area $S(\psi)$ and its derivative $S^{\prime}(\psi)$ are continuous
at the matching value $\psi=\pi-\theta_1$.

Rewrite the integral $J_s$ required for the volume computation in (\ref{volume})
$J_s=4J_3+cI_1+s (J_1+3J_2)$, where
$$
J_1=\int_{t_2}^{t_1}\frac{\sin^4 t\;dt}{\sqrt{\sin^2 t+c}}, \ \
J_2=\int_{t_2}^{t_1}\sin^2 t\sqrt{\sin^2 t+c}\;dt, \ \
J_3=\int_{t_2}^{t_1}\sin^3 t\;dt.
$$
The last integral reads
\be
J_3(t_1,t_2)=\frac{\cos 3t_1-9\cos t_1}{12}-\frac{\cos 3t_2-9\cos t_2}{12}.
\label{J3}
\ee
We find $J_1=J_2-cI_2$, where
\bea
J_2(t_1,t_2) &=&\sqrt{c}\left\{\frac{2+c}{3}[{\overline E}(t_1,k)-{\overline E}
(t_2,k)]-\frac{1+c}{3}[{\overline F}(t_1,k)-{\overline F}(t_2,k)]\right\}
\nonumber\\
&-&\frac{1}{6}\left[\sin 2t_1\sqrt{\sin^2t_1+c}-\sin2t_2\sqrt{\sin^2t_2+c}\;
\right].\label{J2}
\eea
Collecting the expressions for $J_i$ and using $J_1+3J_2=4J_2-cI_2$ we finally
arrive at
\be
J_s=4J_3+cI_1-scI_2+4sJ_2.\label{Jpm}
\ee
We also need the following integral
\be
I_4(t_1,t_2)=\int_{t_2}^{t_1}\frac{\sin^2 t\;dt}{(\sin^2 t+c)^{3/2}},
\label{I_4_def}
\ee
that evaluates to
\bea
I_4(t_1,t_2)&=&\frac{1}{\sqrt{c}}\left\{\left[\;\overline F(t_1,k)-\overline F
(t_2,k)\;\right]-\frac{c}{1+c}\left[\;\overline E(t_1,k)-\overline
E(t_2,k)\;\right]\right\}\nonumber\\
&-&\frac{1}{2(1+c)}\left[\frac{\sin 2t_1}{\sqrt{\sin^2t_1+c}}-\frac{\sin 2t_2}
{\sqrt{\sin^2t_2+c}}\right].\label{I_4}
\eea
\section{Menisci}\label{s4}
In this section we discuss menisci shapes of four types (sphere, catenoid,
nodoid and unduloid) between two non-touching axisymmetric solids, sphere and
plane.

First show that for any meniscus type with a fixed sign $s$ and a fixed number
$n$ of inflection points
\be
H(\psi,d_1)\neq H(\psi,d_2),\quad\mbox{if}\quad d_1\neq d_2\;.\label{nund2}
\ee
Indeed, let by way of contradiction, the opposite holds, i.e., there exist two
different $d_1$ and $d_2$ and at least one value $\psi_{\star}$ of filling
angle such that $H_{\star}(\psi_{\star},d_1)=H_{\star}(\psi_{\star},d_2)$.
However, this contradicts the master equation (\ref{und_curv_general}) which
states
\be
d=\frac{I_s(\psi_{\star},H_{\star})+n\hat I_2(c(\psi_{\star},H_{\star}))-s(1-
\cos\pi n)I_2(\pi/2,\pi-\theta_2)}{2H_{\star}}-1+\cos\psi_{\star}\;,
\label{nund3}
\ee
where $c(\psi_{\star},H_{\star})$ and $I_s(\psi_{\star},H_{\star})$ are given
in (\ref{cdef}) and (\ref{curv}), respectively, and depend explicitly on
$H_{\star}$, $\psi_{\star}$, $\theta_1$ and $\theta_2$, but not on $d$. The
latter means that for these four variables the r.h.s in (\ref{nund3}) reaches
its unique value which implies the relationship (\ref{nund2}).
\subsection{Catenoids ($c=0$)}\label{r41}
In contrast with the ${\sf Nod}$ and ${\sf Und}$ menisci the ${\sf Cat}$
meniscus exists only for fixed $\psi$ values for which $H=0$, i.e., $\alpha=c=
0$. We make use of (\ref{und_curv_general}) in the limit $c\to 0$ in the form
(\ref{curv_sphere_general}) for $s=-1$ where we set the l.h.s. to zero to find
a relation
\be
2n-(1-\cos\pi n)\cos\theta_2 =0,\label{catenoid_general}
\ee
from which it follows that the catenoid can be observed only for $n=0$, i.e.,
at the point of the transition ${\sf Und_0^-}\leftrightarrow {\sf Nod^-}$. Using
the relation (\ref{I2_asympt}) in (\ref{und_curv_general}) for $s=-1$, $n=0$ and
approximation $c=-4H\sin\psi\sin t_1$ valid for small $H$ in (\ref{cdef}), we
find
$$
2H\Psi=-2H M\sin\psi\sin t_1.
$$
This leads to
\be
1+d-\cos\psi+\sin\psi\sin(\theta_1+\psi)\ln\left(\tan\frac{\theta_1+\psi}{2}
\tan\frac{\theta_2}{2}\right)=0\;,\label{b3}
\ee
which can also be derived (see (\ref{catenoid_psi})) from the YL equation for
$H=0$. Solutions to the above equation exist only for $\theta_1+\psi <\pi$.
This condition implies that the logarithmic term in (\ref{b3}) is negative which
leads to a stronger condition $\theta_1+\theta_2+\psi<\pi$. In the special case
of ideal plane wetting $\theta_2=0$ this term diverges and the equation
(\ref{b3}) has no solutions, so that the catenoids are forbidden.

Rewrite the equation (\ref{b3}) in the form
\be
g(\theta_1,\psi,d)=\frac{d}{\sin(\theta_1+\psi)\sin\psi}+\frac{\tan(\psi/2)}
{\sin(\theta_1+\psi)}+\ln\tan\frac{\theta_1+\psi}{2}=\ln\cot\frac{\theta_2}{2}
\;.\label{b31}
\ee
For $d=0$ and fixed values of the contact angles $g(\theta_1,\psi,0)$
monotonically grows and tends to asymptote at $\psi=\pi-\theta_1$ that implies
existence of a single solution of equation (\ref{b3}). This solution exists
when the minimal value of $g(\theta_1,0,0)=\ln\tan(\theta_1/2)$ is smaller than
the r.h.s. of (\ref{b31}), that leads to condition $\theta_1+\theta_2<\pi$. For
nonzero $d$ we note that the function $g$ has an additional asymptote at $\psi
=0$. When $d<0$ the monotonic behavior of the function $g(\theta_1,\psi,d)$ does
not change, so that we still have only one solution to equation (\ref{b3}).
When $d>0$ the function $g(\theta_1,\psi,d)$ has a minimum $g_{min}$ so that
depending on this value compared to $\ln\cot(\theta_2/2)$ equation (\ref{b3})
may have two, one or no solutions.

\begin{figure}[h!]\begin{tabular}{cc}
\psfig{figure=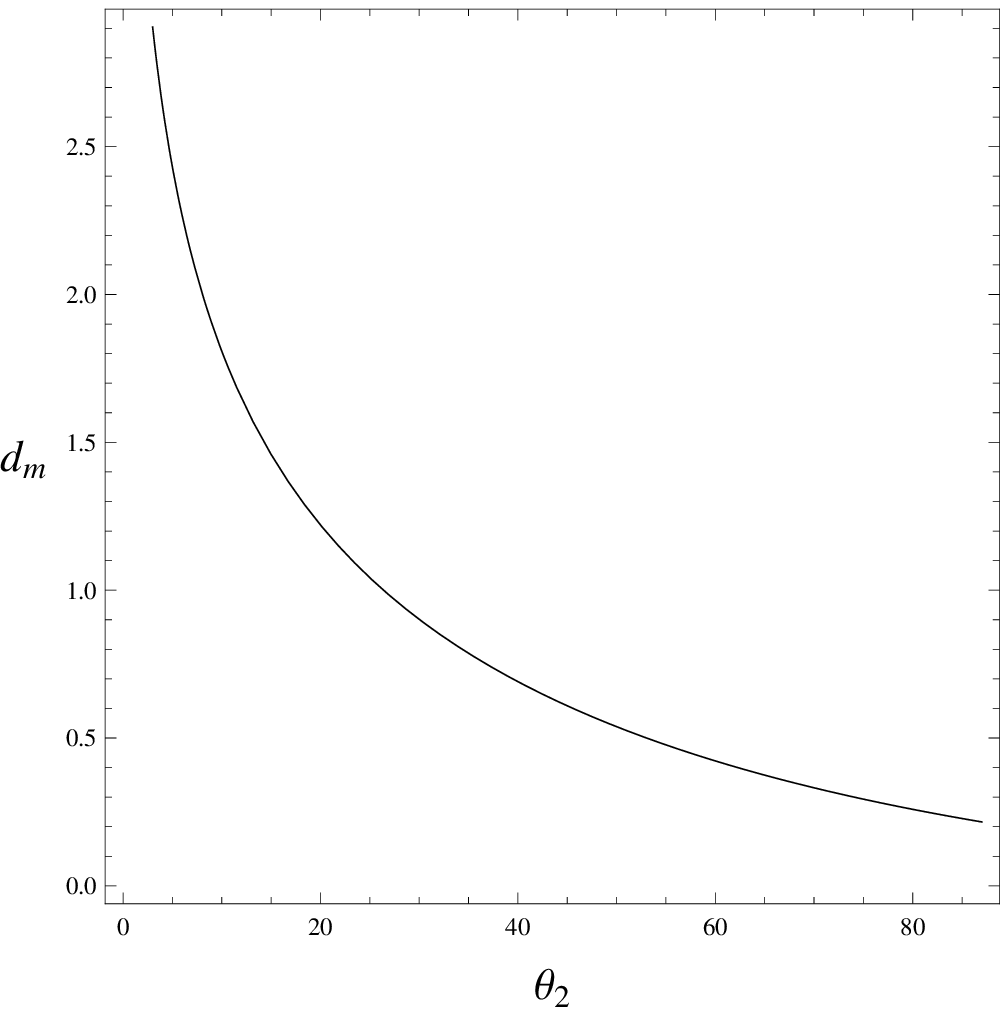,width=7cm} &
\psfig{figure=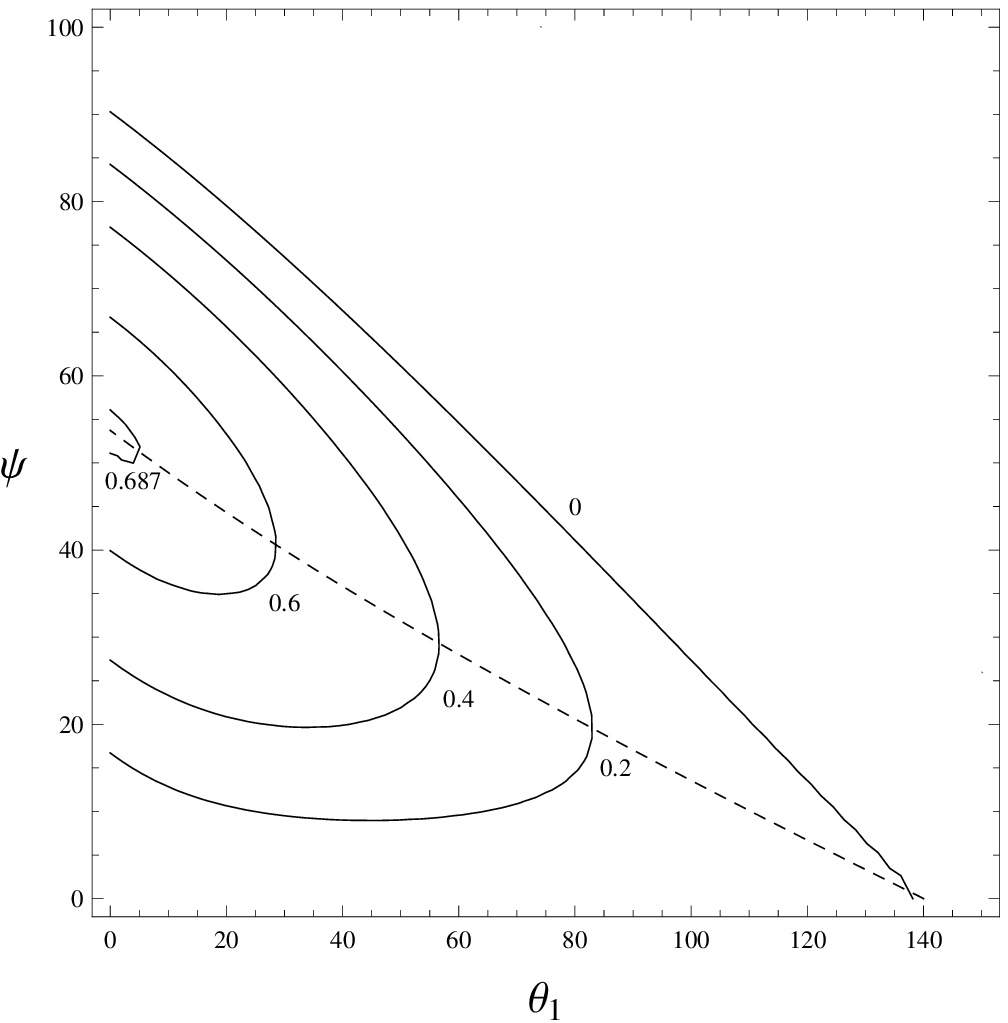,width=7cm}\\
(a) & (b)\end{tabular}
\caption{(a) The dependence of the maximal distance $d_m$ from the sphere to
the plane on the contact angle $\theta_2$ for $\theta_1=0$. (b) The solutions
of the equation (\ref{b3}) for $\theta_2=40^o$ for different values of the
distance $d$ from the sphere to the plane. The dashed curve shows the position
of the degenerate ${\sf Cat}$ menisci.}\label{Catenoid_fig}
\end{figure}

Find the maximal value $d_m$ of the distance $d$ as a function of $\psi$ and
$\theta_1$ for which equation (\ref{b3}) has a single solution. This value is
reached when two following conditions are met: $\partial d/\partial\psi =0$ and
$\partial d/\partial\theta_1 =0$; these conditions lead to $\theta_1 =0$ and we
obtain $d_m=(1-\cos\psi_m)/\cos\psi_m$, where $\psi_m$ satisfies the relation
$$
1+\cos\psi_m\ln\tan\frac{\psi_m}{2}\tan\frac{\theta_2}{2}=0\;,
$$
that leads to
\be
2(1+d_m)+\ln\frac{d_m}{d_m+2}=2\ln\cot\frac{\theta_2}{2}\;.\label{psi_dm}
\ee
The numerical solution of equation (\ref{psi_dm}) for $d_m(\theta_2)$ is shown
in Figure \ref{Catenoid_fig}(a).

\begin{figure}[h!]\begin{tabular}{cc}
\psfig{figure=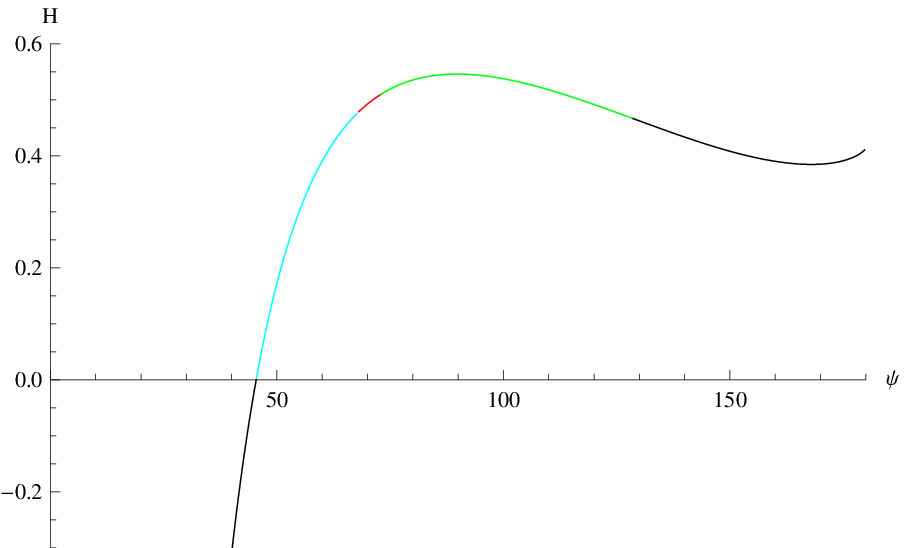,height=4.8cm} &
\psfig{figure=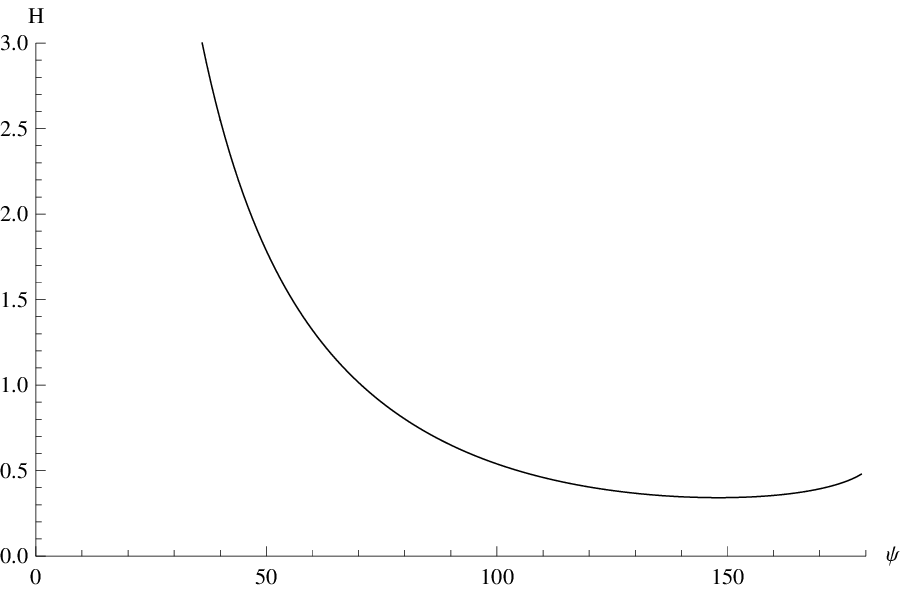,height=4.8cm}\\
(a) & (b) \\
\psfig{figure=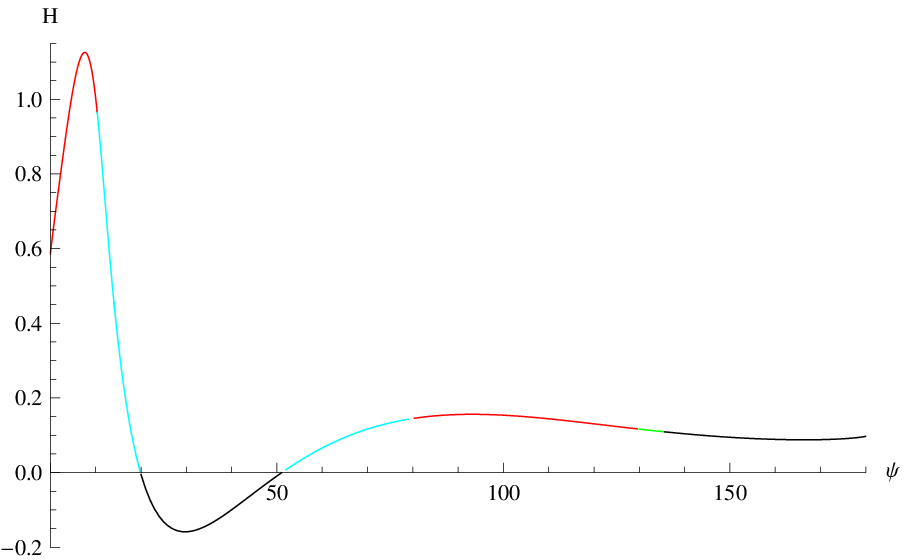,height=4.8cm} &
\psfig{figure=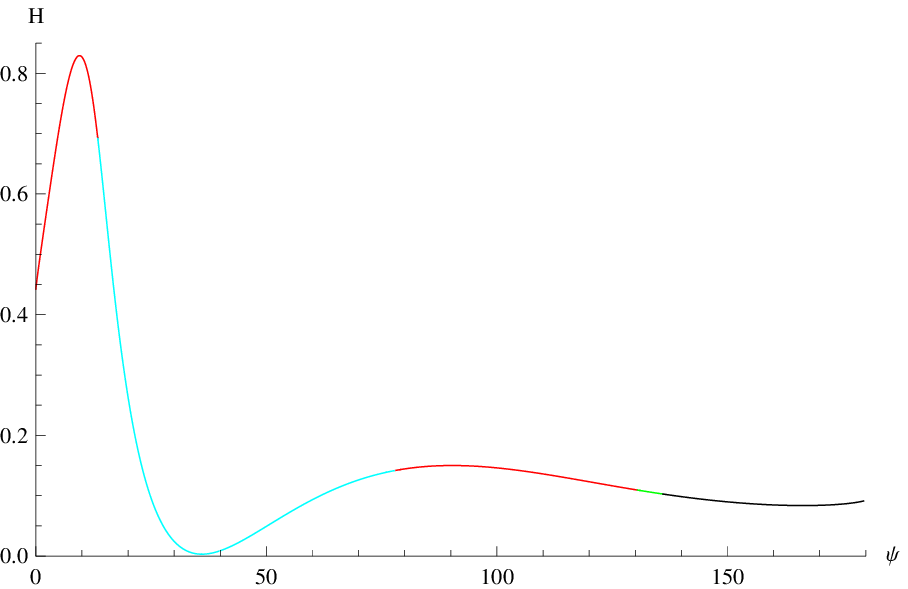,height=4.8cm}\\
(c) & (d)\end{tabular}
\caption{The case (a) presents the typical curve at $d=0$ with $\theta_1=30^o$,
$\theta_2=80^o$, discussed in \cite{Orr1975}, while the case (b) at $d=0$ with
$\theta_1=120^o$, $\theta_2=90^o$, characterized by a requirement $\theta_1+
\theta_2>\pi$, was not considered in \cite{Orr1975}. For positive distance
$d>0$ one observes (c) with $d=0.4$ with $\theta_1=\theta_2=40^o$ two catenoids
that degenerate for larger distance into a single catenoid (d) with $d=0.53$.
Different colors indicate the menisci of different types: ${\sf Und^-_1}$ ({\em
red}), ${\sf Und^-_0}$ ({\em cyan}), ${\sf Und^+_0}$ ({\em green}), ${\sf
Nod^-}$ and ${\sf Nod^+}$ ({\em black}). They correspond to colors of types
shown in Figure \ref{UnduloidTransitionFigure}.}\label{qi8}
\end{figure}

There exists a single nonzero value of the filling angle $\psi$ satisfying
equation (\ref{b3}) at $d=0$ that leads to the Plateau sequence (see Figure
\ref{qi8}(a)). Catenoid cannot be observed for $d>d_m$, but there exists a
range of distances $0<d<d_m$ for which equation (\ref{b3}) is satisfied for
two values of the filling angle as shown in Figure \ref{qi8}(c). The
corresponding sequence of menisci looks like
\bea
{\sf Und_0^-}\rightarrow {\sf Cat}\rightarrow {\sf Nod^-}\rightarrow
{\sf Cat}\rightarrow {\sf Und_0^-}\rightarrow {\sf Und_1^-}\rightarrow
{\sf Und_0^+}\rightarrow {\sf Sph^+_0}\rightarrow {\sf Nod^+}\;.
\nonumber
\eea
The sequence of menisci can pass through a single catenoid but it differs from
the Plateau type as shown in Figure \ref{qi8}(d). This occurs in the
degenerated case when equation (\ref{b3}) has to be complemented by an additional
requirement: $\partial\theta_1/\partial\psi =0$. Taking derivatives with
respect to $\psi$ in both sides of equation (\ref{b3}) and applying the above
condition we arrive at a trigonometric equation
\be
2\sin\psi=\left(\ln\cot\frac{\theta_1+\psi}{2}+\ln\cot\frac{\theta_2}{2}\right)
\sin(\theta_1+2\psi)\;.\label{b5}
\ee
Solution of equation (\ref{b5}) for fixed $\theta_2$ determines a curve
$\theta_1(\psi)$ on which the degenerated ${\sf Cat}$ meniscus appears. It is
shown in Figure \ref{Catenoid_fig}(b) by dashed curve.

\subsection{Unduloids ($c<0$)}\label{r51}
\begin{figure}[h!]\begin{tabular}{cc}
\psfig{figure=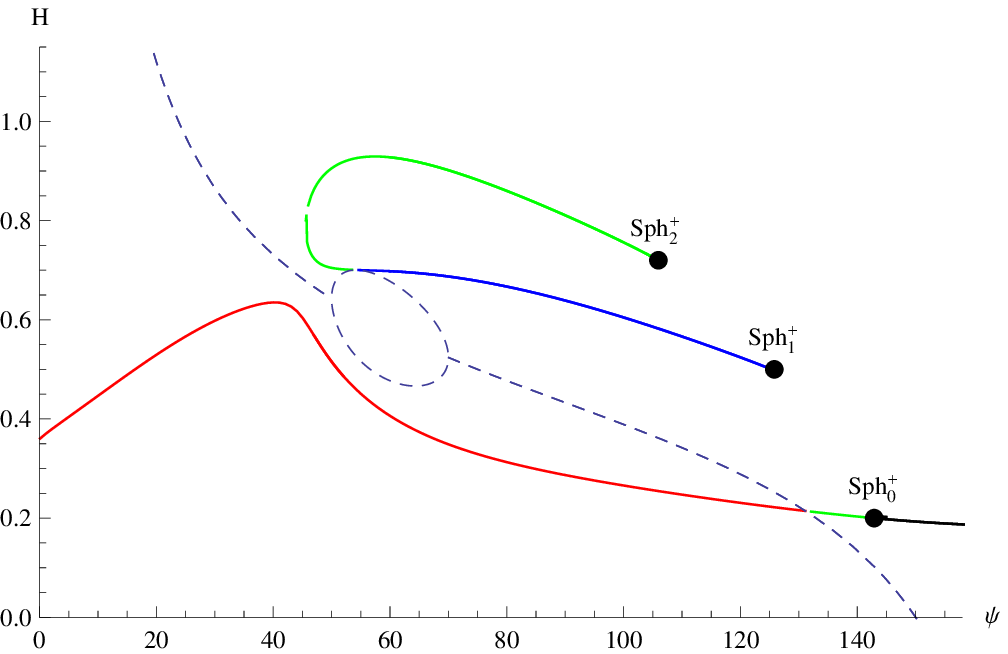,height=5cm}&
\psfig{figure=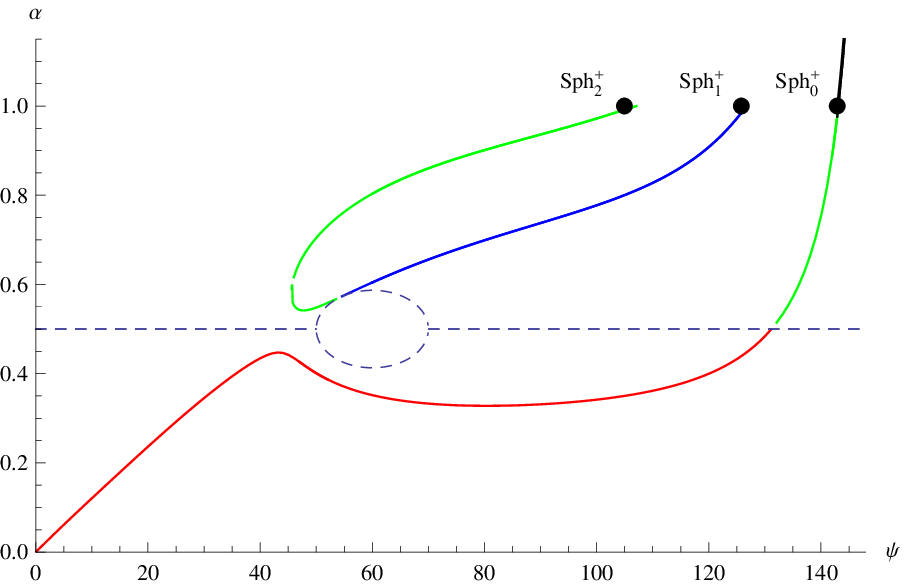,height=5cm}\\
(a) & (b)\end{tabular}
\caption{The dependencies $H_n(\psi)$ in (a) and $\alpha_n(\psi)$ in (b) for
$\theta_1=30^o$, $\theta_2=80^o$ and $d=2.3$ with two disjoined curves. The
balloon ${\mathbb B}$ is shown by dashed line. Spherical menisci ${\sf
Sph^+_0}$, ${\sf Sph^+_1}$ and ${\sf Sph^+_2}$ are denoted by black dots. The
curve segments correspond to colors of the menisci types shown in Figure
\ref{UnduloidTransitionFigure}.}
\label{qu1}
\end{figure}

Before starting to treat the unduloidal solution of equation (\ref{curv}) it is
worth to make
\begin{remark}\label{rem1}
Equation (\ref{YLeq}) with boundary conditions is associated with nonlinear
eigenvalue problem and in the case of unduloids its solution has a discrete
spectrum and, therefore, is enumerated by two indices. The first integer
non-negative index $n$ determines the number of the inflection points on the
meniscus meridional profile. When the part of this profile touching the solid
sphere is convex the second integer index $s$ takes value of $1$, otherwise it
equals to $-1$. Thus, the unduloid meniscus is denoted as ${\sf Und_n^s}$.
\end{remark}
The existence of the ${\sf Und_n^s}$ meniscus requires satisfaction of the
condition $\sin^2t+c\ge 0$ for $t\in\{t_2,t_1\}$. This condition is rewritten
in the form
\be
\sin^2 t\ge [1-(1-2\alpha)^2]\sin^2 t_1\;.\label{exist_cond}
\ee
Denote $\sin t_m=\min(\sin t_1,\sin t_2)$. The last condition leads to $\sin^2
t_m\ge [1-(1-2\alpha)^2]\sin^2 t_1$, and yields
\be
\alpha\le\beta^-(\psi),\quad\alpha\ge\beta^+(\psi),\quad\mbox{where}\quad
\beta^{\pm}(\psi)=\frac{1}{2}\left(1\pm\sqrt{1-\sin^2 t_m/\sin^2 t_1}\right).
\label{alpha_relation_und}
\ee
The restrictions (\ref{alpha_relation_und}) define in the plane $\{\psi,\alpha
\}$ a new object ${\mathbb B}$ which we called {\em a balloon}. It comprises
an oval and two additional lines $\alpha=1/2$ (see detailed description in
section \ref{rr10}). Hereafter the balloon ${\mathbb B}$ becomes a main tool of
our study the dependence $\alpha(\psi)$. In regard to $H(\psi)$ the balloon
undergoes the non linear transformation (\ref{alpha_def}). In Figure \ref{qu1}
we present both dependencies $H_n(\psi)$ and $\alpha_n(\psi)$ with the balloon
${\mathbb B}$.

\begin{figure}[h!]\begin{center}\begin{tabular}{ccc}
\psfig{figure=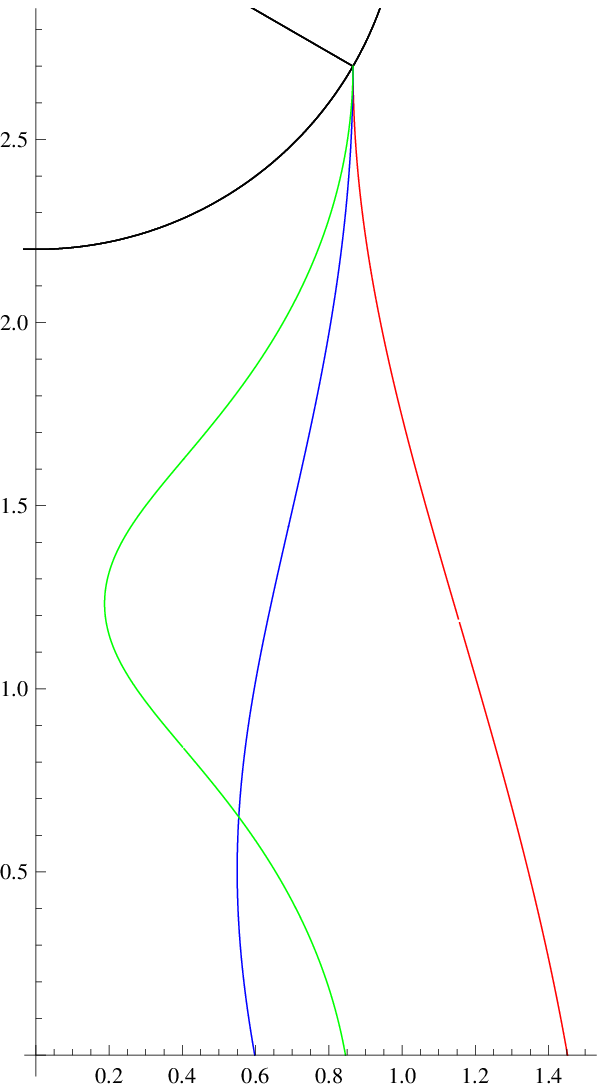,height=5.7cm}
\hspace{1.5cm}&
\hspace{1.5cm}\psfig{figure=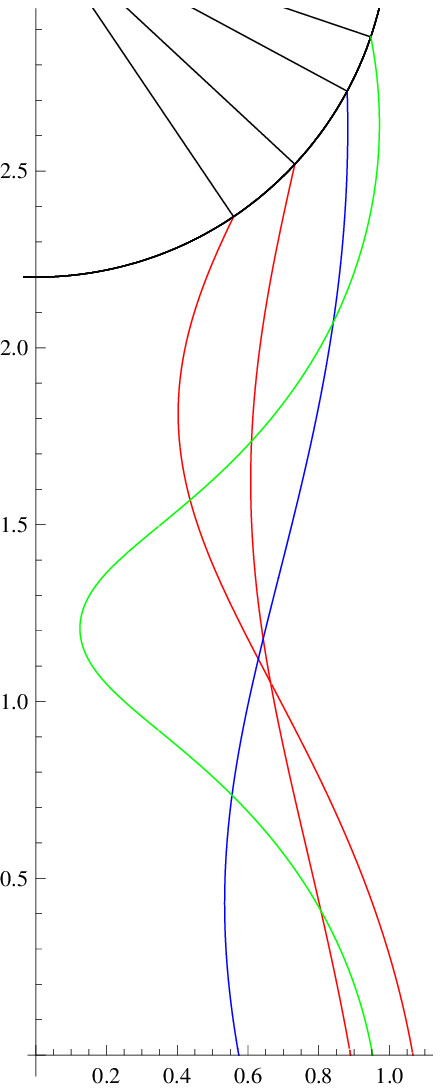,height=5.7cm}
\hspace{1.5cm}&
\hspace{1.5cm}\psfig{figure=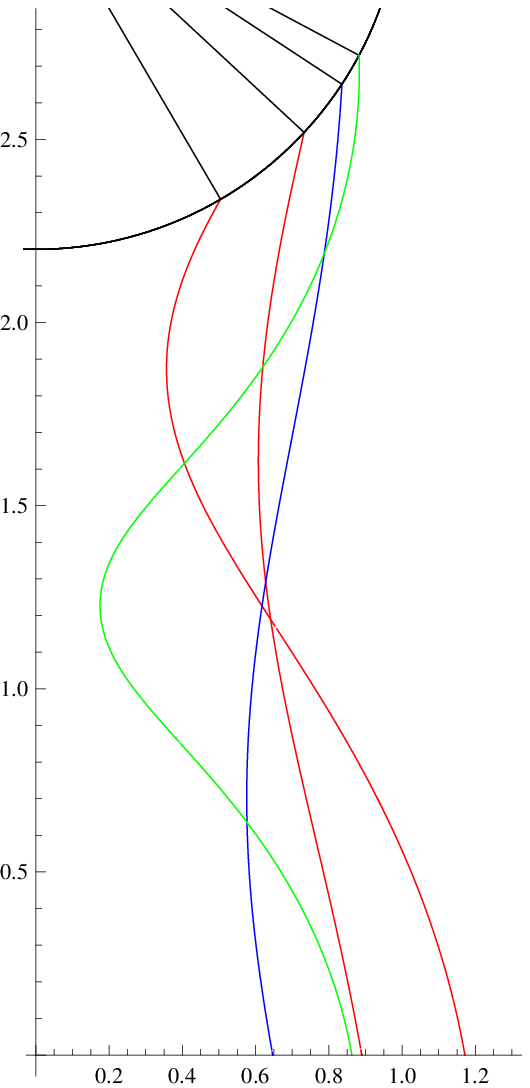,height=5.7cm}\\
(a) & (b) & (c)\end{tabular}
\end{center}
\caption{The menisci ${\sf Und^-_1}$ ({\em red}), ${\sf Und^+_1}$
({\em green}), and ${\sf Und^+_2}$ ({\em blue}) for $\theta_1=30^o$,
$\theta_2=80^o$ and $d=2.2$ computed in three different setups: (a) for
$\psi=60^o$ they have different volumes $V$ and surface areas $S$; (b) for
fixed $V=3.6$ there are 4 menisci with $\psi_1=34^o$ and $\psi_2=47.2^o$ --
${\sf Und^-_1}$, $\psi_3=61.7^o$ -- ${\sf Und^+_1}$, $\psi_4=71.3^o$ -- ${\sf
Und^+_2}$ and different surface areas $S$; (c) for fixed $S=11.3$ there are 4
profiles with $\psi_1=30.3^o$ and $\psi_2=47.1^o$ -- ${\sf Und^-_1}$,
$\psi_3=56.7^o$ -- ${\sf Und^+_1}$, $\psi_4=61.95^o$ -- ${\sf Und^+_2}$ and
different volumes $V$.}\label{qu11}
\end{figure}
The relations (\ref{alpha_relation_und}) are independent of distance $d$ and
remain valid for all types of inflectional unduloids with $0<\alpha<1$. For
all unduloids they define the restrictions on the curvature values that can be
obtained using the definition (\ref{alpha_def}). On the other hand, these
relations correspond to a single condition on $c$ value
\be
-\sin^2 t_m \le c \le 0.\label{c_relation_und}
\ee
In Figure \ref{qu11} we present three families of unduloid menisci for fixed 
filling angle $\psi$ (Figure \ref{qu11}(a)), volume $V$ (Figure \ref{qu11}(b)) 
and surface area $S$ (Figure \ref{qu11}(c)) of pendular ring.
\subsection{Nodoids ($c>0$)}\label{r52}
The ${\sf Nod}$ meniscus has positive $c$ irrespectively to its concave or
convex version. Therefore the above analysis of $\alpha$ fails and it is
replaced by another one, less strong but still universal.

Show that for $d>0$ the ${\sf Nod^s}$ menisci satisfy the following
constraints,
\bea
H(\psi,0)<H(\psi,d)<0\quad ({\sf Nod^-}),\quad\quad 0<H(\psi,d)<H(\psi,0)\quad
({\sf Nod^+})\;,\label{nund1}
\eea
using (\ref{nund2}) and several additional observations listed below.

First, the curvature $H$ for any finite $d$ is always negative for the ${\sf
Nod^-}$ meniscus and always positive for the ${\sf Nod^+}$ meniscus. Next, the
${\sf Nod^-}$ meniscus disappears at finite $d$ when two catenoids annihilate.
Finally, when $\psi\to\pi$ and $d\to\infty$ the curvature of the ${\sf Nod^+}$
meniscus tends to zero. Combining these facts with (\ref{nund2}) we arrive at
constraints (\ref{nund1}).

The main statement stemming from (\ref{nund1}) is that the two different curves
$H(\psi,d_1)$ and $H(\psi,d_2)$, $d_1\neq d_2$, do not intersect in the entire
angular $\psi$ range of the ${\sf Nod^-}$ and ${\sf Nod^+}$ menisci existence.
Although this statement is of high (topological) importance, it does not
provide the quantitative estimates. Therefore for the ${\sf Nod^-}$ meniscus we
give one more estimate for the curvature.

Start with relationship between the curvature $H$ and the surface area $S$ for
${\sf Nod^-}$ meniscus. For this purpose combine formulas (\ref{curv},
\ref{surf}, \ref{I_1}, \ref{I_2_def}, \ref{K}) and obtain
\bea
\frac{2H^2}{\pi}S=K_-=2(I_1-I_2)-I_3=4H\Psi -I_3\;.
\label{nund4}
\eea
Keeping in mind the positiveness of $S$ we arrive at the bound $4H\Psi\geq I_3$.
Using the definition (\ref{I_3}) we obtain $I_3\geq -\delta\sqrt{c}$, where
$\delta=\pi-\theta_1-\theta_2-\psi$, $\delta\geq 0$. Thus, substituting this
relation in the above inequality we find
\bea
H\geq -\frac{\delta}{4\Psi}\sqrt{c}\;.
\label{nund5}
\eea
This bound does not contradict inequalities (\ref{nund1}) for the ${\sf Nod^-}$
meniscus since its curvature is negative. Keeping in mind this fact and
substituting (\ref{cdef}) into (\ref{nund5}) we arrive at
\be
H\geq\frac{\sin\psi\;\sin(\theta_1+\psi)}{\sin^2\psi-4\Psi^2\delta^{-2}}\;.
\label{nund6}
\ee
This inequality is equivalent to
\be
\alpha\geq\frac{\delta^2\sin^2\psi}{\delta^2\sin^2\psi-4\Psi^2}\;.
\label{nund6alpha}
\ee
The last inequalities have one important consequence. Since $H<0$ then the
necessary condition for the ${\sf Nod^-}$ meniscus existence is
\be
\delta\sin\psi <2\Psi\Rightarrow\frac{\pi-\theta_1-\theta_2-\psi}{2}\sin\psi+
\cos\psi<1+d.\label{nund7}
\ee
For convex nodoid ${\sf Nod^+}$ meniscus it is possible to find explicit
expression for the upper bound $H(\psi,0)$. First, note that as it is
demonstrated in Appendix \ref{appendix5} meniscus curvature has a local
minimum, so that the upper bound is given by the largest of two curvature
values -- $H(\phi_0^+)$ at the sphere ${\sf Sph_0^+}$ and $H(\pi)$ at $\psi=
\pi$. These values are
$$
H(\phi_0^+)=\frac{\sin(\theta_1+\phi_0^+)}{\sin\phi_0^+},\quad
H(\pi)=\frac{1-\cos\theta_2}{2+d}.
$$
Show that $H(\phi_0^+)>H(\pi)$. Indeed, this condition implies
$$
(1+d)\sin(\theta_1+\phi_0^+)+\cos\theta_2\sin\phi_0^+>\sin\phi_0^+-\sin(\theta_1
+\phi_0^+).
$$
Using in the above relation the l.h.s. of formula
(\ref{psi_cond_sphere_general}) we obtain
$$
\sin\theta_1 >\sin\phi_0^+-\sin(\theta_1+\phi_0^+),
$$
leading to
\be
\cos\frac{\theta_1}{2}+\cos\left(\phi_0^++\frac{\theta_1}{2}\right)=\cos\frac{
\theta_1+\phi_0^+}{2}\cos\frac{\phi_0^+}{2}>0.\label{sphere1_cond1}
\ee
Recalling that $\theta_1+\phi_0^+ <\pi$ we see that (\ref{sphere1_cond1}) is
always valid. Thus, the upper bound for the ${\sf Nod^+}$ meniscus is given by
$H(\phi_0^+,0)$. To find this value explicitly we use
(\ref{psi_cond_sphere_general}) with $n=0$ and $d=0$ to produce
$$
\sin(\theta_1+\phi_0^+)+\cos\theta_2\sin\phi_0^+-\sin\theta_1=0\Rightarrow\cos
\left(\theta_1+\frac{\phi_0^+}{2}\right)+\cos\theta_2\cos\frac{\phi_0^+}{2}=0.
$$
The last relation implies
\be
\tan\frac{\phi_0^+}{2} =\frac{\cos\theta_1+\cos\theta_2}{\sin\theta_1}.
\label{psi_s_d0}
\ee
On the other hand, we have
$$
H(\phi_0^+,0)=\frac{\sin(\theta_1+\phi_0^+)}{\sin\phi_0^+}=\cos\theta_1+\sin
\theta_1\cot\phi_0^+.
$$
Using (\ref{psi_s_d0}) in last expression we find for the upper bound of
${\sf Nod^+}$ meniscus
\be
H(\phi_0^+,0)=\frac{\sin^2\theta_2}{2(\cos\theta_1+\cos\theta_2)}.
\label{Nod_+_upperbound}
\ee
\subsection{Spheres ($c=0$)}\label{r40}
The spheres can be considered as a limiting case of the unduloid menisci and,
therefore, also labeled by two indices ${\sf Sph^s_n}$. Consider a function
\be
f_n(\psi) =(1+d)\sin(\theta_1+\psi)-(n-\delta)\sin\psi-\sin\theta_1,
\label{funct_psi_n}
\ee
which root $\psi=\phi_n^+$ satisfying the equation $f_n(\phi_n^+)=0$
provides the value of the filling angle at which sphere ${\sf Sph^+_n}$ is
observed (see (\ref{psi_cond_sphere_general})). In (\ref{funct_psi_n}) one has
$\delta=0$ for odd $n$ and $\delta=\cos\theta_2$ for even $n$. It is easy to
check that
$$
f_{n+1}(\psi)-f_n(\psi)=-[1+\cos \pi n \cos\theta_2]\sin\psi <0,
$$
which means that the curve $f_{n+1}$ lies {\it below} the curve $f_n$.

The value of the function $f_n$ at $\psi=0$ reads $f_n(0)=d\sin\theta_1$ which
is non-negative, while $f_n(\pi)=-(2+d)\sin\theta_1$ is always negative. Noting
that the function $f_n$ is a periodic one with the period $2\pi$ we find that
for positive $d$ in the interval $\{0,\pi\}$ the function $f_n$ has a single
root $\phi_n^+$ and $f_n'(\phi_n^+)< 0$. As $f_{n+1}(\phi_n^+)<f_n(\phi_n^+)=0$
and $f_{n+1}'(\phi_{n+1}^+)<0$ we immediately find that $\phi_{n+1}^+<\phi_n^+$.
This means that for positive $d$ the value $\psi_n$ of the filling angle at
which sphere ${\sf Sph^+_n}$ is observed decreases with increase of $n$. For
very large $n$ the value of $\phi_n^+$ tends to zero. Using (\ref{funct_psi_n})
we find in linear approximation
\be
\sin\phi_n^+=\phi_n^+=\frac{d\sin\theta_1}{n-\delta-\cos\theta_1}\approx\frac{
d\sin\theta_1}{n}, \ \ \ n \gg 1.\label{psi_n_approx_large_n}
\ee
For $d\le 0$ we have $f_n(0)\le 0$ and the first derivative reads $f'_n(0)=
\cos\theta_1-(n-\delta)$. For $n>0$ this derivative is negative, so that the
function $f_n$ has no roots. When $n=0$ we have $f'_0(0)=\cos\theta_1+\cos
\theta_2$. This expression is negative for $\theta_1+\theta_2>\pi$, so that no
spheres exist for $d=0$ when the last condition holds. Finally, in case of the
wetting sphere $\theta_1=0$ we have $f_n(0)=f_n(\pi)=0$ and no spheres are
allowed to exist.

Show that for fixed $n$ the value $\phi_n^+(d)$ of the filling angle of sphere
${\sf Sph^+_n}$ increases with growing $d$. Indeed, from (\ref{funct_psi_n})
one finds that the difference
$$
f_n(\phi_n^+(d),d_1)-f_n(\phi_n^+(d),d)=(d_1-d)\sin(\theta_1+\phi_n^+(d)),
$$
is positive for $d_1 > d$. As $f_n'(\phi_n^+)<0$ it immediately follows that
$\phi_n^+(d_1) >\phi_n^+(d)$.

Another type of sphere ${\sf Sph^-_n}\; (n>0)$ discussed in \ref{ar15} can be
considered as a limiting shape of the ${\sf Und^-_n}$ meniscus at $\psi\to 0$,
its curvature is given by (\ref{curv_sphere_neg}).

In Table 1 the characteristic signs of $H$, $c$ are given for different
types of menisci.
\begin{center}
{\bf Table$\;$1}.\\
\vspace{-.5cm}
$$
\begin{array}{|c|c|c|c|c|c|c|c|}\hline
 & {\sf Nod^-} & {\sf Cat} & {\sf Und_n^s} & {\sf Cyl} &
 {\sf Sph_n^s} & {\sf Nod^+} \\\hline\hline
H         & - & 0 & +  & + & + & +  \\\hline
c         & + & 0 & -  & - & 0 & + \\\hline
\end{array}
$$
\label{ta1}
\end{center}
Note that the meniscus ${\sf Und_n^s}$ comes in several different types
depending on the number $n$ of inflection points and the curvature of the
segment of meridional profile is touching the sphere: $s=-1$ for concave
profile and $s=+1$ for convex one.
\section{Unduloid Menisci Transitions}\label{r45}
\begin{figure}[h!]\begin{center}
\psfig{figure=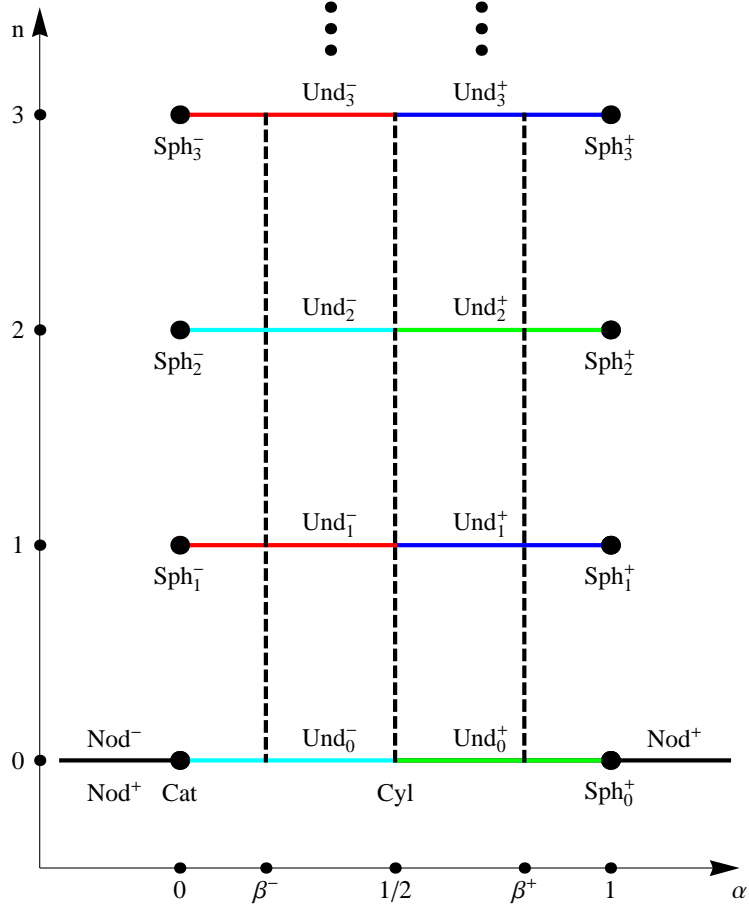,width=10cm}
\end{center}
\caption{Possible transitions between menisci of different types shown as a
function of the parameter $\alpha$. The left vertical dashed line represents
the transitions at $\alpha=\beta^-$ between ${\sf Und_n^-}$ unduloids.
Similarly, the right vertical dashed line corresponds to the transitions at
$\alpha=\beta^+$ between ${\sf Und_n^+}$ unduloids. Finally, the transitions
between unduloids of opposite signs takes place at $\alpha=1/2$ and is
represented by the central vertical dashed line.}
\label{UnduloidTransitionFigure}
\end{figure}

Transitions between unduloids and nodoids of different types can be easily
classified. Namely, there exist transitions between the concave ${\sf Und_0^-}$
(convex ${\sf Und_0^+}$) unduloid to concave ${\sf Nod^-}$ (convex ${\sf
Nod^+}$) nodoid through catenoid ${\sf Cat}$ (sphere ${\sf Sph^+_0}$). The
transitions between different types of unduloids are numerous and schematically
presented in Figure \ref{UnduloidTransitionFigure}.
As shown in \ref{ar4even} and \ref{ar4odd} all inflection points for a given
inflectional unduloid ${\sf Und_n^s}$ have the same abscissa value $x_{\ast}=
\sin t_{\ast}^s/(2H)=\sqrt{-c}/(2H)$. An addition (removal) of an inflection
point may take place only as the result of the inflection point separation from
(merging with) the sphere or the plane. When the inflection point is on the
sphere we have $c+\sin^2 t_1 =0$ to obtain using (\ref{inflect_sphere})
$$
2H\sin\psi=\sin t_1.
$$
Recalling definition (\ref{alpha_def}) we find $\alpha=1/2$. Noting that
menisci ${\sf Und_n^-}$ (${\sf Und_n^+}$) exist at $\alpha\le 1/2$ ($\alpha\ge
1/2$) we find that transitions of the type ${\sf Und_n^s}\leftrightarrow {\sf
Und_{n\pm1}^{-s}}$ occur at $\alpha=1/2$ when the inflection point is on the
sphere.
For the inflection point on the plane we have $c+\sin^2 t_2=0$ to obtain with
(\ref{inflect_plane})
$$
2H\sin\psi=\sin t_1 +s\sqrt{\sin^2 t_1-\sin^2 t_2}\;,
$$
and we find using (\ref{alpha_relation_und}) that in this case the transitions
of the type ${\sf Und_n^s}\leftrightarrow {\sf Und_{n\pm 1}^s}$ occur at
$\alpha=\beta^s$.

Menisci of the ${\sf Und_n^+}$ type exist for $1/2<\alpha< 1$. With growth of
$\alpha$ the abscissa of the leftmost point of the meniscus meridional section
tends to zero and reaches it at $\alpha=1$. At this moment the profile is made
of several segments of a circle and the meniscus touches the axis of rotation
$x=0$. 
The spherical menisci ${\sf Sph_n^+}$ at $\alpha=1$ 
are described in \ref{appendix2}.

The above considerations lead to the following rules of transition between
unduloids:
$$
\begin{array}{ccc}
{\sf Und_n^s}\longleftrightarrow {\sf Und_{n\pm1}^s}&\mbox{at}&
\alpha=\beta^s,\\
{\sf Und_n^s}\longleftrightarrow {\sf Und_{n\pm1}^{-s}}&\mbox{at}&\alpha=1/2.
\end{array}
$$
Thus we show that the balloon introduced above can be viewed as a
set of transition points between unduloids.

Show that for fixed values of $\theta_1,\theta_2$ and $d$ the curves
corresponding to different unduloid types never intersect (except for the
transition points discussed above that arises in case when the unduloid orders
differ by unity). As the curves corresponding to unduloids of opposite signs
cannot intersect in $\{\psi,\alpha\}$ plane, we have to consider only unduloids
of the same sign.

Consider first same sign unduloids of the orders that differ by an even number,
for example, ${\sf Und_{2k}^s}$ and ${\sf Und_{2k'}^s}$, or ${\sf Und_{2k+1}
^s}$ and ${\sf Und_{2k'+1}^s}$, where $k\ne k'$. It immediately follows from
(\ref{curv_infla2k}) and (\ref{curv_infl2kp1}) that the difference between the
curvatures of these menisci is $(k-k')\hat I_2/\Psi\ne 0$.

If the order difference is odd and larger than two we have for unduloids ${\sf
Und_{2k}^s}$ and ${\sf Und_{2k'+1}^s}$ with $k'\ne k$, the curvature difference
reads $[(k'-k)\hat I_2+I_2(t_{\ast}^s,t_2)]/\Psi$. Consider the integrals $I_2(t_{
\ast}^s,t_2)$ which are finite and real. Note that $t_{\ast}^+<t_2<t_{\ast}^-$,
which leads to
$$
|I_2(t_{\ast}^s,t_2)|<I_2(t_{\ast}^-,t_{\ast}^+)=\hat I_2.
$$
The last relation implies that the curvature differences mentioned above are
nonzero that finishes the proof.

From (\ref{curv_sphere_general}) it follows that the curvature of spherical
menisci ${\sf Sph^+_n}$ grows monotonically with order increase without any
restriction to the order value. This means also that the corresponding
unduloids ${\sf Und_n^+}$ can be observed without any restrictions to the
order $n$.

The asymptotic behavior of the curvature at small filling angles $\psi\ll 1$ is
discussed in Appendix \ref{appendix2}. For $d> 0$ at zero filling angle $\psi=0$
only concave unduloids ${\sf Und_n^-}\;(n>0)$ exist. From
(\ref{curv_sphere_neg}) it follows that for $n=0$ the curvature $H(0)=0$ leading
to contradiction as the curvature can turn to zero only for catenoid. It is
important to underline that there are no other restrictions to existence of
${\sf Und_n^-}\;(n>0)$ unduloids at zero filling angle.

Using the formula (\ref{und_curv_general}) for unduloid curvature we have
\be
2H_n^s\Psi=I_1(t_1,t_2)+s I_2(t_1,t_2)+n\hat I_2-s(1-\cos\pi n)I_2(\pi/2,t_2),
\quad s=\mbox{sgn}(2\alpha_n-1)\;,\label{curv_und_pm}
\ee
where the sign $s=\pm 1$ corresponds to two menisci which differ by the sign of
the meridional curvature at the meniscus-sphere contact point. Equation
(\ref{curv_und_pm}) defines the function $\alpha_n(\psi)$ in two different
regions, $\alpha_n>1/2$ and $\alpha_n<1/2$, while $\alpha_n(\psi)$ is a smooth
at $\alpha_n=1/2$.

\noindent
Rewrite (\ref{curv_und_pm}) as follows, $\Phi_n^s(\alpha_n,\psi)=0$, where
\be
\Phi_n^s(\alpha_n,\psi)=-2\alpha_n\Psi\sin t_1+\left[I_1(t_1,t_2)+sI_2(t_1,t_2)
+n\hat I_2-s(1-\cos\pi n)I_2(\pi/2,t_2)\right]\sin\psi,\label{alpha_und_pm}
\ee
One can define the derivative $\alpha'_n(\psi)$ having a unique value determined
from the equation
\be
A_n^s\alpha'_n(\psi)+B_n^s=0, \ \ \ A_n^s=\frac{\partial\Phi_n^s}{\partial\alpha
_n}, \ \ \ B_n^s=\frac{\partial\Phi_n^s}{\partial\psi},\label{alpha_diff_eq}
\ee
where the both functions $A_n^s(\alpha_n,\psi)$ and $B_n^s(\alpha_n,\psi)$ do
not vanish simultaneously.

Direct computation gives the following expressions for $A_n^s(\alpha_n,\psi)$
and $B_n^s(\alpha_n,\psi)$
\bea
A_n^s(\alpha_n,\psi)&=&-2\Psi\sin t_1-\nonumber\\
&&2(2\alpha_n-1)\sin^2t_1\sin\psi\left[s I_4(t_1,t_2)-2n\hat I_2'(c)-s(1-\cos
\pi n)I_4(\pi/2,t_2)\right],\hspace{.8cm}\label{A_def}\\
B_n^s(\alpha_n,\psi)&=&2\alpha_n\Psi\frac{\sin\theta_1}{\sin\psi}-\frac{
4\alpha_n^2-2\alpha_n-1}{2\alpha_n-1}\sin t_1\sin\psi+\sin^2\psi-\nonumber\\
&&2\alpha_n(\alpha_n-1)\sin2t_1\sin\psi\left[sI_4(t_1,t_2)-2n\hat
I_2'(c)-s(1-\cos\pi n)I_4(\pi/2,t_2)\right],\label{B_def}
\eea
where the integral $I_4(t_1,t_2)$ is computed in (\ref{I_4}) and the derivative
$\hat I_2'(c)$ is given by (\ref{dI_2_hat/dc}).
\subsection{Transitions ${\sf Und_n^-}\leftrightarrow {\sf Und_{n+s}^+}$}
\label{alpha=1/2_left}
Consider first the transitions on the line $\alpha=1/2$ between unduloids of
opposite signs. This transition takes place when the point $t_1=t_{\ast}^s$
separates from the sphere, where $t_{\ast}^+=\arcsin\sqrt{-c}$ and $t_{\ast}^-
=\pi-\arcsin\sqrt{-c}$. As $H=\sin t_1/(2\sin\psi)$ we use (\ref{curv_und_pm})
to obtain
\be
\Psi\sqrt{-c}=\left[I_1(t_{\ast}^s,t_2)-I_2^{\ast}(t_{\ast}^s,t_2)+n\hat I_2+(1
-\cos\pi n)I_2(\pi/2,t_2)\right]\sin(t_{\ast}^s-\theta_1),
\label{curv_alpha=1/2_left}
\ee
where we introduce a special case of the integral $I_2$
\be
I_2^{\ast}(t_1,t_2)=\int_{t_2}^{t_1}\frac{\sin^2t\;dt}{\sqrt{\sin^2t-\sin^2
t_1}}.\label{I_2^star_def}
\ee
The general expressions for the abscissa of the meniscus meridional profile
contain the term $\sqrt{\sin^2 t+c}$, which at the transition point transforms
into $\sqrt{\sin^2 t-\sin^2 t_{\ast}^{\pm}}$. It leads to a condition $t_2<t_{
\ast}^-$ producing
$$
\begin{array}{lcll}
t_2 <\pi-t_1&\longrightarrow & \psi<\theta_2-\theta_1, & s=+1, \\
t_2 <t_1&\longrightarrow & \psi>\pi-\theta_2-\theta_1, & s=-1,
\end{array}
$$
implying that the transition takes place to the left (right) of the balloon for
$s=1$ ($s=-1$).

Show that the transition considered in this subsection is smooth, i.e.,
unduloids ${\sf Und_n^-}$ and ${\sf Und_{n+s}^+}$ meet smoothly at $\alpha=1/2$.
It means that the value of the derivative $\alpha'(\psi)$ computed on both sides
of the transition point is the same. As at the transition point we have $c+
\sin^2t_1=0$ the integral $I_4(t_1,t_2)$ in (\ref{A_def},\ref{B_def}) diverges.
We show below that nevertheless both $A$ and $B$ have finite value at the
transition point. In the vicinity of $\alpha=1/2$ we introduce $\alpha=1/2+s
\epsilon$ and find $c=(4\epsilon^2-1)\sin^2t_1$ and $\sqrt{c+\sin^2t_1}=2
\epsilon\sin t_1$. Making use of (\ref{I_4}) decompose the integral $I_4(t_1,
t_2)$ into diverging $I_{4d}(t_1,t_2)$ and non-diverging $I_{4c}(t_1,t_2)$
parts,
\be
I_4(t_1,t_2)=I_{4c}(t_1,t_2) + I_{4d}(t_1,t_2)\;,\quad
I_{4d}(t_1,t_2)=-\frac{1}{s(2\alpha-1)\cos t_1}\;.\label{new1}
\ee
Using this relation in (\ref{A_def}) we find
\be
A_n^s=-2\Psi\sin t_1-\frac{2\sin^2t_1\sin\psi}{\cos t_1}=2\tan t_1[\cos
\theta_1-(1+d)\cos t_1].\label{A_alpha=1/2}
\ee
Substitute (\ref{new1}) into (\ref{B_def}) and note that two diverging
terms cancel each other in vicinity of $\alpha=1/2$,
\bea
&-&\frac{4\alpha^2-2\alpha-1}{2\alpha-1}\sin t_1\sin\psi-2s\alpha(\alpha-1)
\sin2t_1\sin\psi I_{4d}(t_1,t_2)\nonumber\\
&=& \frac{1}{2\alpha-1}\sin t_1\sin\psi-\frac{1}{2(2\alpha-1)}\frac{\sin2t_1
\sin\psi}{\cos t_1}=0.\nonumber
\eea
The remaining terms read
\bea
B_n^s=\Psi\frac{\sin\theta_1}{\sin\psi}+\sin^2\psi+\sin t_1\cos t_1\sin\psi
\left[s I_{4c}(t_1,t_2)-2n\hat I_2'(c)-s(1-\cos\pi n)I_4(\pi/2,t_2)\right].
\nonumber
\eea
Show that this expression is conserved at the transition ${\sf Und_n^-}
\leftrightarrow {\sf Und_{n+s}^+}$. It is sufficient to show that it is valid
for the expression in the square brackets in (\ref{B_alpha=1/2}) that leads to
the relation
$$
I_{4c}(t_1,t_2)-I_4(\pi/2,t_2)=s_1\hat I_2'(c),
$$
where $s_1=1$ ($s_1=-1$) corresponds to the transition to the left (right) of
the balloon with $t_1=t_{\ast}^{s_1}$. The last relation can be written as
follows:
\be
I_{4c}(t_{\ast}^{s_1},\pi/2)=s_1\hat I_2'(c),\label{I4_and_I2hatD1}
\ee
It is easy to see that
$$
I_{4c}(t_{\ast}^s,\pi/2)=I_{4c}(\pi/2,\pi-t_{\ast}^s)=\frac{1}{2}I_{4c}
(t_{\ast}^s,\pi-t_{\ast}^s),
$$
and using (\ref{I4_hat}) we establish the validity of (\ref{I4_and_I2hatD1}).
Thus we find at $\alpha=1/2$
\be
B_n^s=\Psi\frac{\sin\theta_1}{\sin\psi}+\sin^2\psi-\sin t_1\cos t_1\sin\psi[
(2n-ss_1)\hat I_2'(c)-s\cos\pi n \;I_4(\pi/2,t_2)].\label{B_alpha=1/2}
\ee
Selecting here $s=-1$ we arrive at the final expression for $B$
\be
B_n^-=\Psi\frac{\sin\theta_1}{\sin\psi}+\sin^2\psi-\sin t_1\cos t_1\sin\psi[
(2n+s_1)\hat I_2'(c)+\cos\pi n \;I_4(\pi/2,t_2)],\label{B_alpha=1/2_final}
\ee
where $s_1=1$ ($s_1=-1$) corresponds to the transition ${\sf Und_n^-}
\leftrightarrow {\sf Und_{n+s_1}^+}$ to the left (right) of the balloon.
\subsection{Transitions ${\sf Und_n^s}\leftrightarrow {\sf Und_{n+1}^s}$}
\label{alpha_lower}
This transition for odd $n$ takes place at $\alpha=\beta^s=(1+s\sqrt{1+c/
\sin^2t_1})/2$ when the point $t_2=t_{\ast}^{-s}$ separates from the plane and
we obtain
\be
2H_n^s\Psi=I_1(t_1,t_{\ast}^{-s})+sI_2(t_1,t_{\ast}^{-s})+(n+1)\hat I_2.
\label{curv_alpha=alpha_m1}
\ee
As $c=-\sin^2t_2$ we find
\be
I_2(t_1,t_2)=\int_{t_2}^{t_1}\frac{\sin^2 t}{\sqrt{\sin^2t-\sin^2t_2}}=-
\int_{t_1}^{t_2}\frac{\sin^2 t}{\sqrt{\sin^2t-\sin^2t_2}}=-I_2^{\ast}(t_2,t_1),
\label{I_2_star_rel1}
\ee
and arrive at
$$
\Psi=\left[I_1(t_1,t_{\ast}^{-s})-sI_2^{\ast}(t_{\ast}^{-s},t_1)+(n+1)\hat I_2
\right]\frac{\sin\psi}{2\beta^s\sin t_1}.
$$
For even $n$ this transition is observed with separation of the point $t_2=
t_{\ast}^s$ from the plane for which we obtain
$$
\Psi=\left[I_1(t_1,t_{\ast}^s)-sI_2^{\ast}(t_{\ast}^s,t_1)+n\hat I_2\right]
\frac{\sin\psi}{2\beta^s\sin t_1}.
$$
Discuss one more question: how smooth are transitions
${\sf Und_n^+}\leftrightarrow {\sf Und_{n+1}^+}$ at upper ($\alpha=\beta^+$)
and ${\sf Und_n^-}\leftrightarrow {\sf Und_{n+1}^-}$ at lower ($\alpha=
\beta^-$) arcs of balloon, respectively. To this end, consider equation
(\ref{alpha_diff_eq}) in vicinity of the transition point ($\psi_*,\beta^s
(\psi_*)$) belonging to the balloon and estimate the leading terms in
(\ref{A_def}) and (\ref{B_def}) when $c\to -\sin^2t_2$. The only divergent
terms are integrals $I_4(t_1,t_2)$ and $I_4(\pi/2,t_2)$ which by (\ref{I_4})
behave as follows,
\be
I_4(t_1,t_2),I_4(\pi/2,t_2)\stackrel{c+\sin^2t_2\;\to\;0}{\simeq}\frac{\sin2
t_2}{2(1+c)\sqrt{\sin^2t_2+c}}.\label{approx_I4}
\ee
Substituting (\ref{approx_I4}) into (\ref{A_def}) and (\ref{B_def}) we arrive
in the limiting case $c+\sin^2t_2\to 0$ to explicit formula
\bea
\alpha'(\psi_*)=\frac{2\beta^s(\beta^s-1)}{2\beta^s-1}\cot(\theta_1+\psi_*),
\label{lim_deriv_alpha}
\eea
which is finite for $\beta^s\ne 1/2$ and independent on order $n$, i.e., both
curves ${\sf Und_n^s}$ and ${\sf Und_{n+1}^s}$ meet smoothly at the balloon.
\subsection{Transitions ${\sf Und_1^-}\leftrightarrow{\sf Und_1^+}$, ${\sf Und_
{2n}^-}\leftrightarrow{\sf Und_{2n}^+}$ and ${\sf Und_{2n}^-}\leftrightarrow
{\sf Und_{2n+2}^+}$}\label{r451}
In previous sections we have discussed in details conditions and rules of
regular transitions between unduloids that accompanied by addition or removal
of one inflection point. It is also possible to observe degenerate transitions
when the number of inflection points changes by two or does not change at all.
Find the conditions for such degenerate transitions.

An addition of two inflection points takes place only when one of these points
separates from the solid sphere and the other one from the plane. The first
event corresponds to a condition $c+\sin^2 t_1=0$, while the second one
requires $c+\sin^2 t_2=0$. These relations imply $\sin t_1=\sin t_2$ leading to
$\psi_1^{\ast}=\min\{\theta_2-\theta_1,\pi-\theta_1-\theta_2\}$ and $\psi_2^{
\ast}=\max\{\theta_2-\theta_1,\pi-\theta_1-\theta_2\}$. These critical values
define the extremal points of the balloon where it transforms into the segments
of the line $\alpha=1/2$. When $\psi_1^{\ast}$ is negative only a part of the
balloon is observed. When both values $\psi_i^{\ast}$ are negative the balloon
does not exists. Finally, for $\theta_2=\pi/2$ the balloon reduces to a point
at $\psi^{\ast}=\pi/2-\theta_1$.

Consider the transition ${\sf Und_{2k}^-}\leftrightarrow{\sf Und_{2k+2}^+}$ at
$\psi=\psi_1^{\ast}$. Using (\ref{curv_infla2k}) and noting that $t_1=\theta_2
=t_{\ast}^-$ we find
$$
\frac{\sin\theta_2}{\sin\psi}\Psi=I_1(\theta_2,\pi-\theta_2)-I_2(\theta_2,\pi-
\theta_2)+2k\hat I_2=-2\cos\theta_2+(2k+1)\hat I_2,
$$
from which we obtain the distance $d_n$ where this transition occurs
\be
d_n=-1+\cos(\theta_2-\theta_1)+\frac{\sin(\theta_2-\theta_1)}{\sin\theta_2}\;
\left[(n+1)\hat I_2-2\cos\theta_2\right].\label{transition_2k_left}
\ee
The case $k=0$ should formally correspond to the transition ${\sf Und_0^-}
\leftrightarrow{\sf Und_2^+}$. It can be checked that ${\sf Und_0^-}$ unduloid
can exist only below of the balloon, and not to the left or right of it, so
that the above transition is forbidden. The relation (\ref{transition_2k_left})
at $k=0$ leads to a special case ${\sf Und_1^-}\leftrightarrow {\sf Und_0^-}
\leftrightarrow{\sf Und_1^+}$ when the segment corresponding to ${\sf Und_0^-}$
reduced to a point. The sequence of distances given by
(\ref{transition_2k_left}) is a periodic one with the period equal to
$$
d_{n+2}-d_n=\frac{2\hat I_2\sin(\theta_2-\theta_1)}{\sin\theta_2}=4\sin(
\theta_2-\theta_1) E\left(1-\frac1{\sin^2\theta_2}\right).
$$
It can be shown that the transition ${\sf Und_{2k+1}^-}\leftrightarrow {\sf
Und_{2k+3}^+}$ is forbidden at $\psi=\psi_1^{\ast}$. As the number of inflection
points is odd both new points to be added should correspond to either $t_{
\ast}^+$ or $t_{\ast}^-$. It means that $t_1=t_2$ which contradicts (for
$\theta_2\ne\pi/2$) to the $\psi_1^{\ast}$ value.

The second critical point $\psi=\psi_2^{\ast}$ initiates the segment of the
line $\alpha=1/2$ to the right of the balloon. Only the transition
${\sf Und_n^-}\leftrightarrow {\sf Und_{n-1}^+}$ is allowed on this line when
the inflection point merges the sphere. It means that at $\psi=\psi_2^{\ast}$
the inflection point with $t=t_1$ merges the sphere while the point with $t=
t_2$ separates from the plane. As the result total number of inflection points
remains constant. For the ${\sf Und_{2k}^-}$ unduloid we find $t_1=t_2$ that
corresponds to $\psi=\psi_2^{\ast}$. Thus, the transition ${\sf Und_{2k}^-}
\Leftrightarrow {\sf Und_{2k}^+}$ is allowed, while ${\sf Und_{2k+1}^-}
\Leftrightarrow {\sf Und_{2k+1}^+}$ is forbidden again (for $\theta_2\ne\pi/2$).
In this case we have
$$
\frac{\sin\theta_2}{\sin\psi}\Psi=2k\hat I_2,
$$
from which we obtain
\be
d_n=-1-\cos(\theta_2+\theta_1)+n\frac{\sin(\theta_2+\theta_1)}{\sin\theta_2}
\hat I_2\;.\label{transition_2k_right}
\ee
The sequence of distances given by (\ref{transition_2k_right}) is also a
periodic one with the period equal to
$$
d_{n+2}-d_n=\frac{2\hat I_2\sin(\theta_2+\theta_1)}{\sin\theta_2}=4\sin(
\theta_2+\theta_1) E\left(1-\frac1{\sin^2\theta_2}\right).
$$
\section{Topology of Unduloid Menisci Transitions}\label{rr10}
This section is mostly topological and deals with qualitative behavior of
different branches of function $\alpha(\psi)$ enumerated by the number $n$ of
inflection points at corresponding menisci. We list the most general properties
of curves $\alpha_n(\psi)$ in the plane $\left\{\psi,\alpha\right\}$ and study
ramification of these curves around the balloon. This global geometrical
representation allows to classify possible trajectories and their intersection
points.
\subsection{Balloons $\beta^{\pm}(\psi)$}\label{rr1}
Consider the rectangle $\Delta:=\left\{-\theta_1\leq\psi\leq\pi-\theta_1,\;0
\leq\alpha\leq 1\right\}$ in the plane $\left\{\psi,\alpha\right\}$ and define
{\it a balloon} ${\mathbb B}$
\bea
{\mathbb B}={\cal B}_l\cup{\cal B}_r\cup{\cal B}_d\cup{\cal B}_u\;,\quad
\mbox{where} \label{t2}
\eea
\bea
&&{\cal B}_l=\left\{-\theta_1\leq\psi\leq\Theta_{min}\;,\;\alpha=\frac1{2}
\right\},\hspace{.9cm}
{\cal B}_d=\left\{\Theta_{min}\leq\psi\leq\Theta_{max}\;,\;\alpha=\beta^-
(\psi)\right\},\nonumber\\
&&{\cal B}_r=\left\{\Theta_{max}\leq\psi\leq\pi-\theta_1\;,\;\alpha=\frac1{2}
\right\},\quad
{\cal B}_u=\left\{\Theta_{min}\leq\psi\leq\Theta_{max}\;,\;\alpha=\beta^+
(\psi)\right\},\nonumber
\eea
and $\Theta_{min}\!=\!\min\left\{\theta_2-\theta_1,\pi-\theta_2-\theta_1
\right\}$, $\Theta_{max}\!=\!\max\left\{\theta_2-\theta_1,\pi-\theta_2-\theta_1
\right\}$, $0\leq\theta_1,\theta_2\leq\pi$.
Two functions $\beta^+(\psi)$ and $\beta^-(\psi)$ give the upper and
lower parts of convex symmetric oval,
\bea
\beta^{\pm}(\psi)=\frac{1}{2}\left(1\pm\sqrt{1-\frac{\sin^2\theta_2}{\sin^2
\left(\theta_1+\psi\right)}}\;\right),\quad\beta^{\pm}(\psi)=\beta^{\pm}
(\pi-2\theta_1-\psi)\;.\nonumber
\eea
Subscripts {\it l}, {\it r}, {\it d} and {\it u} stand for the {\it left-,
right-, down-} and {\it upward} directions on ${\mathbb B}$. Denote by
$\overline{{\cal B}}$ the balloon ${\mathbb B}$ with its open interior
${\mathfrak B}$,
\bea
\overline{{\cal B}}={\mathbb B}\cup{\mathfrak B}\;,\quad{\mathfrak B}:=\left\{
\Theta_{min}\leq\psi\leq\Theta_{max}\;,\;\beta^-(\psi)<\alpha<\beta^+
(\psi)\right\}\;.\label{t3}
\eea
In special case $\theta_2=\pi/2$ we have ${\mathbb B}_{\pi/2}=\left\{-\theta_1
\leq\psi\leq\pi-\theta_1\;,\;\alpha=1/2\right\}$ while the part ${\cal B}_d\cup
{\cal B}_u$ of balloon is reduced into a point ${\cal O}=\{\psi=\pi/2-\theta_1,
\alpha=1/2\}$ such that ${\cal O}\in{\mathbb B}_{\pi/2}$.
\subsection{Trajectories $\alpha_n(\psi)$}\label{rr2}
\begin{figure}[h!]\begin{tabular}{cc}
\psfig{figure=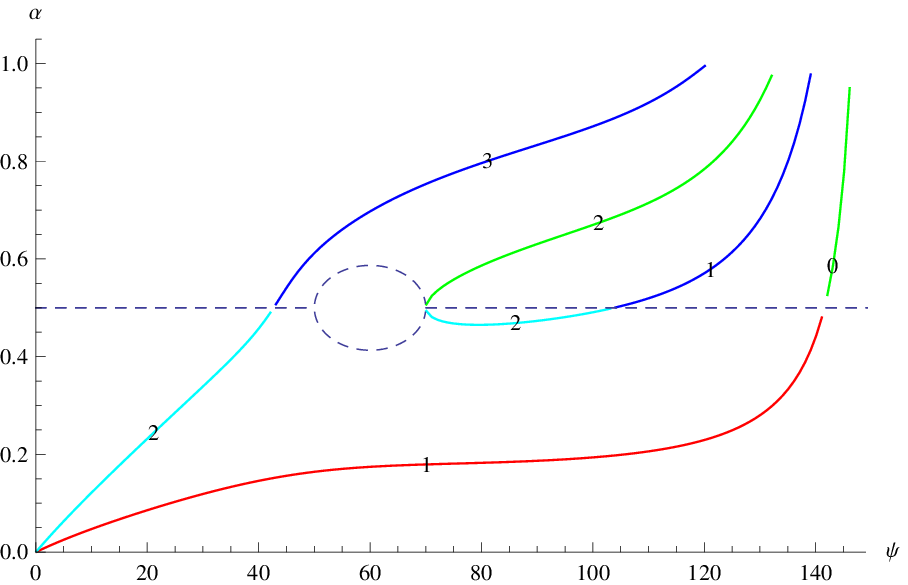,height=5cm} &
\psfig{figure=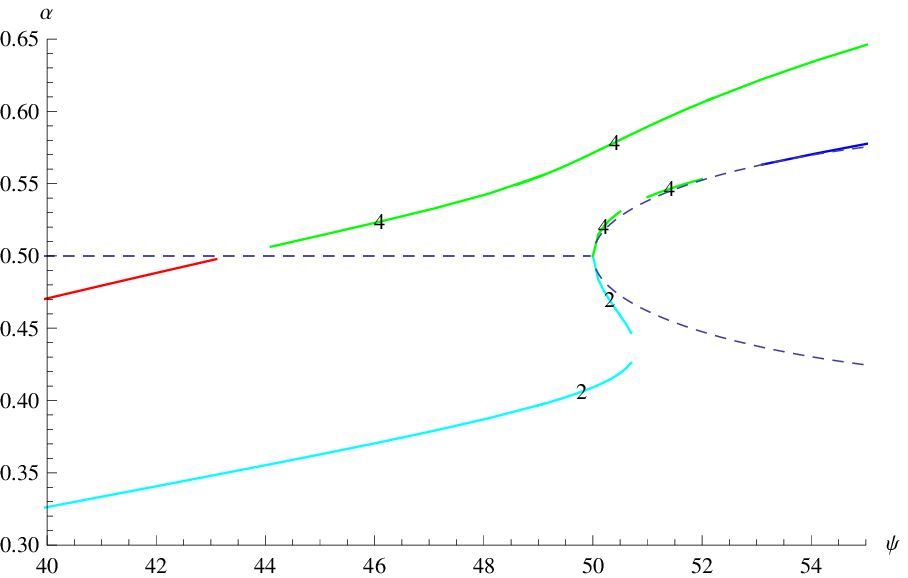,height=5cm}\\
(a) & (b)\end{tabular}
\caption{Plots $\alpha_n(\psi)$ for $\theta_1=30^o$, $\theta_2=80^o$ and
(a) $d=5.2919$, (b) $d=6.6482$.}\label{qi4}
\end{figure}

\begin{figure}[h!]\begin{tabular}{cc}
\psfig{figure=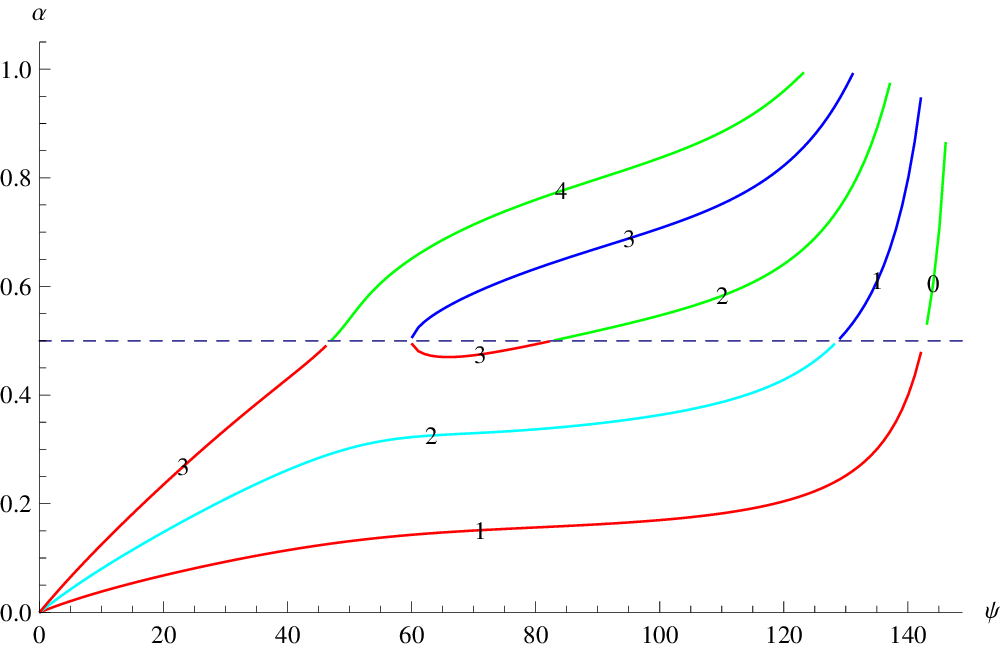,height=5cm} &
\psfig{figure=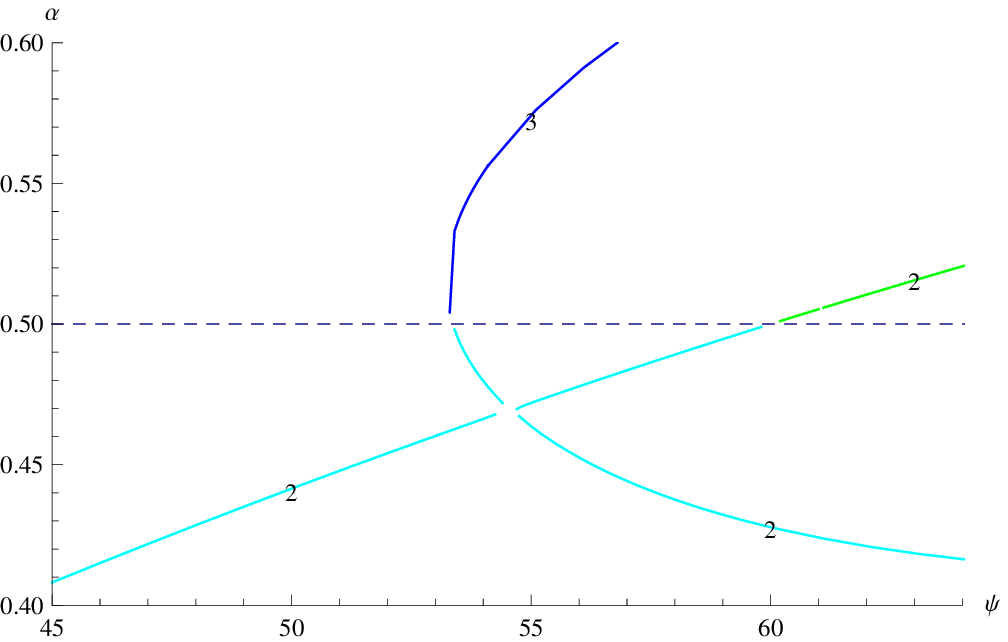,height=5cm}\\
(a) & (b) \\ \end{tabular}
\caption{Plots $\alpha_n(\psi)$ for $\theta_1=30^o$, $\theta_2=90^o$ at (a) $d=
7.66209$ and (b) $d=5.82697$. In (b) a vicinity of the saddle point ${\sf
B_2^-}$ is shown.}
\label{qi3}
\end{figure}

Below we give a list of rules for topological behavior of $\alpha_n(\psi)$ in
the presence of ${\mathbb B}$.
\begin{enumerate}
\item $\alpha_n(\psi)$ is completely defined by three parameters: $0\leq
\theta_1,\theta_2\leq\pi$ and $d\geq -2$.
\item $\alpha_n(\psi)$ is a real function representable in the $\left\{\psi,
\alpha\right\}$ plane by a nonorientable trajectory $\Gamma$ without
self-intersections. All trajectories are continuous smooth curves and located
in domain $\Delta^{\prime}\setminus{\mathfrak B}$, where $\Delta^{\prime}=
\left\{0\leq\psi\leq\pi-\theta_1,\;0\leq\alpha\leq1\right\}$.
\item The final points of $\Gamma$ are associated with spheres ${\sf Sph^-_{
n_1}}$ and ${\sf Sph^+_{n_2}}$. Equip $\Gamma$ with indices according to the
final spheres designation in such a way that a left lower index does not exceed
a right lower, i.e., $^-_{n_1}\Gamma^+_{n_2}:=\left\{{\sf Sph^-_{n_1}}
\rightarrow{\sf Sph^+_{n_2}}\right\}$, $n_1\leq n_2$. The following coincidence
property holds: $^-_n\Gamma^+_n=\;^+_n\Gamma^-_n$.
\item
In the $\left\{\psi,\alpha\right\}$ plane the spheres satisfy
\bea
&&{\sf Sph^-_n}\in\{\psi=0,\;\alpha_n=0\}\;,\quad{\sf Sph^+_n}\in\{0\leq\psi
\leq\pi,\;\alpha_n=1\},\quad \theta_1>0\;,\label{t4}\\
&&{\sf Sph^-_n}\in\{\psi=0,\;0\leq\alpha_n\leq 1\}\;,\quad{\sf Sph^+_n}\in\{
\psi=\pi,\;0\leq\alpha_n\leq 1\},\quad \theta_1=0\;.\nonumber
\eea
\item Angular coordinates $\psi=\phi^+_n$ of spheres ${\sf Sph^+_n}$ are
arranged leftward in ascending order (see \ref{sphere1}),
\bea
0<\ldots<\phi^+_n<\phi^+_{n-1}<\ldots <\phi^+_1<\phi^+_0<\pi-\theta_1\;.
\label{t44a}
\eea
\item There exist generic trajectories of four topological types,
\bea
^-_n\Gamma^-_{n+1},\quad ^-_n\Gamma^+_{n+1},\quad ^+_{n-1}\Gamma^-_n,\quad
^+_{n-1}\Gamma^+_n,\quad n\geq 1\;.\label{t4a}
\eea
\item Different parts of trajectories are labeled by different sub- and
superscripts ${\sf Und} _{n_1}^+$ and ${\sf Und}_{n_2}^-$ where the upper index
is equal to $s=\mbox{sgn}(2\alpha_n-1)$.
\item In vicinity of the point ($\psi=0,\;\alpha_n=0$) the sheaf of trajectories
$^-_n\Gamma^-_{n+1}$, $^-_n\Gamma^+_{n+1}$ and $^+_{n-1}\Gamma^-_n$ is build in
such a way that slopes $\xi^-_n$ of the ${\sf Und}_n^-$ parts are arranged 
clockwise in descending order (see section \ref{ar15}),
\bea
\frac{\pi}{2}>\ldots >\xi^-_n>\xi^-_{n-1}>\ldots>\xi^-_2>\xi^-_1>0\;.
\label{t4b}
\eea
\item For given $\theta_1$, $\theta_2$ there exists a unique $d=d^s_n$ such
that there appears one of five types of intersection (saddle) points:
\bea
{\sf A_n^-}=\;^-_n\Gamma^+_n\otimes\;^+_{n-1}\Gamma^-_{n+1}\;,
&{\sf A_n^+}=\;^-_n\Gamma^+_n\otimes\;^-_{n-1}\Gamma^+_{n+1}\;,
&\nonumber\\
{\sf B_n^-}=\;^-_n\Gamma^+_n\otimes\;^+_{n-1}\Gamma^+_{n+1}\;,
&{\sf B_n^+}=\;^-_{n-1}\Gamma^+_{n-1}\otimes\;^+_{n-2}\Gamma^+_n\;,
&{\sf B_n^0}=\;^-_n\Gamma^+_n\otimes\;^+_{n-1}\Gamma^+_{n+1}\;,
\label{t4e}
\eea
where operation $\Gamma_1\otimes\Gamma_2$ denotes intersection of two
trajectories $\Gamma_1$ and $\Gamma_2$. The indices of a saddle point
correspond to unduloid ${\sf Und}_n^s$ observed at this point. The saddle
points ${\sf B_n^0}$ of mixed type are located on a line $\alpha=1/2\;(s=0)$.

\begin{figure}[h!]\begin{center}\begin{tabular}{cc}
\psfig{figure=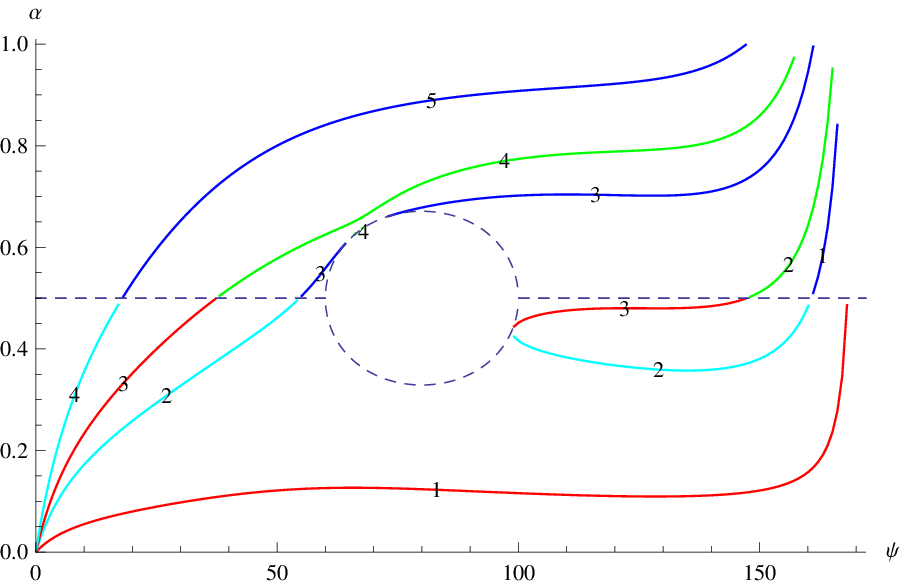,height=5cm} &
\psfig{figure=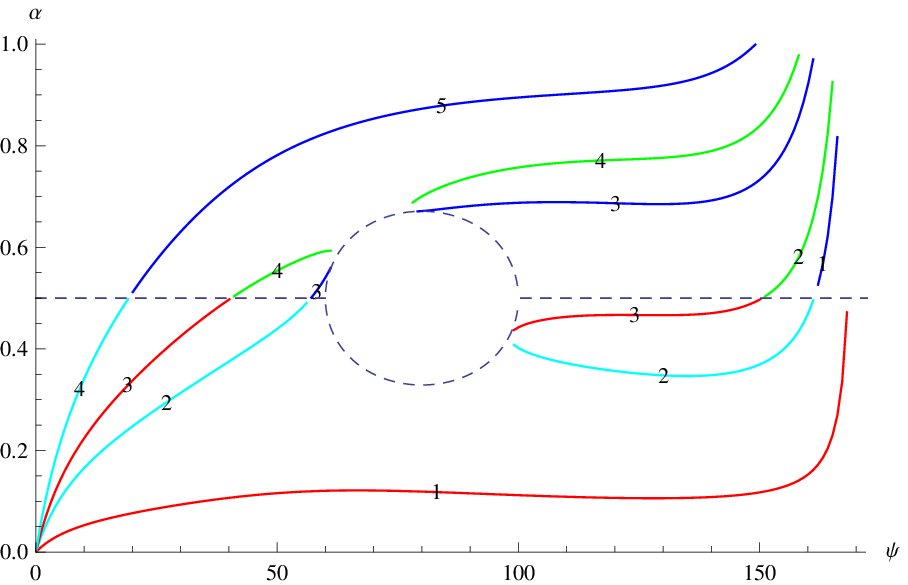,height=5cm}\\(a) & (b)
\end{tabular}\end{center}
\vspace{-.5cm}
\caption{Plots $\alpha_n(\psi)$ for (a) $\theta_1=10^o$, $\theta_2=70^o$,
$d=6.45$ and (b) $\theta_1=10^o$, $\theta_2=70^o$, $d=6.7$.}\label{qi1}
\end{figure}
\item Every trajectory of the types $^-_n\Gamma^-_{n+1}$, $^+_n\Gamma^+_{n+1}$,
$^-_n\Gamma^+_{n+1}$ and $^+_{n-1}\Gamma^-_n$ is necessary smooth at $\alpha=
1/2$ (see section \ref{alpha=1/2_left}).
\item Every trajectory of the types $^-_n\Gamma^-_{n+1}$ and $^+_n\Gamma^+_{
n+1}$ is necessary tangent to ${\cal B}_d\cup{\cal B}_u$ (see section
\ref{alpha_lower}).
\item The changes of indices in ${\sf Und}_{n_i}^s$ occur at balloon ${\mathbb
B}$ in accordance with Table 2 (see Figure \ref{qi1}).
\begin{enumerate}
\item The change of the upper indices ($s\leftrightarrow -s$) and lower indices
($n_1\leftrightarrow n_2$) occurs at the point $T\in{\cal B}_l\cup{\cal B}_r$.
\item The change of the lower index ($n_1\leftrightarrow n_2$) only occurs at
the point $T\in{\cal B}_d\cup{\cal B}_u$.

In the nondegenerate case one has $|n_1-n_2|=1$.
\end{enumerate}
\end{enumerate}
\begin{center}
{\bf Table$\;$2}\\
\vspace{-.6cm}
$$
\begin{array}{|c||c|c|c|c|c|c|c|c|c|c|c|}\hline
 & ^-_n\Gamma^-_{n+1} &^-_n\Gamma^+_{n+1} & ^+_{n-1}\Gamma^-_n &
^+_{n-1}\Gamma^+_n \\\hline\hline
{\cal B}_l & _n^-T_{n+1}^+,\quad _{n+1}^-T_{n+2}^+ & _n^-T_{n+1}^+ & -
 & _{n-1}^-T_n^+\\\hline
{\cal B}_r & - & - & _{n-1}^+T_n^- & _{n-1}^+T_n^-,\quad
_n^+T_{n+1}^-\\\hline
{\cal B}_d & _n^-T_{n+1}^- & - & _{n-1}^-T_n^- & _n^-T_{n+1}^-\\\hline
{\cal B}_u & _{n+1}^+T_{n+2}^+ & _{n+1}^+T_{n+2}^+ & - &
_{n-1}^+T_n^+ \\\hline
\end{array}
$$
\label{tar1}
\end{center}
In Table 2 a symbol $^-_{n_1}T^+_{n_2}$ denotes a point belonging to two parts
${\sf Und}_{n_1}^-$ and ${\sf Und}^+_{n_2}$ of trajectory. Empty boxes mean that
corresponding transitions do not exist.
The trajectories can be tangent to balloon ${\cal B}_d\cup{\cal B}_u$ at its
left and right points (see Figure \ref{qi4}),
\bea
^-_1T^+_1,\quad ^-_{2n}T^+_{2n+2}\in\;{\cal B}_l\cap\left\{{\cal B}_d\cup{\cal
B}_u\right\},\quad\mbox{and}\quad ^-_{2n}T^+_{2n}\in\;{\cal B}_r\cap\left\{
{\cal B}_d\cup{\cal B}_u\right\}\;.\label{t9}
\eea
When $\theta_2=\pi/2$ the allowed transitions are the following (see Figure
\ref{qi3}),
\bea
^-_{2n}T^+_{2n},\quad ^-_{2n+1}T^+_{2n+1}\in\;{\cal B}_r\cap {\cal B}_l\;.
\label{t9a}
\eea
\section{Saddle Points}\label{r511}
For given values of the contact angles $\theta_1,\theta_2$ a change in the
distance $d$ value leads to changes of the trajectories shape in $\{\psi,
\alpha\}$ plane. Sometimes such transitions are accompanied by drastic changes
of the trajectories' topology characterized by an appearance of the saddle
points. The saddle point can be defined as a point that belongs to two
trajectories simultaneously.

It is instructive to find the coordinates of the saddle point $\{\psi_n^c,
\alpha_n^c\}$ as well as the distance $d_n^c$ at which the saddle point is
observed. Below we describe a procedure for such computation for each type of
the saddle points belonging to a single meniscus type ${\sf Und_n^s}$.
\subsection{Saddle Points of Simple Type}
First note that the saddle point may be observed at the intersection of two 
segments of the curve $\alpha_n(\psi)$ determined by the sign $s$ and order $n$
of unduloid meniscus ${\sf Und_n^s}$ that completely defined by the relation
(\ref{alpha_und_pm}). At every point of these segments (except the saddle
point) one can define the derivative $\alpha_n'(\psi)$ having a unique value
determined from the equation (\ref{alpha_diff_eq}) where both $A_n^s$ and $B_n^s$
do not vanish simultaneously. At the saddle point the derivative $\alpha_n'(
\psi)$ is not unique and in this case $A_n^s=B_n^s=0$. Thus the saddle point
$\{\psi_n^c,\alpha_n^c\}$ at $d=d_n^c$ is determined from the condition
$A_n^s=B_n^s=0$ together with (\ref{alpha_und_pm}).

\begin{figure}[h!]\begin{tabular}{cc}
\psfig{figure=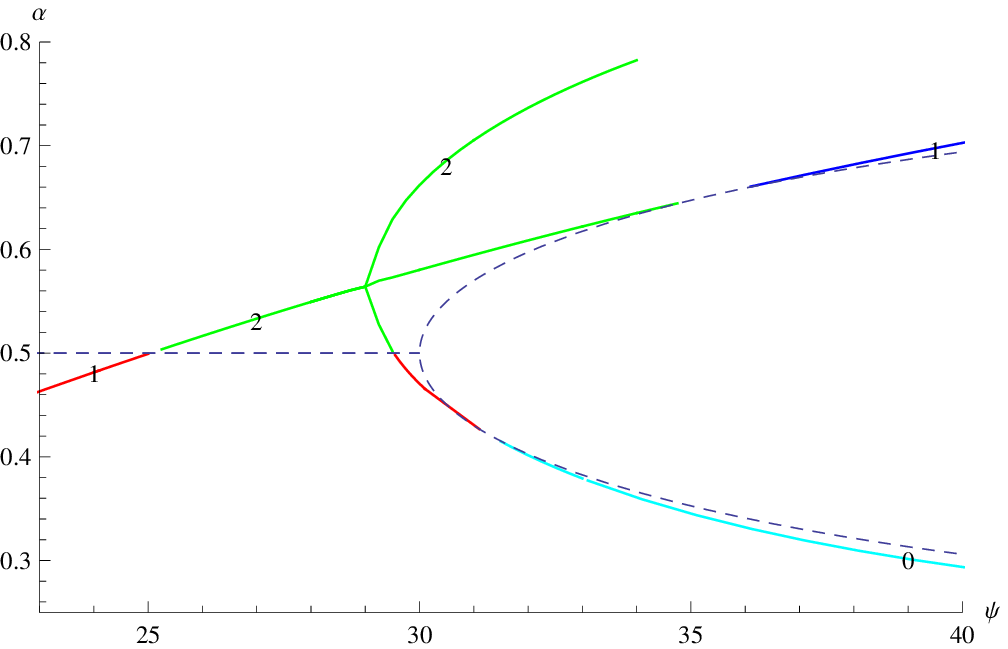,width=7.5cm} &
\psfig{figure=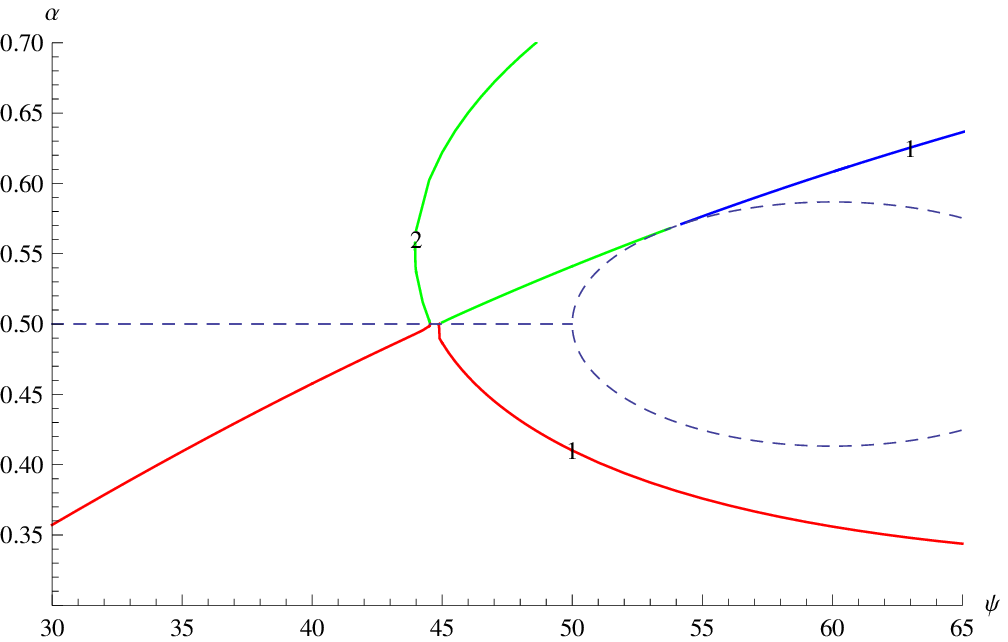,width=7.5cm}\\
(a) & (b)\end{tabular}
\caption{The saddle points at intersection of $^-_n\Gamma^+_n$ and $^+_{n-1}
\Gamma^+_{n+1}$ trajectories for $n=1$. (a) The trajectories in the vicinity of
${\sf B_2^+}$ saddle point for $\theta_1=30^o,\theta_2=60^o$ and $d=1.05187$. 
(b) The trajectories in the vicinity of ${\sf B_1^0}$ mixed type saddle point 
observed for $\theta_1=30^o,\theta_2=80^o$ and $d=2.2465$. Colors and numbers 
correspond to indices of respective unduloids.}\label{saddle}
\end{figure}

Using the condition $A_n^s=0$ we find the expression in the square brackets in
(\ref{A_def}) and substituting it into (\ref{B_def}) we obtain at the saddle
point
\be
B_n^s=2\alpha_n\Psi\frac{\sin\theta_1}{\sin\psi}-\frac{4\alpha_n(1-\alpha_n)}
{2\alpha_n-1}\Psi\cos t_1-\frac{4\alpha_n^2-2\alpha_n-1}{2\alpha_n-1}\sin t_1
\sin\psi+\sin^2\psi.\label{B_expr1}
\ee
Use (\ref{B_expr1}) in the condition $B_n^s=0$ to express $\Psi$ at the saddle
point
\be
\Psi_n^c=\frac{\sin^2\psi\left[\left(4\alpha_n^2-2\alpha_n-1\right)\sin t_1-(
2\alpha_n-1)\sin\psi\right]}{2\alpha_n\left[(2\alpha_n-1)\sin\theta_1-2
(1-\alpha_n)\cos t_1\sin\psi\right]},\label{Psi_new_def}
\ee
and eliminate
\be
d_n^c=\Psi_n^c+\cos\psi_n^c-1,\label{d_n^s}
\ee
from the saddle point conditions. Thus we arrive at the final equations
determining the saddle point position in the $\{\psi,\alpha\}$ plane:
\bea
2\alpha_n\Psi_n^c\sin t_1-\left[I_1(t_1,t_2)+sI_2(t_1,t_2)+n\hat I_2-s(1-\cos\pi
n)I_2(\pi/2,t_2)\right]\sin\psi &=&0,\label{saddle_eq0}\\
\Psi_n^c+(2\alpha_n-1)\sin t_1\sin\psi\left[s I_4(t_1,t_2)-2n\hat I_2'(c)-s(1-
\cos\pi n)I_4(\pi/2,t_2)\right]&=&0.\label{saddle_eq1}
\eea
Solving the equations (\ref{saddle_eq0},\ref{saddle_eq1}) we find the saddle
point. An example of trajectories in a vicinity of such a point is shown in
Figure \ref{saddle}(a).
\subsection{Saddle Points of Mixed Types}
The saddle points considered above belong to a single type of unduloid meniscus
and are characterized, in particular, by the sign of $\alpha_n-1/2$. When at the
saddle point $\alpha_n=1/2$ this point belongs to two different types of menisci
of different orders and signs. These points do not have unduloid sign
characteristics and we designate them by a smaller (of two) order only. Consider
first such saddle points located to the left of the balloon ($\Theta_{min},
\Theta_{max}>0$). This transition described in Section \ref{alpha=1/2_left} 
happens when $t_1=t_{\ast}^+=\arcsin\sqrt{-c}$. Equation (\ref{saddle_eq1}) 
cannot be used directly at $\alpha_n=1/2$ as integral $I_4$ diverges at $t_1=
t_{\ast}^+$. In the limit $\alpha\to 1/2$ we have by (\ref{I_4}),
$$
I_4(t_{\ast}^+,t_2) \sim -\frac{1}{|1-2\alpha|\sqrt{1+c}}
$$
and obtain from (\ref{saddle_eq1})
\be
\Psi_{\ast}=\sin\psi\tan t_{\ast}^+.\label{Psi_ast}
\ee
Note that this value does not depend on both order $n$ and contact angle
$\theta_2$. Substitution of (\ref{Psi_ast}) into (\ref{saddle_eq0}) produces a
condition on $c$ value
\be
I_1(t_{\ast}^+,t_2)-I_2(t_{\ast}^+,t_2)+n\hat I_2+(1-\cos\pi n)I_2(\pi/2,t_2)+
\frac{c}{\sqrt{1+c}}=0,\label{saddle_eq_left}
\ee
where order $n$ corresponds to the meniscus with $\alpha_n <1/2$. For given
values of order $n$ and contact angle $\theta_2$ we find $c$ value verifying
the last condition that leads to determination of $\psi_{\ast}=\arcsin
\sqrt{-c}-\theta_1$. Using it in (\ref{Psi_ast}) and (\ref{d_n^s}) we arrive at
the distance $d_{\ast}$ value for which the mixed saddle point is observed.
An example of such a point is shown in Figure \ref{saddle}(b).

Consider a mixed type saddle points located to the right of the balloon. This
transition described in (\ref{alpha=1/2_left}) happens for $s=-1$ when $t_1=
t_{\ast}^-=\pi-\arcsin\sqrt{-c}$. Equation (\ref{saddle_eq1}) leads to
$$
\Psi_{\ast}=\sin\psi\tan t_{\ast}^-.
$$
Using it in (\ref{saddle_eq0}) we obtain a condition on $c$ value
\be
I_1(t_{\ast}^-,t_2) - I_2(t_{\ast}^-,t_2)+n\hat I_2 +(1-\cos\pi n)I_2(\pi/2,
t_2)-\frac{c}{\sqrt{1+c}}=0,\label{saddle_eq_right}
\ee
where positive order $n$ corresponds to the meniscus with $\alpha_n < 1/2$.
Show that the equation (\ref{saddle_eq_right}) does not have solutions for
$n>0$. Using the relation
$$
I_2(t_{\ast}^-,t_2)=\frac{\hat I_2}{2}+I_2(\pi/2,t_2),
$$
rewrite the left hand side of (\ref{saddle_eq_right}) as
$$
\frac{1}{\sqrt{1+c}}+\cos t_2+(n-1/2)\hat I_2-\cos\pi n I_2(\pi/2,t_2),
$$
where sum $s_1$ of the first two terms is always positive. For odd $n=2k-1$ we
have
$
s_1+(2k-1/2)\hat I_2+I_2(\pi/2,t_2) >0.
$
For even $n=2k$ we have
$
s_1 +(2k-1/2)\hat I_2-I_2(\pi/2,t_2) >0,
$
as $I_2(\pi/2,t_2)\le I_2(\pi/2,t_{\ast}^{\pm})=\hat I_2/2$.
Thus the saddle points of the mixed type cannot be observed to the
right of the balloon.
\subsection{Saddle Points Sequences}
\begin{figure}[h!]\begin{tabular}{cc}
\psfig{figure=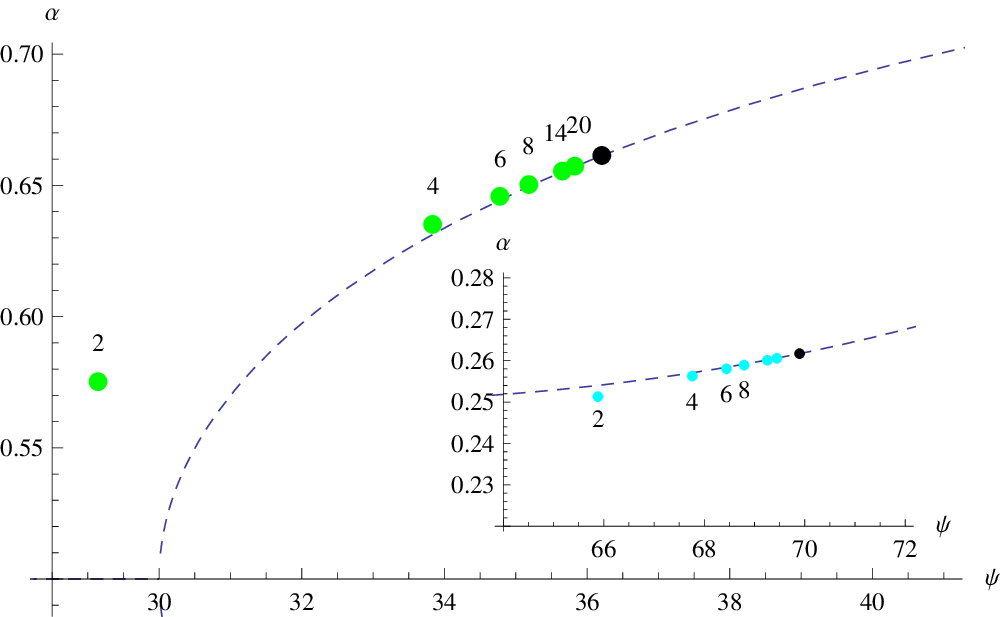,width=7.5cm}
&
\psfig{figure=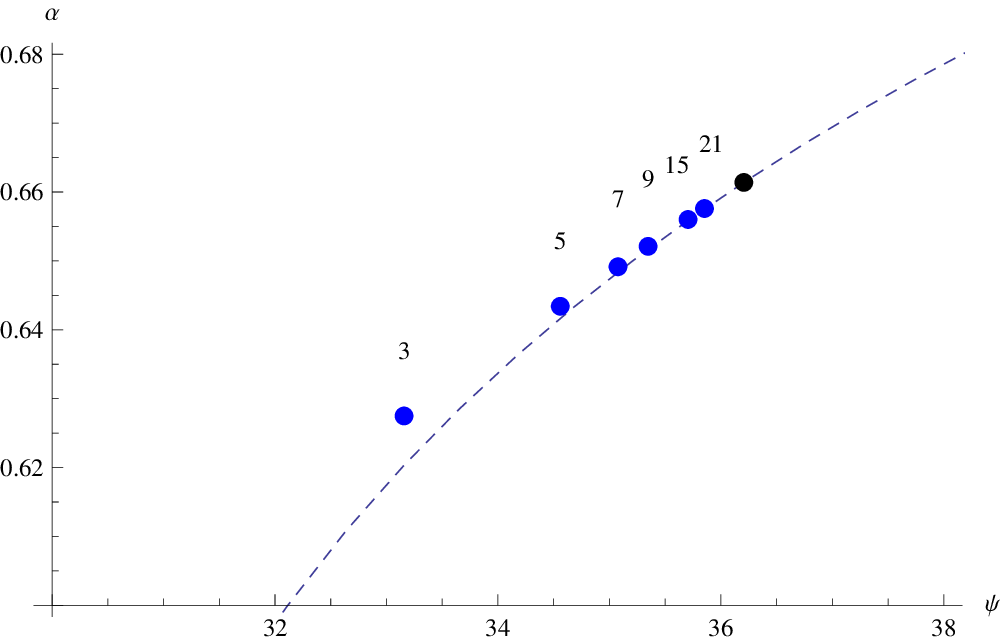,width=7.5cm}\\
(a) & (b)\\
\psfig{figure=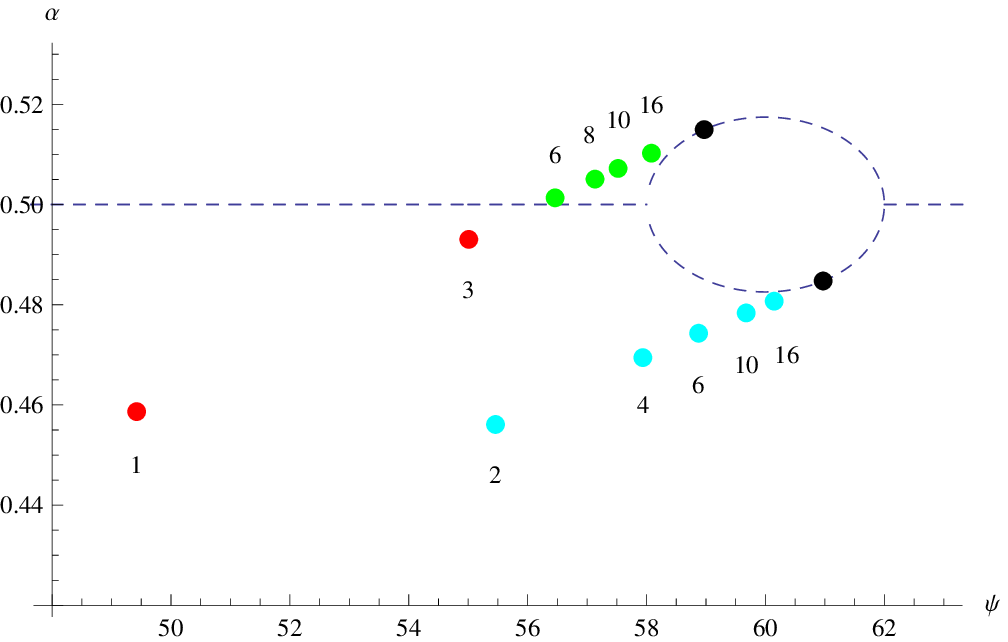,width=7.5cm}
&
\psfig{figure=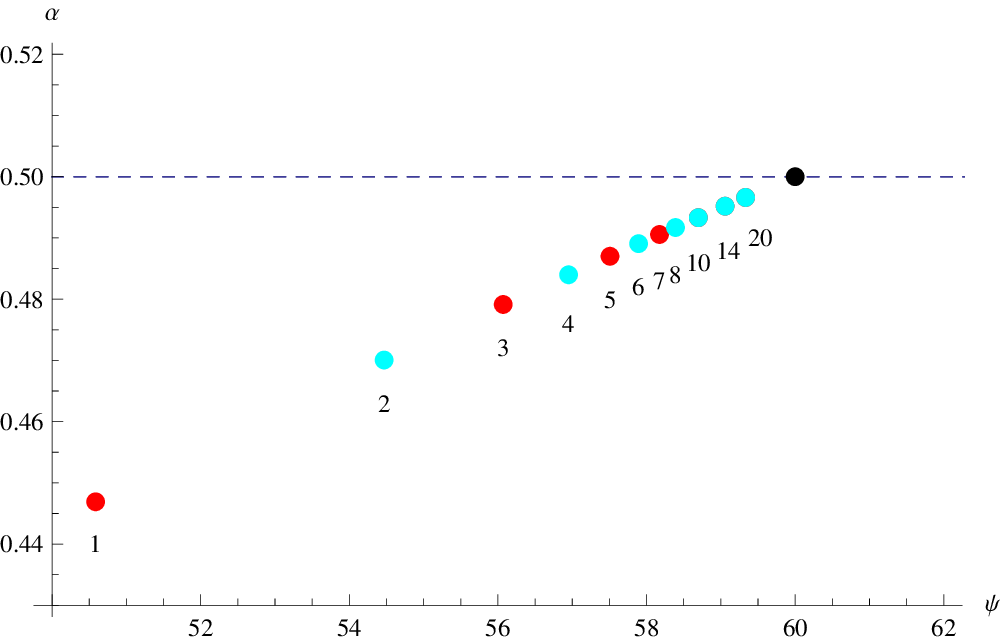,width=7.5cm}\\
(c) & (d)\end{tabular}
\caption{(a) The saddle points (green) ${\sf B_2^+}$ and ${\sf A_{2k}^+}$ for
$\theta_1=30^o, \theta_2=60^o$ and $k=2,3,4,7,10$ approach the accumulation
point (black) on the balloon with growth of the index $2k$ marking the saddle
points. The inset shows saddle points (cyan) ${\sf A_{2k}^-}$ for $k=1,2,3,4,7,
10$ approaching the accumulation point. (b) The saddle points (blue) ${\sf A_{
2k-1}^+}$ for $\theta_1=30^o, \theta_2=120^o$ and $k=2,3,4,5,8,11$ approach the
accumulation point (black). (c) For $\theta_1=30^o, \theta_2=88^o$ the saddle
points ${\sf B_n^-}$ for $n=1,3$ ({\em red}), $n=2,4,6,10,16$ ({\em cyan}) and
${\sf B_n^+}$ ({\em green}) for $n=6,8,10,16$ approach the corresponding
accumulation points ({\em black}). (d) For $\theta_1=30^o, \theta_2=90^o$ the
saddle points ${\sf B_n^-}$ for $n=1,3,5,7$ ({\em red}) and $n=2,4,6,8,10,14,
20$ ({\em cyan}) approach the accumulation point ({\em black}).}
\label{saddle_sequence}
\end{figure}

The computation of the saddle point position for fixed values of the contact
angles $\theta_1,\theta_2$ and increasing $n$ shows that for large $n$ a
sequence $\{\psi_n^c,\alpha_n^c\}$ accumulates in a small vicinity of a point
$\{\psi_{\ast}^c,\alpha_{\ast}^c\}$ belonging to the balloon (not reaching it),
i.e., $\alpha_{\ast}^c=\beta^s(\psi_{\ast}^c)$ as shown in Figure
\ref{saddle_sequence}.

It is instructive to determine the position of the accumulation point $\{\psi_{
\ast}^c,\alpha_{\ast}^c\}$. First note that in (\ref{saddle_eq0}) for $n\gg 1$
the dependence of $\Psi_n^c$ on $n$ is determined by a relation
$$
2\alpha_n^c\Psi_n^c\sin t_1=n\hat I_2,
$$
implying that both $\Psi_n^c$ and $d_n^c$ grow linearly in $n$. As the integral
$\hat I_2'(c)$ is always negative for $c<0$ then the leading term for $n\gg 1$
in (\ref{saddle_eq1}) is contributed by the integrals $I_4$. From (\ref{I_4})
it follows that the proper divergence is provided by the term $\sin2t_2/[2(1+c)
\sqrt{\sin^2t_2+c}]$ when $c\to -\sin^2t_2$. Substitution of $\beta^s(\psi)$
in (\ref{cdef_alpha}) shows that $c+\sin^2\theta_2=0$ holds on the balloon. Thus the
divergence of $\Psi_n^c$ used in (\ref{Psi_new_def}) leads to the condition
\be
(2\alpha_{\ast}^c-1)\sin\theta_1=2(1-\alpha_{\ast}^c)\sin\psi_{\ast}^c\cos(
\theta_1+\psi_{\ast}^c),
\label{accumulation_cond}
\ee
where
$$
\alpha_{\ast}^c=\frac{1}{2}\left(1+s\sqrt{1-\frac{\sin^2\theta_2}{\sin^2(
\theta_1+\psi_{\ast}^c)}}\;\right)=\beta^s(\psi_{\ast}^c).
$$
It follows from (\ref{accumulation_cond}) that the accumulation point with
$\alpha_{\ast}^c< 1/2$ can be observed for $\psi_{\ast}^c>\pi/2-\theta_1$, while
at the point with $\alpha_{\ast}^c> 1/2$ we have $\psi_{\ast}^c<\pi/2-\theta_1$.
The relation (\ref{accumulation_cond}) also implies
$$
2(1-\alpha_{\ast}^c)\sin(\theta_1+\psi_{\ast}^c)\cos\psi_{\ast}^c=\sin\theta_1,
\ \ \ 2\alpha_{\ast}^c\sin(\theta_1+\psi_{\ast}^c)\cos\psi_{\ast}^c=\sin
(\theta_1+2\psi_{\ast}^c).
$$
Multiplying these relations and using the definition (\ref{cdef_alpha}) we find
the sign independent condition on the accumulation point
\be
\sin\theta_1\sin(\theta_1+2\psi_{\ast}^c)=\sin^2\theta_2\cos^2\psi_{\ast}^c.
\label{accumulation_cond1}
\ee
It can be shown that the equation (\ref{accumulation_cond1}) has two solutions
$\psi_{\ast}^c$ corresponding to the accumulation points belonging to $\beta^s
(\psi)$. Computing $\psi_{\ast}^c$ from (\ref{accumulation_cond}) or
(\ref{accumulation_cond1}) we find a growth rate of $d_n^c$ for large $n$ which
is given by
$$
\frac{\sin\psi_{\ast}^c}{2\alpha_{\ast}^c\sin\left(\theta_1+\psi_{\ast}^c
\right)}\hat I_2\left(-\sin^2\theta_2\right)=\frac{\sin2\psi_{\ast}^c}{2\sin
\left(\theta_1+2\psi_{\ast}^c\right)}\hat I_2\left(-\sin^2\theta_2\right).
$$
\section{Touching and Intersecting Bodies}\label{r13}
When the distance $d$ between the solids is non-positive it leads to strong
simplification of the topological structure of the solutions of (\ref{YLeq}).
First, only a single branch of the solution that always includes ${\sf Nod^+}$
meniscus survives. It follows from the statement made in section \ref{r40} that
spheres ${\sf Sph^+_n}$, $n>0$, cannot exist when $d\le 0$, so that only the
trajectory that contains ${\sf Sph^+_0}$ can be observed in $\{\psi,\alpha\}$
plane.

For $d=0$ (the solid sphere on the plane) we show in Appendix \ref{nod_as0} that
the curvature of the ${\sf Nod^-}$ is negative and diverges as $H\sim\psi^{-2}$ 
as $\psi\to 0$, confirming the result reported in \cite{Orr1975}. In case 
$\theta_1+\theta_1>\pi$ one can observe similar divergence of positive curvature
for ${\sf Nod^+}$ at small $\psi$. The same ${\sf Nod^+}$ meniscus can be found
in special case $\theta_1+\theta_1=\pi$ when the curvature diverges as $H\sim
\psi^{-1}$. In two last cases the whole trajectory is represented by ${\sf
Nod^+}$ meniscus (see Figure \ref{qi8}(b)).
\begin{figure}[h!]
\begin{tabular}{cc}
\psfig{figure=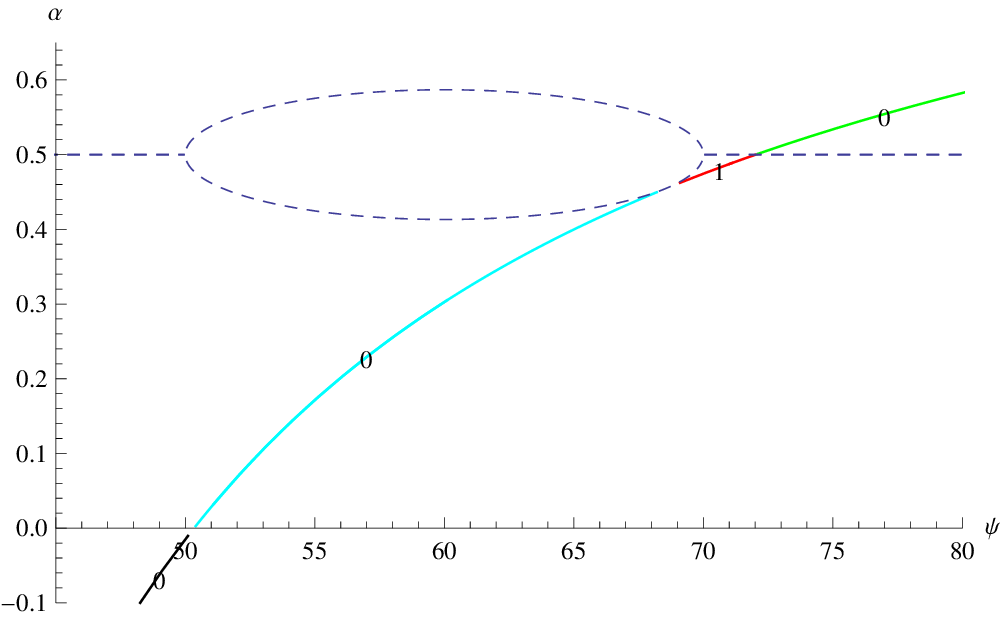,width=7cm,height=5cm}&\hspace{1cm}
\psfig{figure=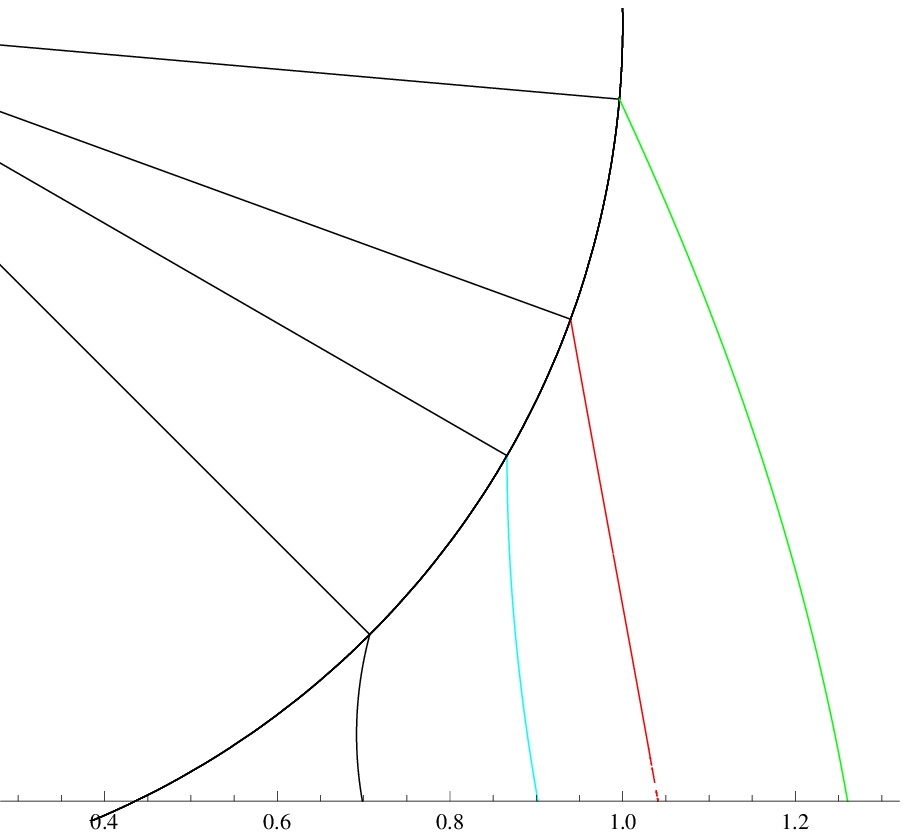,width=7cm,height=5cm}\\
(a) & (b)
\end{tabular}
\caption{(a) Plot $\alpha(\psi)$ for $\theta_1=30^o$, $\theta_2=80^o$ and $d=-
0.1$, i.e., $\psi_*=25.84^o$, and (b) four menisci ${\sf Nod^-}$ ({\em black}),
${\sf Und^-_0}$ ({\em cyan}), ${\sf Und^-_1}$ ({\em red}) and ${\sf Und^+_0}$
({\em green}) for $\psi=45^o$, $\psi=60^o$, $\psi=70^o$ and $\psi=85^o$,
respectively.}
\label{intersect_menis}
\end{figure}

Another divergent behavior of the curvature is observed when $d<0$ (the solid
sphere intersecting the plane). In this case the menisci can exist only for
$\psi\ge\psi_{\ast}=\arccos(1+d)$. Formula (\ref{nod_as13}) shows that the
curvature diverges as $H\sim (\psi-\psi_{\ast})^{-1}$ as $\psi\to\psi_{\ast}$.
Depending on the parameters values this divergence is observed for menisci,
\bea
{\sf Nod^-}:\quad H<0,\;\;\theta_1+\theta_1+\psi_{\ast}<\pi,\quad\mbox{and}\quad
{\sf Nod^+}:\quad H>0,\;\;\theta_1+\theta_1+\psi_{\ast}>\pi.\label{asym1}
\eea
When $\theta_1+\theta_1+\psi_{\ast}=\pi$ we show in \ref{nod_as0} that the
${\sf Nod^-}$ meniscus is forbidden while the ${\sf Nod^+}$ meniscus is observed
in the whole range $\psi>\psi_{\ast}$ and its curvature does not diverge.
In Figure \ref{intersect_menis} the $\alpha(\psi)$ trajectory for contact angles
$\theta_1=30^o$, $\theta_2=80^o$ and $d=-0.1$ is drawn. We present also the
four different menisci observed in this case.

The case of intersecting bodies $d<0$ does not lead to change of formula
(\ref{curv}) for meniscus curvature $H$ and (\ref{surf}) for surface area $S$.
However formula for the volume $V_-(d)$ of meniscus with $d<0$ reads
\bea
V_-(d)=\frac{\pi}{8H^3}J_s-\frac{\pi}{3}\left(2-3\cos\psi+\cos^3\psi\right)+
\frac{\pi d^2}{3}(3+d),\quad d<0,\label{asym2}
\eea
where integral $J_s$ is given in (\ref{volume}).
\section{Concluding Remarks and Open Problems}\label{conclus}
Extending the rigorous approach used \cite{Orr1975} to describe the menisci
shapes between two touching ($d=0$) axisymmetric solids, sphere and plane, we
develop a theory of pendular rings in its general form for the separated ($d>0$)
or intersecting ($d<0$) solids. The main results are listed below.
\begin{enumerate}
\item The YL equation (\ref{YLeq}) with boundary conditions can be viewed as a
nonlinear eigenvalue problem. Its unduloidal solutions exhibit a discrete
spectrum and are enumerated by two indices: the number $n\in{\mathbb Z}_+\cup
\{0\}$ of inflection points on the meniscus meridional profile ${\cal M}$ and
the index $s=\mbox{sgn}(2\alpha_n-1)$ determined by the shape of a segment of
the curve ${\cal M}$ touching the solid sphere: the shape is either convex,
$s=1$, or concave, $s=-1$.

Menisci shapes $z_n^s(r)$ and their curvatures $H_n^s$ play a role of
eigenfunctions and eigenvalues of equation (\ref{YLeq}), respectively. The
Neeman boundary conditions, two contact angles $\theta_1$ and $\theta_2$, and a
single governing parameter $d>-2$ together with one more parameter, the filling
angle $\psi$, completely determine the meniscus shape $z_n^s(r)$ and curvature
evolution $H=H_n^s(\psi)$.
\item For the fixed $\theta_1$ and $\theta_2$ the set of the functions
$H_n^s(\psi)$ behaves in such a way that in the plane $\{\psi,H\}$ there exists
a bounded domain ${\mathbb B}$ where $H_n^s(\psi)$ do not exist for any $d$.

Under non-linear transformation $\alpha_n(\psi)=H_n^s(\psi)\sin\psi/\sin(
\theta_1+\psi)$ this domain ${\mathbb B}$ in the plane $\{\psi,\alpha\}$ takes
a simple shape with a smooth boundary $\beta^{\pm}(\psi)$ which we call a {\em
balloon}. At this boundary and at the line $\alpha=1/2$ there occur all
transitions between different types of unduloidal menisci. The other two lines
$\alpha=0,1$ are also the locations of special menisci: all spherical menisci
${\sf Sph^+_n}$ at $\alpha=1$, all spherical menisci ${\sf Sph^-_n}$ at
$\alpha=0$ and also catenoidal menisci ${\sf Cat}$ at $\alpha=0$.

A behavior of $\alpha_n(\psi)$ curves reminds in some cases the 2-dim dynamical
system with trajectories ramified in a {\em non-simply connected} domain
$\Delta^{\prime}$. This global representation allows to classify possible
trajectories and introduce a saddle point notion into the PR problem. We observe
several types of saddle points and give their classification.
\item If the distance between the solids is non-positive, $d\leq 0$, then a
single (possibly disconnected) sequence of solutions (menisci), that always
includes the ${\sf Nod^+}$ type, survives. We describe the asymptotic behavior
of the mean curvature $H_0^-(\psi)$ of nodoidal meniscus in vicinity of a
singular point $\psi_*=\arccos(1+d)$; such singular point does not exist for 
$d>0$.
\end{enumerate}
Beyond the scope of the present paper we have left several questions which are
related to the theory developed here. Below we mention two of them.
\begin{enumerate}
\item The theory of pendular rings in the special cases of boundary conditions:
\begin{itemize}
\item $\theta_1=0$, this implies $\alpha=H$ and asymptotics (\ref{nod_as8}) of
the ${\sf Nod^+}$ meniscus curvature fails. This is the case of completely
wetted sphere.
\item $\theta_2=\pi/2$, the balloon ${\mathbb B}$ is reduced to a line $\alpha
=1/2$ with one singular point $\psi=\pi/2-\theta_1$. This is the case of two
solid spheres of equal radii.
\item $\theta_1=\theta_2$, the balloon ${\mathbb B}$ is located in region
$0\leq\psi\leq\pi-2\theta_1$ and the domain $\Delta^{\prime}$ becomes simply
connected. This is the case of two solids of the same material.
\item $\theta_1+\theta_2=\pi$, $\theta_2>\theta_1$, the balloon ${\mathbb B}$ is
located in region $0\leq\psi\leq\theta_2-\theta_1$ and the domain $\Delta^{
\prime}$ becomes simply connected. The physical meaning of this case is unclear.
\end{itemize}
\item Stability of pendular rings. In this regard, we know only two papers where
the stability was studied for the ${\sf Nod^+}$ and ${\sf Und_0^+}$ menisci
\cite{Vo06} and those menisci which occur as volume decreases from a convex
bridge \cite{Vo12} between two solid spheres of equal radii. The rest of
meniscus types including those with inflection points are open for analysis.
\end{enumerate}
\section*{Acknowledgement}\label{a9}
The useful discussions with R. Finn and T. Vogel are appreciated. We thank T.
Vogel for sending us the preprint \cite{Vo12} submitted for publication. The
research was supported in part (LGF) by the Kamea Fellowship.
\section*{Appendices}
\appendix
\section{Menisci Formulae}\label{appendix1}
\setcounter{equation}{0}
\subsection{Catenoids ${\sf Cat}$}\label{ar1}
The catenoid case is the simplest one that requires a solution of the equation
(\ref{main_eq}) with $H=0$. This shape corresponds to the transition from the
concave nodoid to concave unduloid. The solution for positive $x$ and $C$
reads $x=C/\sin t$. Using the boundary condition we find
\be
C=\sin t_1\sin\psi\;.\label{c_catenoid}
\ee
Employing (\ref{y(u)}) we obtain the vertical component of the catenoid
meridional profile
\be
y=\int_{t_2}^{t}\tan t\;dx=-C\int_{t_2}^{t}\frac{dt}{\sin t}=C\left(\ln\cot
\frac{t}{2}-\ln\cot\frac{t_2}{2}\right)\;.\label{y_catenoid}
\ee
Using the boundary condition (\ref{BC}) we find a relation
\be
1+d-\cos\psi=\sin t_1\sin\psi\left(\ln\cot\frac{t_1}{2}-\ln\cot\frac{t_2}{2}
\right)=\sin t_1\sin\psi\;\ln\frac{\sin t_2(1+\cos t_1)}{\sin t_1(1+\cos t_2)}
\;,\label{psi_cond_catenoid}
\ee
that implicitly defines the value of the filling angle $\psi$ at which catenoid
is found. The last condition can be rewritten in the form
\be
\tan\frac{t_1}{2}\cot\frac{t_2}{2}=\exp\left(-\frac{1+d-\cos\psi}{\sin\psi\sin
t_1}\right)\;.\label{catenoid_psi}
\ee
Shape of catenoid is found in parametric form
\be
x(t)=\frac{C}{\sin t}\;,\quad y(t)=C\left(\ln\cot\frac{t}{2}-\ln\cot\frac{t_2}
{2}\right)\;.\label{shape_catenoid}
\ee
The solid of rotation volume is given by the formula (\ref{volume}) with 
$x=C/\sin t$. From (\ref{y(u)}) we find $dy=-C dt/\sin t$, so that the ring 
volume reads
$$
-\pi C^3\int_{t_2}^{t_1}\frac{dt}{\sin^3 t}=\frac{\pi C^3}{2}\left(\frac{\cos
t_1}{\sin^2t_1}-\frac{\cos t_2}{\sin^2t_2}+\ln\cot\frac{t_1}{2}-\ln\cot\frac{
t_2}{2}\right).
$$
The volume of the ${\sf Cat}$ meniscus reads
\be
V=\frac{\pi C^3}{2}\left(\frac{\cos t_1}{\sin^2t_1}-\frac{\cos t_2}{\sin^2t_2}
+\ln\cot\frac{t_1}{2}-\ln\cot\frac{t_2}{2}\right)-\frac{\pi}{3}(2-3\cos\psi+
\cos^3\psi)\;.\label{volume_catenoid}
\ee
The surface area of the ${\sf Cat}$ meniscus is given by (\ref{surf})
with $x=C/\sin t$ producing
\be
S=-2\pi C^2\int_{t_2}^{t_1}\frac{dt}{\sin^3 t}=\pi C^2\left(\frac{\cos t_1}
{\sin^2t_1}-\frac{\cos t_2}{\sin^2t_2}+\ln\cot\frac{t_1}{2}-\ln\cot\frac{t_2}
{2}\right)\;.\label{surface_catenoid}
\ee
\subsection{Nodoids ${\sf Nod^{\pm}}$}\label{ar3}
The curvature expression reads
\be
2H\Psi=I_1(t_1,t_2)+sI_2(t_1,t_2)\;.\label{curvature_nodoid}
\ee
The meniscus nodoid shape is given by
\bea
x(t)=\frac{1}{2H}\left(\sin t+s\sqrt{\sin^2t+c}\right)\;,\quad
y(t)=\frac{1}{2H}\left[I_1(t,t_2)+s I_2(t,t_2)\right]\;.\label{x(t)nod}
\eea
Using the formulas (\ref{volume}) and (\ref{Jpm}) we have for the nodoid volume
\be
V=\frac{\pi}{8H^3}\left[4J_3(t_1,t_2)+cI_1(t_1,t_2)-scI_2(t_1,t_2)+4sJ_2(
t_1,t_2)\right]-\frac{\pi}{3}(2-3\cos\psi+\cos^3\psi)\;.\label{volume_nodoid}
\ee
Using the formulas (\ref{surf}) and (\ref{K}) we find the nodoid surface
area
\be
S=\frac{\pi}{2H^2}K_s(t_1,t_2)\;.\label{surface_nodoid}
\ee
\subsection{Unduloids ${\sf Und_0^{\pm}}$}\label{ar2}
The curvature expression reads
\be
2H_0^s\Psi=I_1(t_1,t_2)+sI_2(t_1,t_2)\;.\label{curvature_unduloid}
\ee
The meniscus unduloid shape is given by (\ref{solx}) as
\bea
x(t)=\frac{1}{2H_0^s}\left(\sin t+s\sqrt{\sin^2t+c}\;\right)\;,\quad
y(t)=\frac{1}{2H_0^s}\left[I_1(t,t_2)+sI_2(t,t_2)\right]\;.\label{x(t)und}
\eea
Using the formulae (\ref{volume}) and (\ref{Jpm}) we have for the unduloid
volume
\be
V_0^s=\frac{\pi}{8(H_0^s)^3}\left[4J_3(t_1,t_2)+cI_1(t_1,t_2)-scI_2(t_1,t_2)+4s
J_2(t_1,t_2)\right]-\frac{\pi}{3}(2-3\cos\psi+\cos^3\psi)\;.
\label{volume_unduloid}
\ee
Using the formulae (\ref{surf}) and (\ref{K}) we have for the unduloid surface
area
\be
S_0^s=\frac{\pi}{2(H_0^s)^2}K_s(t_1,t_2)\;.\label{surface_unduloid}
\ee
\subsection{Inflectional Unduloids ${\sf Und_1^{\pm}}$ with Single Inflection
Point}\label{ar4}
The inflection point $u_{\ast}$ of unduloid satisfies a condition $u_{\ast}
^2+c=0$ for negative $c$. The inflection point at the sphere surface
corresponds to
\be
H_n^s=\frac{\sin t_1}{2\sin\psi},\label{inflect_sphere}
\ee
while when this point is at the plane
\be
H_n^s=\frac{\sin t_1+s\sqrt{\sin^2t_1-\sin^2 t_2}}{2\sin\psi}.
\label{inflect_plane}
\ee
Substitution of (\ref{inflect_sphere}, \ref{inflect_plane}) into
(\ref{curvature_unduloid}) for given $s$ generates equations for the
critical values $\psi_1^{\ast}$ and $\psi_2^{\ast}$ of the filling angle at
which the inflection point is at the sphere and at the plane, respectively. In
case $\psi_1^{\ast}=\psi_2^{\ast}$ inflectional unduloid reduces to the
cylinder reached for $t_2=\pi/2$ at $t_1^{\ast}=\pi/2$ for all
$\theta_1<\pi/2$. It has curvature equal to $H_1^s=(2\cos\theta_1)^{-1}$.

The integrals in (\ref{soly}) and (\ref{curv}) in case of inflectional unduloid
should be broken into two integrals. The meridional profile is made of two
unduloid profiles matching at the point $\{x_{\ast},y_{\ast}\}$, i.e., $u=u_{
\ast}$. Consider the case of the ${\sf Und_1^-}$ meniscus when the profiles
touching the plane and the sphere have positive and negative curvature,
respectively. Using (\ref{x(t)und}) we write for the convex unduloid part
\bea
x(t)=\frac{1}{2H_1^-}\left(\sin t+\sqrt{\sin^2t+c}\;\right)\;,\quad y(t)=
\frac{1}{2H_1^-}\left[I_1(t,t_2) + I_2(t,t_2)\right]\;,\quad t\in\{t_2,
t_{\ast}\}\;.\label{x_lower_pos}
\eea
The upper concave unduloid part is given by
\bea
x(t)=\frac{1}{2H_1^-}\left(\sin t-\sqrt{\sin^2t+c}+A_x\right),\quad y(t)=\frac{
1}{2H_1^-}\left[I_1(t,t_2)-I_2(t,t_2)+A_y\right]\;,\quad t\in\{t_1,t_{\ast}\}.
\nonumber
\eea
The values of $A_x$ and $A_y$ have to be found from the matching conditions at
$t=t_{\ast}$. Using $u_{\ast}=\sin t_{\ast}=\sqrt{-c}$ we get $t_{\ast}^-=\pi-
\arcsin\sqrt{-c}$ and $\cos t_{\ast}^-=-\sqrt{1+c}$. At the inflection point
we find
\bea
x_{\ast}&=&\frac{\sqrt{-c}}{2H_1^-}=\frac{\sqrt{-c}+A_x}{2H_1^-},\nonumber\\
y_{\ast}&=&\frac{1}{2H_1^-}\left[I_1(t_{\ast}^-,t_2)+I_2(t_{\ast}^-,t_2)\right]=
\frac{1}{2H_1^-}[I_1(t_{\ast}^-,t_2)-I_2(t_{\ast}^-,t_2)+A_y].\label{y_infl}
\eea
The matching conditions produce $A_x=0,\ A_y=2I_2(t_{\ast}^-,t_2),$ leading to
the following shape of upper concave unduloid
\bea
x(t)=\frac{1}{2H_1^-}\left(\sin t-\sqrt{\sin^2t+c}\right),\quad
y(t)=\frac{1}{2H_1^-}\left[I_1(t,t_2)-I_2(t,t_2)+2I_2(t_{\ast}^-,t_2)\right].
\label{y_upper_neg}
\eea
Using the second equation in (\ref{y_upper_neg}) we have for $t=t_1$
\be
2H_1^-\Psi=I_1(t_1,t_2)-I_2(t_1,t_2)+2I_2(t_{\ast}^-,t_2).\label{curv_infl1}
\ee
The case of the ${\sf Und_1^+}$ meniscus when the profile touching the plate has
negative curvature, and the profile touching the sphere has positive curvature
is treated similarly and we obtain the general expression for the curvature:
\be
2H_1^s\Psi=I_1(t_1,t_2)+s\left[I_2(t_1,t_2)-2I_2(t_{\ast}^s,t_2)\right].
\label{curv_infl}
\ee
\subsubsection{Shape}\label{ar41}
The meniscus shape is given by the following general expressions:
\bea
x(t)=\frac{1}{2H_1^s}\left(\sin t-s\sqrt{\sin^2t+c}\right),\;
y(t)=\frac{1}{2H_1^s}\left[I_1(t,t_2)-sI_2(t,t_2)\right],\;
t\in\{t_2,t_{\ast}^s\},\nonumber
\eea
\bea
x(t)=\frac{1}{2H_1^s}\left(\sin t+s\sqrt{\sin^2t+c}\right),\
y(t)=\frac{1}{2H_1^s}\left[I_1(t,t_2)+s(I_2(t,t_2)-2I_2(t_{\ast}^s,t_2))\right]
,\ t\in\{t_1,t_{\ast}^s\}.\nonumber
\eea
\subsubsection{Volume}\label{ar42}
As the ${\sf Und_1^-}$ meniscus is made of two menisci having shape of
concave (upper) and convex (lower) unduloids, a solid of rotation volume equals
sum of volumes $V_l$ and $V_u$ of lower and upper parts, respectively,
\bea
V_l&=&\frac{\pi}{8(H_1^-)^3}\left[4J_3(t_{\ast}^-,t_2)+cI_1(t_{\ast}^-,t_2)-
cI_2(t_{\ast}^-,t_2)+4J_2(t_{\ast}^-,t_2)\right],\nonumber\\
V_u&=&\frac{\pi}{8(H_1^-)^3}\left[4J_3(t_1,t_{\ast}^-)+cI_1(t_1,t_{\ast}^-)+
cI_2(t_1,t_{\ast}^-)-4J_2(t_1,t_{\ast}^-)\right].\nonumber
\eea
Adding up the above expressions we have for the meniscus volume
\bea
V_1^-&=&\frac{\pi}{8(H_1^-)^3}\left\{4J_3(t_1,t_2)+cI_1(t_1,t_2)-c\left[I_2(
t_{\ast}^-,t_2)-I_2(t_1,t_{\ast}^-)\right]+4\left[J_2(t_{\ast}^-,t_2)-
J_2(t_1,t_{\ast}^-)\right]\right\}\nonumber\\
&-&\frac{\pi}{3}\left(2-3\cos\psi+\cos^3\psi\right)\;.
\label{volume_unduloid_infl_spneg}
\eea
The general formula for the volume of inflectional unduloid reads
\bea
V_1^s&=&\frac{\pi}{8(H_1^-)^3}\{4J_3(t_1,t_2)+cI_1(t_1,t_2)+sc\left[I_2(t_{
\ast}^s,t_2)-I_2(t_1,t_{\ast}^s)\right]\nonumber\\
&-&4s\left[J_2(t_{\ast}^s,t_2)-J_2(t_1,t_{\ast}^s)\right]\}-\frac{\pi}{3}
\left(2-3\cos\psi+\cos^3\psi\right)\;.\label{volume_unduloid_infl}
\eea
\subsubsection{Surface Area}\label{ar43}
Apply the same approach to calculation of the surface area of the ${\sf 
Und_1^-}$ meniscus. The area $S_1^-$ equals the sum of the surface areas 
$S_l$ and $S_u$ of lower (convex) and upper (concave) parts, respectively,
$$
S_l=\frac{\pi}{2H^2}K_+(t_{\ast}^-,t_2)\;, \ \
S_u=\frac{\pi}{2H^2}K_-(t_1,t_{\ast}^-).
$$
Adding up the above expressions we have for the meniscus surface area
\be
S_1^-=\frac{\pi}{2(H_1^-)^2}[K_+(t_{\ast}^-,t_2)+K_-(t_1,t_{\ast}^-)],
\label{surface_unduloid_infl_spneg}
\ee
and we find the formula for the inflectional unduloid surface area
\be
S_1^s=\frac{\pi}{2(H_1^s)^2}[K_{-s}(t_{\ast}^s,t_2)+K_s(t_1,t_{\ast}^s)].
\label{surface_unduloid_infl}
\ee
\subsection{Inflectional Unduloids ${\sf Und_2^{\pm}}$ with Two Inflection
Points}\label{ar4a}
In previous section we consider the simplest basic inflectional unduloid
structure characterized by a single inflection point. We show that if the
inflection point $t_{\ast}^-$ originates at the plane the ${\sf Und_1^-}$
meniscus emerges, while separation of the  inflection point $t_{\ast}^+$ from
the sphere generates the ${\sf Und_1^+}$ meniscus.

It is shown in section \ref{r45} that the value $\alpha=1/2$ is a critical
point at which a transition between the ${\sf Und_1^-}$ and the ${\sf Und_0^+}$
menisci takes place when $t_{\ast}^-=t_1$ and the single inflection point
reaches the solid sphere. What does happen when $t_{\ast}^-=\pi-t_1$ and the
inflection point is inside the meridional profile of the ${\sf Und_1^-}$ 
meniscus? In this case the second inflection point, namely, with $t_{\ast}^+=
t_1$, separates from the sphere and we observe a meniscus ${\sf Und_2^+}$ having
two inflection points $t_{\ast}^+$ and $t_{\ast}^-$. The profile of such 
meniscus is made of three unduloid segments -- two convex (touching both the 
sphere and the plane) and a concave one between them.

Consider derivation of the equation for the curvature for this meniscus. Using
(\ref{x(t)und}) we write for the lower convex unduloid part touching the plane
$$
x(t)=\frac{1}{2H_2^+}\left(\sin t+\sqrt{\sin^2t+c}\;\right),\ y(t)=
\frac{1}{2H_2^+}\left[I_1(t,t_2)+I_2(t,t_2)\right],\ t\in\{t_2,t_{\ast}^-\}.
$$
The middle concave unduloid part is given by
$$
x(t)=\frac{1}{2H_2^+}\left(\sin t-\sqrt{\sin^2t+c}+A_{x1}\;\right),\ y(t)=
\frac{1}{2H_2^+}\left[I_1(t,t_2)-I_2(t,t_2)+A_{y1}\right],\ t\in\{t_{\ast}^-,
t_{\ast}^+\}.
$$
Finally, for the upper convex unduloid part touching the sphere we write
$$
x(t)=\frac{1}{2H_2^+}\left(\sin t+\sqrt{\sin^2t+c}+A_{x2}\;\right),\
y(t)=\frac{1}{2H_2^+}\left[I_1(t,t_2)+I_2(t,t_2)+A_{y2}\right],\ t\in\{t_1,
t_{\ast}^+\}.
$$
The values of $A_{xi}$ and $A_{yi}$ have to be found from the matching
conditions at $t=t_{\ast}^{\mp}$ producing
$$
A_{x1}=A_{x2}=0,\quad A_{y1}=2I_2(t_{\ast}^-,t_2),\quad A_{y2}=2I_2(t_{\ast}
^-,t_2)-2I_2(t_{\ast}^+,t_2)=2I_2(t_{\ast}^-,t_{\ast}^+),
$$
leading to the following shape of upper convex unduloid
\bea
x(t)=\frac{1}{2H_2^+}\left(\sin t+\sqrt{\sin^2t+c}\;\right)\;,\ y(t)=\frac{1}
{2H_2^+}\left[I_1(t,t_2)+I_2(t,t_2)+2I_2(t_{\ast}^-,t_{\ast}^+)\right]\;.
\label{y_upper_pos2}
\eea
Using the second equation in (\ref{y_upper_pos2}) we have for $t=t_1$
\be
2H_2^+\Psi=I_1(t_1,t_2)+I_2(t_1,t_2)+2I_2(t_{\ast}^-,t_{\ast}^+).
\label{curv_infl1a}
\ee
The case of the ${\sf Und_2^-}$ meniscus when the profiles touching both the
plate and the sphere have negative curvature is treated similarly to obtain
\be
2H_2^-\Psi=I_1(t_1,t_2)-I_2(t_1,t_2)+2I_2(t_{\ast}^-,t_{\ast}^+).
\label{curv_infl2a}
\ee
Thus, we obtain the general expression for the curvature for the ${\sf Und_2^{
\pm}}$ menisci
\be
2H_2^s\Psi=I_1(t_1,t_2)+sI_2(t_1,t_2)+2I_2(t_{\ast}^-,t_{\ast}^+),
\label{curv_infla}
\ee
As shown in \ref{app32} the value $\hat I_2$ of integral $I_2(t_{\ast}^-,
t_{\ast}^+)$  reads $\hat I_2=2\sqrt{-c}E(1+1/c)$.
\subsubsection{Shape}\label{ar41a}
The meniscus shape is given by the following general expressions:
\bea
x(t)=\frac{1}{2H_2^s}\left(\sin t+s\sqrt{\sin^2t+c}\;\right),\ y(t)=\frac{1}{2
H_2^s}\left[I_1(t,t_2) +s I_2(t,t_2)\;\right],\ t\in\{t_2,t_{\ast}^{-s}\},
\label{y_plate_a}
\eea
\bea
x(t)=\frac{1}{2H_2^s}\left(\sin t-s\sqrt{\sin^2t+c}\;\right), \
y(t)=\frac{1}{2H_2^s}\left[I_1(t,t_2)-s(I_2(t,t_2)-2I_2(t_{\ast}^s,t_2))
\right], \ t\in\{t_{\ast}^-,t_{\ast}^+\},\nonumber
\eea
\bea
x(t)=\frac{1}{2H_2^s}\left(\sin t+s\sqrt{\sin^2t+c}\;\right),\
y(t)=\frac{1}{2H_2^s}\left[I_1(t,t_2)+sI_2(t,t_2)+2\hat I_2\right],\
t\in\{t_1,t_{\ast}^s\}.\nonumber
\eea
\subsubsection{Volume}\label{ar42a}
Inflectional unduloid ${\sf Und_2^+}$ meniscus is made of three menisci
having shape of concave (middle segment) and convex (upper and lower segments)
unduloids; its volume equals the sum of the volumes $V_l, V_m$ and
$V_u$ of lower, middle and upper parts, respectively:
\bea
V_l&=&\frac{\pi}{8(H_2^+)^3}[4J_3(t_{\ast}^-,t_2)+cI_1(t_{\ast}^-,t_2)-cI_2
(t_{\ast}^-,t_2)+4J_2(t_{\ast}^-,t_2)]\;,\nonumber\\
V_m&=&\frac{\pi}{8(H_2^+)^3}[4J_3(t_{\ast}^+,t_{\ast}^-)+cI_1(t_{\ast}^+,t_{
\ast}^-)+cI_2(t_{\ast}^+,t_{\ast}^-)-4J_2(t_{\ast}^+,t_{\ast}^-)]\;,\nonumber\\
V_u&=&\frac{\pi}{8(H_2^+)^3}[4J_3(t_1,t_{\ast}^+)+cI_1(t_1,t_{\ast}^+)-cI_2(
t_1,t_{\ast}^+)+4J_2(t_1,t_{\ast}^+)]\;.\nonumber
\eea
Adding up the above expressions we have for the meniscus volume
\bea
V_2^+&=&\frac{\pi}{8(H_2^+)^3}\{4J_3(t_1,t_2)+cI_1(t_1,t_2)-c[I_2(t_{\ast}^-,
t_2)-I_2(t_{\ast}^+,t_{\ast}^-)+I_2(t_1,t_{\ast}^+)]\nonumber\\
&+&4[J_2(t_{\ast}^-,t_2)-J_2(t_{\ast}^+,t_{\ast}^-)+J_2(t_1,t_{\ast}^+)]\}-
\frac{\pi}{3}(2-3\cos\psi+\cos^3\psi)\;.\label{volume_unduloid_infl_sppos2}
\eea
Using the properties of the integrals $I_2$ and $J_2$ we find:
\bea
V_2^+&=&\frac{\pi}{8(H_2^+)^3}\{4J_3(t_1,t_2)+cI_1(t_1,t_2)-c[I_2(t_1,t_2)
-2\hat I_2]\nonumber\\
&+&4[J_2(t_1,t_2)-2\hat J_2]\}-\frac{\pi}{3}(2-3\cos\psi+
\cos^3\psi)\;,\label{volume_unduloid_infl_sppos2a}
\eea
where the value $\hat J_2$ of integral $J_2(t_{\ast}^-,t_{\ast}^+)$ is computed
in (\ref{J_2_hat}). The general expression for the menisci volume reads
\bea
V_2^s&=&\frac{\pi}{8(H_2^s)^3}\{4J_3(t_1,t_2)+cI_1(t_1,t_2)-sc[I_2(t_1,t_2)+
2s\hat I_2]\nonumber\\
&+&4s[J_2(t_1,t_2)+2s\hat J_2]\}-\frac{\pi}{3}(2-3\cos\psi+\cos^3\psi)\;.
\label{volume_unduloid_infl_sp2a}
\eea
\subsubsection{Surface Area}\label{ar43a}
Similar approach is applied for calculation of the surface area of the ${\sf
Und_2^+}$ meniscus. The area $S_2^+$ equals the sum of the surface areas $S_l$,
$S_m$ and $S_u$ of lower, middle and upper segments, respectively,
$$
S_l=\frac{\pi}{2(H_2^+)^2}K_+(t_{\ast}^-,t_2)\;, \ \
S_m=\frac{\pi}{2(H_2^+)^2}K_-(t_{\ast}^+,t_{\ast}^-)\;, \ \
S_u=\frac{\pi}{2(H_2^+)^2}K_+(t_1,t_{\ast}^+).
$$
Adding up the above expressions we have for the meniscus surface area
\be
S_2^+=\frac{\pi}{2(H_2^+)^2}\left[K_+(t_{\ast}^-,t_2)+K_-(t_{\ast}^+,t_{\ast}^-)
+K_+(t_1,t_{\ast}^+)\right].\label{surface_unduloid_infl_sppos2a}
\ee
We give a general expression for the menisci surface area
\be
S_2^s=\frac{\pi}{2(H_2^s)^2}\left[K_s(t_{\ast}^{-s},t_2)-s\hat K_{-s}+K_s(t_1,
t_{\ast}^s)\right].\label{surface_unduloid_infl_sp2a}
\ee
\subsection{Inflectional Unduloids ${\sf Und_3^{\pm}}$ with Three Inflection
Points}\label{ar4b}
In \ref{ar4a} we consider the inflectional unduloid ${\sf Und_2^{\pm}}$ with two
inflection points for the values of inflection points parameter $0< t_{\ast}^{
\pm}<\pi$. There exist also menisci with larger number of inflection
points.

Consider first an inflectional meniscus ${\sf Und_3^-}$ with three inflection
points. A natural way to generate it is to consider the ${\sf Und_2^+}$ unduloid
having two inflection points $t_{\ast,0}^+=t_1$ and $t_{\ast,0}^-=\pi-t_1$
and allow at $\alpha=1/2$ a third inflection point $t_{\ast,1}^-=\pi-t_1$ to
separate from the sphere. This point appears due to vertical translational
periodicity of the meridional profile. The profile of such a meniscus is made
of four unduloids -- two convex (one of them touches the sphere) and two concave
(one touches the plane). In all formulas below we drop additional indices $k$
of $t_{\ast,k}^{\pm}$. Derivation of the curvature equation for this meniscus
is similar to the one of ${\sf Und_1^s}$ and ${\sf Und_2^s}$,
\be
2H_3^-\Psi=I_1(t_1,t_2)-I_2(t_1,t_2)+2\hat I_2+2I_2(t_{\ast}^-,t_2).
\label{curv_infl3m}
\ee
A dual inflectional meniscus ${\sf Und_3^+}$ is generated from the ${\sf Und_2
^+}$ unduloid at $\alpha=\beta^+$ by separation of a third inflection point
$t_{\ast}^+=t_1$ from the plane:
\be
2H_3^+\Psi=I_1(t_1,t_2)+I_2(t_1,t_2)+2\hat I_2-2I_2(t_{\ast}^+,t_2).
\label{curv_infl3p}
\ee
Merging (\ref{curv_infl3m},\ref{curv_infl3p}) we arrive at the general
formula for the curvature
\be
2H_3^s\Psi=I_1(t_1,t_2)+s I_2(t_1,t_2)-2sI_2(t_{\ast}^s,t_2)+2\hat I_2.
\label{curv_infl3}
\ee
\subsubsection{Shape}\label{ar41b}
The shape of the ${\sf Und_3^s}$ meniscus is given by the following general
expressions
\bea
x(t)=\frac{1}{2H_3^s}\left(\sin t-s\sqrt{\sin^2t+c}\right),\;y(t)=\frac{1}{
2H_3^s}[I_1(t,t_2)-sI_2(t,t_2)],\;t\in\{t_2,t_{\ast}^s\},\label{y_plate_d}
\eea
\bea
x(t)=\frac{1}{2H_3^s}\left(\sin t+s\sqrt{\sin^2t+c}\right),\;y(t)=\frac{1}{
2H_3^s}[I_1(t,t_2)+sI_2(t,t_2)-2sI_2(t_{\ast}^s,t_2)],\;t\in\{t_{\ast}^s,
t_{\ast}^{-s}\}.\nonumber
\eea
\bea
x(t)=\frac{1}{2H_3^s}\left(\sin t-s\sqrt{\sin^2t+c}\right),\;y(t)=\frac{1}{
2H_3^s}[I_1(t,t_2)-sI_2(t,t_2)+ 2\hat I_2],\;t\in\{t_{\ast}^{-s},t_{\ast}^s\},
\label{y_middle_dd}
\eea
\bea
x(t)&=&\frac{1}{2H_3^s}\left(\sin t+s\sqrt{\sin^2t+c}\right),\nonumber\\
y(t)&=&\frac{1}{2H_3^s}[I_1(t,t_2)+s I_2(t,t_2)+ 2\hat I_2- 2sI_2(t_{\ast}^s,
t_2)],t\in\{t_1,t_{\ast}^s\}.\label{y_sphere_d}
\eea
\subsubsection{Volume}\label{ar42b}
Inflectional unduloid ${\sf Und_3^s}$ meniscus is made of four menisci having
shape of concave and convex unduloids; its volume $V_3^s$ equals the sum of
the volumes of corresponding parts:
\bea
V_3^s&=&\frac{\pi}{8(H_3^s)^3}\{4J_3(t_1,t_2)+cI_1(t_1,t_2)+sc[I_2(t_1,t_2)-
2s\hat I_2-2I_2(t_1,t_{\ast}^s)]\nonumber\\
&-& 4s[J_2(t_1,t_2)-2s\hat J_2-2J_2(t_1,t_{\ast}^s)]\}-\frac{\pi}{3}(2-3\cos
\psi+\cos^3\psi).\label{volume_unduloid_infl_sp3a}
\eea
\subsubsection{Surface Area}\label{ar43b}
The surface area $S_3^s$ of the ${\sf Und_3^s}$ meniscus equals the sum of
the surface areas of corresponding segments:
\be
S_3^s=\frac{\pi}{2(H_3^s)^2}\left[K_{-s}(t_{\ast}^s,t_2)+\hat K_+-\hat K_-+
K_s(t_1,t_{\ast}^s)\right].\label{surface_unduloid_infl_sp3}
\ee
\subsection{Inflectional Unduloids ${\sf Und_{2k}^{\pm}}$ with Even Number of
Inflection Points}\label{ar4even}
Generalization of the menisci ${\sf Und_0^s}$ and ${\sf Und_2^s}$ to arbitrary
even number of inflection points is straightforward, and we present here the
final formulas for these menisci. Shape of upper unduloid segment touching the
sphere for the ${\sf Und_{2k}^s}$ meniscus reads
\bea
x(t)=\frac{1}{2H_{2k}^s}\left(\sin t+s\sqrt{\sin^2t+c}\right)\;,\quad y(t)=
\frac{1}{2H_{2k}^s}[I_1(t,t_2)+sI_2(t,t_2)+2k\hat I_2]\;.\label{y_upper_pos22k}
\eea
Using the second equation in (\ref{y_upper_pos22k}) we have for $t=t_1$
\be
2H_{2k}^s\Psi=I_1(t_1,t_2)+sI_2(t_1,t_2)+2k\hat I_2.\label{curv_infla2k}
\ee
\subsubsection{Shape}\label{ar41a2k}
The meniscus shape is given for $0\le n\le k-1$ by the following general
expressions :
\bea
x(t)=\frac{1}{2H_{2k}^s}\left(\sin t +s\sqrt{\sin^2t+c}\right),\;
y(t)=\frac{1}{2H_{2k}^s}\left[I_1(t,t_2) +s I_2(t,t_2)+2n\hat I_2\right],\;t\in
\left\{\tilde t_2,t_{\ast}^{-s}\right\},\label{y_plate_a2k}
\eea
\bea
x(t)&=&\frac{1}{2H_{2k}^s}\left(\sin t-s\sqrt{\sin^2t+c}\right),\ \
t\in\left\{t_{\ast}^-,t_{\ast}^+\right\},\nonumber\\
y(t)&=&\frac{1}{2H_{2k}^s}\left[I_1(t,t_2)-s(I_2(t,t_2)-2I_2(t_{\ast}^{\mp},
t_2))+2n\hat I_2\right],\label{y_middle_a2k}
\eea
\bea
x(t)=\frac{1}{2H_{2k}^s}\left(\sin t+s\sqrt{\sin^2t+c}\right),\;y(t)=\frac{1}{
2H_{2k}^s}\left[I_1(t,t_2)+sI_2(t,t_2)+2k\hat I_2\right],\;t\in\left\{t_1,
t_{\ast}^s\right\},\label{y_sphere_a2k}
\eea
where
$$
\tilde t_2=t_2\delta_{n0}+t_{\ast}^s(1-\delta_{n0}),
$$
$\delta_{ij}$ denotes the Kronecker delta.
\subsubsection{Volume}\label{ar42a2k}
Volume of inflectional unduloid ${\sf Und_{2k}^s}$ meniscus is computed as
\bea
V_{2k}^s&=&\frac{\pi}{8(H_{2k}^s)^3}\{4J_3(t_1,t_2)+cI_1(t_1,t_2)-sc[I_2(t_1,
t_2)-2ks\hat I_2]\nonumber\\
&+&4s[J_2(t_1,t_2)+2ks\hat J_2]\}-\frac{\pi}{3}(2-3\cos\psi+\cos^3\psi)\;,
\label{volume_unduloid_infl_sp2k}
\eea
\subsubsection{Surface Area}\label{ar43a2k}
Surface area of the ${\sf Und_{2k}^s}$ meniscus reads
\be
S_{2k}^s=\frac{\pi}{2(H_{2k}^s)^2}\left[K_s(t_{\ast}^{-s},t_2)+k(\hat K_+-
\hat K_-)+K_s(t_1,t_{\ast}^{-s})\right].\label{surface_unduloid_infl_sp2k}
\ee
\subsection{Inflectional Unduloids ${\sf Und_{2k+1}^{\pm}}$ with Odd Number of
Inflection Points}\label{ar4odd}
Generalization of the menisci ${\sf Und_1^s}$ and ${\sf Und_3^s}$ to arbitrary
odd number of inflection points is straightforward, and we present here the
final formulas for these menisci. We find for the curvature of the ${\sf 
Und_{2k+1}^s}$ meniscus
\be
2H_{2k+1}^s\Psi=I_1(t_1,t_2)+sI_2(t_1,t_2)-2sI_2(t_{\ast}^s,t_2)+2k\hat I_2.
\label{curv_infl2kp1}
\ee
\subsubsection{Shape}\label{ar412kp1}
The shape of the ${\sf Und_{2k+1}^s}$ meniscus is given by the following  
general expressions for $0\le n\le k$
\bea
x(t)&=&\frac{1}{2H_{2k+1}^s}\left(\sin t-s\sqrt{\sin^2t+c}\right), \ \ \
t\in\left\{\tilde t_2,t_{\ast}^s\right\},\nonumber\\
y(t)&=&\frac{1}{2H_{2k+1}^s}[I_1(t,t_2)-sI_2(t,t_2)+2n\hat I_2],
\label{y_plate_2kp1}
\eea
\bea
x(t)&=&\frac{1}{2H_{2k+1}^s}\left(\sin t+s\sqrt{\sin^2t+c}\right),\ \ \
t\in\left\{\tilde t_1,t_{\ast}^s\right\},\nonumber\\
y(t)&=&\frac{1}{2H_{2k+1}^s}\left[I_1(t,t_2)+sI_2(t,t_2)+ 2n\hat I_2-2sI_2(
t_{\ast}^s,t_2)\right],\label{y_sphere_2kp1}
\eea
where
$$
\tilde t_2=t_2\delta_{n0} + t_{\ast}^{-s}(1-\delta_{n0}),\ \ \
\tilde t_1=t_1\delta_{nk} + t_{\ast}^{-s}(1-\delta_{nk}).
$$
\subsubsection{Volume}\label{ar422kp1}
Volume of inflectional unduloid ${\sf Und_{2k+1}^s}$ meniscus reads
\bea
V_{2k+1}^s&=&\frac{\pi}{8(H_{2k+1}^s)^3}\{4J_3(t_1,t_2)+cI_1(t_1,t_2)+sc[I_2(
t_1,t_2)-2ks\hat I_2-2I_2(t_1,t_{\ast}^s)]\nonumber\\
&-&4s[J_2(t_1,t_2)-2ks\hat J_2-2J_2(t_1,t_{\ast}^s)]\}-\frac{\pi}{3}(2-3\cos
\psi+\cos^3\psi)\;.\label{volume_unduloid_infl_sp2kp1}
\eea
\subsubsection{Surface Area}\label{ar432kp1}
Surface area of inflectional unduloid ${\sf Und_{2k+1}^s}$ meniscus is computed
as
\be
S_{2k+1}^s=\frac{\pi}{2(H_{2k+1}^s)^2}\left[K_{-s}(t_{\ast}^s,t_2)+k(\hat K_+-
\hat K_-)+K_s(t_1,t_{\ast}^s)\right].\label{surface_unduloid_infl_sp2kp1}
\ee
\subsection{Unduloid General Formulas}\label{ar78}
Merging the expressions (\ref{curv_infla2k}, \ref{curv_infl2kp1})
we write the general expression for the curvature of
${\sf Und_n^s}$ unduloid
\be
2H_n^s\Psi=I_1(t_1,t_2)+sI_2(t_1,t_2)+n\hat I_2-s\frac{1-\cos\pi n}{2}\left[
I_2(t_{\ast}^+,t_2)+I_2(t_{\ast}^-,t_2)\right].\label{und_curv_general0}
\ee
It can be checked by direct computation that the expression in the square
brackets in (\ref{und_curv_general0}) evaluates to $2I_2(\pi/2,t_2)$, and
we have
\be
2H_n^s\Psi=I_1(t_1,t_2)+sI_2(t_1,t_2)+n\hat I_2-s(1-\cos\pi n)I_2(\pi/2,t_2),
\label{und_curv_general}
\ee
Replacing in (\ref{volume_unduloid_infl_sp2kp1}) $k$ by $(n-1)/2$ we find
for the expression in curly brackets
$$
4J_3(t_1,t_2)+cI_1(t_1,t_2)+s cI_2(t_1,t_2)-nc\hat I_2+2scI_2(\pi/2,t_1)-
4sJ_2(t_1,t_2)+4n\hat J_2-8sJ_2(\pi/2,t_1).
$$
Combining it with (\ref{volume_unduloid_infl_sp2k}) we find
\bea
V_n^s&=&\frac{\pi}{8(H_n^s)^3}\{4J_3(t_1,t_2)+cI_1(t_1,t_2)+4n\hat J_2-
\sigma(n)[cI_2(t_1,t_2)-4J_2(t_1,t_2)]+nc\hat I_2\cos\pi n\nonumber\\
&+&s(1-\cos\pi n)[cI_2(\pi/2,t_1)-4J_2(\pi/2,t_1)]\}-\frac{\pi}{3}(2-3\cos\psi
+\cos^3\psi)\;,\label{und_volume_general}
\eea
where $\sigma(n)=s\cdot\cos\pi n$. From (\ref{surface_unduloid_infl_sp2k}) and
(\ref{surface_unduloid_infl_sp2kp1}) one finds
\be
S_n^s=\frac{\pi}{2(H_n^s)^2}\left[K_{\sigma(n)}\left(t_{\ast}^{-\sigma(n)},
t_2\right)+\lfloor n/2\rfloor(\hat K_+-\hat K_-)+K_s\left(t_1,t_{\ast}^{-
\sigma(n)}\right)\right],\label{und_surf_general}
\ee
where $\lfloor x\rfloor$ denotes the floor function.
\section{Spheres ${\sf Sph_n^{\pm}}$}\label{ar5}
In the classical menisci sequence the transition from convex unduloid to convex
nodoid takes place through formation of a spherical surface ${\sf Sph^+_0}$.
The inflectional unduloids ${\sf Und_n^+}$, $n >0$, in the limit $\alpha_n\to 1$
transform into the surfaces ${\sf Sph^+_n}$ made of several spherical
segments. The same time the unduloid menisci ${\sf Und_n^-}$ at small filling
angles approach another type of spherical menisci ${\sf Sph^-_n}$. Below we treat them
both as a limiting case of corresponding ${\sf Und_n^{\pm}}$ menisci.
\subsection{Asymptotic Behavior of ${\sf Und_n^{\pm}}$ Menisci in Vicinity of
${\sf Sph_n^{\pm}}$}\label{appendix2}
Consider the relation (\ref{und_curv_general}) and find dependence $H=H(\psi)$
for the ${\sf Und_n^s}$ meniscus in vicinity of $c=0$ that can be reached for
either $\alpha=0$ (for $s=-1,\psi=\phi_n^-=0$) or $\alpha=1$ (for $s=1,\psi=
\phi_n^+$). Using (\ref{cdef}) find asymptotic for $c(\psi)$
\be
c(\psi)=c'(\phi_n^s)(\psi-\phi_n^s)\;,\quad c'(\phi_n^s)=4s|2\alpha_n-1|
\alpha_n'(\phi_n^s)\sin^2(\theta_1+\phi_n^s)\;,\quad \mbox{sgn}\;c'(\phi_n^s)
=s,\label{w1}
\ee
where $\alpha_n'(\phi_n^s)>0$.

To find the asymptotics of the general terms $\sqrt{c}\;\overline{E}(t,\sqrt{
-1/c})$ and $\sqrt{c}\;\overline{F}(t,\sqrt{-1/c})$ used in (\ref{curv_infl1})
we make use of asymptotic expansions for the elliptic integrals based on
relations found at \cite{wl09a} and obtain for $t\le\pi$,
\bea
&&\sqrt{c}\;\overline{E}(t,\sqrt{-1/c})\stackrel{c\to 0}{\simeq}\tilde E(t,c)=
1-\cos t-\frac{c}{4}\left(\ln\cot^2\frac{t}{2}-4\ln 2+\ln |c|-1\right)\;,
\label{expand_E_1}\\
&&\sqrt{c}\;\overline{F}(t,\sqrt{-1/c})\stackrel{c\to 0}{\simeq}\tilde F(t,c)=-
\frac{c}{2}\left(\ln\cot^2\frac{t}{2}-4\ln 2+\ln |c|\right)\;.\label{expand_F_1}
\eea
Using the above expressions we find an approximation
\be
I_2(t_1,t_2)=I_1(t_1,t_2)-\frac{c}{2} M, \ \ \
M=\ln\left(\tan\frac{t_1}{2}\cot\frac{t_2}{2}\right).\label{I2_asympt}
\ee
The last relation leads to
\be
\hat I_2=I_2(t_{\ast}^-,t_{\ast}^+)=2-c\ln 2+\frac{c}{2}\ln(-c), \ \
I_2(\pi/2,t_2)=\cos t_2-\frac{c}{2}\ln\left(\cot\frac{t_2}{2}\right).
\label{I2_part_asympt}
\ee
Substitution of (\ref{I2_asympt}, \ref{I2_part_asympt}) into
(\ref{und_curv_general}) produces
\be
2H_n^s\Psi(\phi_n^s)=2n-(1+s)\cos(\theta_1+\phi_n^s)-(1+s\cos\pi n)\cos
\theta_2+\frac{cn}{2}\ln(-c).\label{curv_asympt}
\ee
The curvature at the sphere is found as
\be
H_n^s(\phi_n^s)=\frac{2n-(1+s)\cos(\theta_1+\phi_n^s)-(1+s\cos\pi n)\cos
\theta_2}{2\Psi(\phi_n^s)}.\label{curv_sphere_general}
\ee
It follows from (\ref{curv_sphere_general}) that for the ${\sf Sph^-_n}$ sphere 
the curvature is independent of $\theta_1$ and reads
\be
H_n^-(0)=\frac{2n-(1-\cos\pi n)\cos\theta_2}{2d}.\label{curv_sphere_neg}
\ee
For $s=1$ we find a condition on the angle $\phi_n^+$ at which sphere
${\sf Sph^+_n}$ is observed
$$
2(1+d-\cos\phi_n^+)\sin(\theta_1+\phi_n^+)=\sin\phi_n^+[2n-2\cos(\theta_1+
\phi_n^+)-(1+\cos\pi n)\cos\theta_2],
$$
which reduces to
\be
2(1+d)\sin(\theta_1+\phi_n^+)-2\sin\theta_1-\sin\phi_n^+[2n-(1+\cos\pi n)\cos
\theta_2]=0,\label{psi_cond_sphere_general}
\ee
Differentiating the relation (\ref{curv_asympt}) we obtain in the leading order
\be
\frac{dH_n^s(\psi)}{d\psi}_{|\;\psi=\phi_n^s}=\frac{nc'(\phi_n^s)}{4\Psi(
\phi_n^s)}\;\left[-1+\ln(-c)\right]\;.\label{curv_diff_general}
\ee
Combining (\ref{curv_diff_general}) with the last formula in (\ref{w1}) we find
$\mbox{sgn}\left(dH_n^s/d\psi\right)\!=\!-s$ at $\psi=\phi_n^s$. General
expression for derivative of $\alpha_n$ reads
\be
\alpha_n'(\psi)=\frac{dH_n^s(\psi)}{d\psi}\frac{\sin\psi}{\sin t_1}+H_n^s
\frac{\sin\theta_1}{\sin^2 t_1}\;.\label{alpha_diff}
\ee
Using it with (\ref{curv_sphere_neg}) we find
\be
\alpha_n'(0)=\xi^-_n=\frac{H_n^-(0)}{\sin\theta_1}=\frac{2n-(1-\cos\pi n)
\cos\theta_2}{2d\sin\theta_1},\label{alpha_diff_psi_0}
\ee
implying that for small $\psi$ the slope $\xi^-_n$ of $\alpha$ increases with
growth of index $n$ and decreases with growth of the distance $d$. Applying
(\ref{expand_E_1},\ref{expand_F_1}) to integral (\ref{I_3}) we find its
asymptotics $I_3(t_1,t_2)=cM$ and obtain
$$
K_s(t_1,t_2)=2(1+s)(\cos t_2-\cos t_1).
$$
Thus the expression in square brackets in the general formula for unduloid
surface area (\ref{und_surf_general}) is independent of $c$ and reads
$$
\bar K=4n-2(1+s\cos\pi n)\cos\theta_2-2(1+s)\cos(\theta_1+\phi_n^s)=4H_n^s
(\phi_n^s)\Psi(\phi_n^s).
$$
Thus the surface area reads in the leading logarithmic order
\be
S_n^s=\frac{\pi\bar K}{2(H_n^s)^2}=\frac{2\pi \Psi(\phi_n^s)}{H_n^s(\phi_n^s)},
\label{surf_asympt}
\ee
and we find that $\mbox{sgn}\;(dS_n^s(\psi)/d\psi)=s$ at $\psi=\phi_n^s$.
The explicit expression for the surface area of ${\sf Sph^s_n}$ reads
\be
S_n^s(\phi_n^s)=\frac{4\pi\Psi^2(\phi_n^s)}{2n-(1+s)\cos(\theta_1+\phi_n^s)-
(1+s\cos\pi n)\cos\theta_2},\label{surf_sphere_general}
\ee
and we find for ${\sf Sph^-_n}$
\be
S_n^-(0)=\frac{4\pi d^2}{2n-(1-\cos\pi n)\cos\theta_2}\;.
\label{surf_sphere_neg}
\ee
Turning to computation of the menisci volume asymptotics we first use
(\ref{expand_E_1}, \ref{expand_F_1}) in integral (\ref{J2}) to find
\be
J_2(t_1,t_2)=J_3(t_1,t_2), \ \ \ \hat J_2=4/3.
\label{J2_asympt}
\ee
Using this relation we obtain for the expression $\bar V$ in curly brackets in
(\ref{und_volume_general})
$$
\bar V=\frac{4}{3}\left[4n+3J_3(t_1,t_2)+3sJ_3(t_1,\pi/2)+3sJ_3(\pi/2,t_2)\cos
\pi n\right],
$$
which is independent of $c$. Thus the volume reads in the leading logarithmic
order
\be
V_n^s=\frac{\pi\bar V}{8(H_n^s)^3}-V_{ss}=\frac{\pi\bar V}{8(H_n^s)^3}-
\frac{\pi}{3}(2-3\cos\psi+\cos^3\psi)\;,
\label{volume_asympt}
\ee
and we immediately find that $\mbox{sgn}\;(dV_n^s(\psi)/d\psi)=s$ at
$\psi=\phi_n^s$.
The explicit expression for the volume of ${\sf Sph^s_n}$ reads
\be
V_n^s(\phi_n^s)=\frac{4\pi\Psi^3(\phi_n^s)[4n+3J_3(t_1,t_2)+3sJ_3(t_1,\pi/2)+
3sJ_3(\pi/2,t_2)\cos\pi n]}{3[2n-(1+s)\cos t_1-(1+s\cos\pi n)\cos\theta_2]^3}
-V_{ss}(\phi_n^s),\label{volume_sphere_general0}
\ee
with $t_1=\theta_1+\phi_n^s$ and $t_2=\pi-\theta_2$. Recalling that
\bea
J_3(t_1,t_2)=[G(t_2)-G(t_1)]/3\;,\quad\mbox{where}\quad G(t)=3\cos t-\cos^3t\;,
\quad G(\pi/2)=0\;,\nonumber
\eea
we have
\be
V_n^s(\phi_n^s)=\frac{4\pi\Psi^3(\phi_n^s)[4n-(1+s)G(\theta_1+\phi_n^s)-
(1+s\cos\pi n)G(\theta_2)]}{3[2n-(1+s)\cos(\theta_1+\phi_n^s)-(1+s\cos\pi n)
\cos\theta_2]^3}-V_{ss}(\phi_n^s),\label{volume_sphere_general}
\ee
and we find that for ${\sf Sph^-_n}$
\be
V_n^-(0)=\frac{4\pi d^3[4n-(1-\cos\pi n)G(\theta_2)]}{3[2n-(1-\cos\pi n)\cos
\theta_2]^3}\label{volume_sphere_neg}
\ee
the volume does not depend on $\theta_1$. Using (\ref{w1}) and
(\ref{curv_diff_general}) in (\ref{alpha_diff}) we find
\be
\alpha'_n(\phi_n^+)=\frac{\Psi}{\Psi-n\sin\phi_n^+\sin t_1\ln(-c)}\cdot\frac{
\sin\theta_1}{\sin\phi_n^+\sin t_1}\;.\label{alpha_diff_psi}
\ee
which produces two important formulas for $n=0$
\bea
\alpha'_0(\phi_0^+)=\frac{\sin\theta_1}{\sin\phi_0^+\sin(\theta_1+\phi_0^+)}\;,
\label{alpha_diff_psi_00}
\eea
and for $n\gg 1$,
\bea
\alpha'_n(\phi_n^+)\simeq -\frac{\sin\theta_1}{\sin^2\phi_n^+\sin^2(\theta_1+
\phi_n^+)}\cdot\frac{\Psi}{n\ln(-c)}\simeq -\frac{n}{d\sin^3\theta_1}\frac1{
\ln(-c)}\;.\label{alpha_diff_psi_n>>1}
\eea
The last equality in (\ref{alpha_diff_psi_n>>1}) makes use of
(\ref{psi_n_approx_large_n}), i.e., $n\phi_n^+\simeq d\sin\theta_1$, and $\Psi
\simeq d$ when $n\to\infty$.
\subsection{Sphere ${\sf Sph^-_n}$}\label{ar15}
The curvature of ${\sf Sph^-_n},\ n>0$ menisci is given by
(\ref{curv_sphere_neg}). Using the general formula (\ref{y_plate_a2k} -
\ref{y_sphere_a2k}) for the unduloid ${\sf Und_{2k}^-}$ we find in the limit
$\psi\to 0$ that the meniscus is presented by a sequence of $k$ coaxial spheres
of the radius $r_{2k}=1/H_{2k}^-=d/(2k)$ with the centers located at $\{0,(2i+1)
r_{2k}\}$ for $i=0,1,\ldots,k-1$.

Similarly, for odd number $n=2k+1$ of inflection points we have $k$ full spheres
of the radius $r_{2k+1}=1/H_{2k+1}^-=d/(2k+1-\cos\theta_2)$ and a spherical cap
on the plane with the same radius. The centers of the full spheres are at
$\{0,d-(2i+1)r_{2k+1}\}$ for $i=0,1,\ldots,k-1$ and the center of the spherical
cap is at $\{0,-r_{2k+1}\cos\theta_2\}$.
\subsection{Sphere ${\sf Sph^+_n}$}\label{sphere1}
The curvature of ${\sf Sph^+_n}$ menisci is given by (\ref{curv_sphere_general})
with $\psi=\phi_n^+$. In case $n=2k$ the meniscus is represented by the segments
of the spherical surface touching each other at the vertical axis at the points
with zero abscissa and ordinates $(2i+1+\cos t_2)/H_{n}^+,\;i=0,1,\ldots, k$.

In case $n=2k+1$ the meniscus is made of the segments of the spherical surface
touching each other at the vertical axis at the points with zero abscissa and
ordinates $2i/H_{n}^+$. It follows from (\ref{psi_cond_sphere_general}) that the
value of the filling angle $\phi_{2k+1}^+$ does not depend on the value of the
angle $\theta_2$.

In a particular case of wetting sphere $\theta_1=0$ one finds from
(\ref{psi_cond_sphere_general}) that ${\sf Sph^+_n}$ meniscus exists at $\psi=
\pi$ that gives for the curvature $H_{n}^+=[n+1+(1-\cos\pi n))\cos t_2]/(2+d)$.
Taking into account that $H_{n}^+=\alpha_{n}\le 1$ we find an existence
condition for this meniscus
\be
d\ge n-1+(1-\cos\pi n)\cos t_2.\label{Sph_cond_theta1=0}
\ee
The condition (\ref{psi_cond_sphere_general}) at $\theta_1=0$ is also satisfied
for all $0\le\psi\le\pi$ when $d=n-1+(1-\cos\pi n)\cos t_2$.
\section{Special Properties of Nodoids ${\sf Nod^{\pm}}$}\label{appendix5}
In this appendix we discuss the extremal properties of nodoidal menisci (local
minimum of curvature and local maxima of the surface area and volume) and the
asymptotic behavior of the nodoidal curvature in vicinity of singular point
$\psi_{\ast}=\arccos(1+d)$.
\subsection{Non-monotonic Behavior of ${\sf Nod^+}$ Meniscus Characteristics}
\label{appendix5a}
Show that the ${\sf Nod^+}$ meniscus always has a local minimum of curvature and
local maxima of the surface area and volume. First, we show that the curvature
always grows when the filling angle reaches $\pi$, so that the derivative $dH/
d\psi$ at $\psi=\pi$ is positive. In this range the convex nodoid is observed,
so that we start with the asymptotics of the general terms $\sqrt{c}\;\overline{
E}(t,\sqrt{-1/c})$ and $\sqrt{c}\;\overline{F}(t,\sqrt{-1/c})$ in
(\ref{expand_E_1}, \ref{expand_F_1}) valid for $t\le\pi$ and also find for $t>
\pi$
\bea
&&\sqrt{c}\;\overline{E}(t,\sqrt{-1/c})\stackrel{c\to 0}{\simeq}-\tilde E(t,c)+
4-c\left(-1-4\ln2+\ln |c|\right),\label{expand_E_1a}\\
&&\sqrt{c}\;\overline{F}(t,\sqrt{-1/c})\stackrel{c\to 0}{\simeq}-\tilde F(t,c)-
2c\left(-4\ln2+\ln |c|\right).\label{expand_F_1a}
\eea

We use (\ref{curvature_nodoid}) with $s=1$ corresponding to convex nodoid and
substitute into it the expressions (\ref{expand_E_1}, \ref{expand_F_1}) for
$t=t_2=\pi-\theta_2 <\pi$ and (\ref{expand_E_1a}, \ref{expand_F_1a}) for $t=t_1
=\pi+\theta_1 >\pi$. Retaining the leading terms only we arrive at
\be
2(d+2)H\approx 2(1-\cos\theta_2)+\frac{c}{4}\left(2-8\ln2+2\ln c+\ln\tan^2
\frac{\theta_1}{2}\tan^2 \frac{\theta_2}{2}\right),\label{curv_nodoid_c}
\ee
where $c\approx 4H(\pi-\psi)\sin\theta_1$ is positive for $\psi<\pi$. As $c'(
\psi)<0$ the leading term in the above expression is $c\ln c$ which for positive
$c$ guarantees curvature growth in the vicinity of $\psi=\pi$. As $H(0)>H(\pi)$
the curvature dependence on the filling angle cannot be monotonous one, and the
curvature should have a local maximum and a local minimum. Thus, as the
curvature at the spherical meniscus ${\sf Sph_0^+}$ is a decreasing function of
the filling angle (see Appendix \ref{appendix2}) and the same time in the
vicinity of $\psi=\pi$ it always grows, it always has a local minimum at the
convex nodoid meniscus.

Volume behavior analysis gives in the leading order
\be
V(\psi)=\frac{64\pi (d+2)^3(2-\cos t_2)(1+\cos t_2)^2}{3[4(1+\cos t_2)+c\ln c]
^3}-\frac{4\pi}{3},\hspace{1cm} V(\pi)=\frac{\pi (d+2)^3(2-\cos t_2)}{3(1+\cos
t_2)}-\frac{4\pi}{3},\label{volume_psi_pi}
\ee
and its derivative in $\psi$ reads in the leading order
$$
\frac{dV(\psi)}{d\psi}=-\frac{\pi (d+2)^3(2-\cos t_2)\ln c}{4(1+\cos t_2)^2}
\;c'(\psi),
$$
and the volume decreases at $\psi=\pi$.
Comparing the volume at two extreme values of the filling angle we find that
$V(0)<V(\pi)$, which implies that its behavior is non-monotonic and it should
have at least one local maximum and one local minimum. As the volume grows at
the spherical meniscus (see \ref{appendix2}) its local maximum is observed on
convex nodoid meniscus.

Analysis of the surface area of the convex nodoid in the vicinity of $\psi=
\pi$ is done similarly to that of performed at small filling angles in Appendix
\ref{appendix2}. The area is given by (\ref{surface_nodoid}) with $s=1$
that gives in the leading logarithmic order
\be
S(\psi)=\frac{32\pi (d+2)^2(1+\cos t_2)}{[4(1+\cos t_2)+c\ln c]^2},\hspace{1cm}
S(\pi)=\frac{2\pi (d+2)^2}{1+\cos t_2},\label{surface_psi_pi}
\ee
and its derivative in $\psi$ reads in the leading order
$$
\frac{dS(\psi)}{d\psi}=-\frac{\pi (d+2)^2\ln c}{(1+\cos t_2)^2}\;c'(\psi).
$$
Noting that $c'(\psi) < 0$ we find that the derivative of the surface area
w.r.t. the filling angle is negative for $\psi=\pi$ and the surface area
decreases. Comparing the surface area at two extreme values of the filling
angle we find that $S(0) < S(\pi)$, which implies that its behavior is
non-monotonic and it should have at least one local maximum and one local
minimum. As the surface area grows at the spherical meniscus (see
\ref{appendix2}) its local maximum is observed on convex nodoid meniscus.
\subsection{Asymptotics of Nodoid Curvature}\label{nod_as0}
Here we discuss divergence of the curvatures of convex and concave nodoids and
corresponding asymptotics. Consider equation (\ref{curv}) for the ${\sf Nod^s}$
menisci,
\be
2H\Psi=I_1+s I_2\;,\label{nod_as1}
\ee
where $\Psi$, $I_1$ and $I_2$ are defined in (\ref{curv}), (\ref{I_1}) and
(\ref{I_2_def}). For both nodoids $c>0$, therefore integrals $I_1$ and $I_2$
are always convergent and divergence appears only when $\Psi$ is vanishing.
This happens when $-2<d\leq 0$, i.e., the divergence does not occur when the
solid bodies are separated.

Consider first the case $d=0$ for which a singular point is $\psi_{\ast}=0$ and
in its vicinity $\psi\ll 1$ we obtain $\Psi\simeq\psi^2/2$. Choose the power
law of divergence, $H\simeq U\psi^{-\beta_0}$, where $U,\beta_0>0$, then in
accordance with (\ref{cdef}) we find
\bea
c\simeq\left\{\begin{array}{rll}-4H\sin\psi\sin t_1\simeq U_1\psi^{1-\beta_0}\;
,&\beta_0<1,& U_1=-4U\sin\theta_1>0,\\
4H\sin\psi(H\sin\psi-\sin t_1)=U_2\;,&\beta_0=1,& U_2=4U(U-\sin\theta_1)>0,\\
4H^2\sin^2\psi\simeq U_3\psi^{2(1-\beta_0)}\;,&\beta_0>1,& U_3=4U^2>0,
\end{array}\right.\label{nod_as2}
\eea
that for $\psi\to 0$ implies $c\simeq 0,\;\beta_0<1$, $c=4U(U-\sin\theta_1),\;
\beta_0=1$ and $c \simeq \infty,\;\beta_0>1$.

Consider the integrals $I_1$ and $I_2$. For the first of them we have $I_1
\simeq I_1^{\ast}+\psi\sin\theta_1$, where $I_1^{\ast}=-(\cos\theta_1+\cos
\theta_2)$. Regarding $I_2$, denote $P(t,c)=\sin^2 t/\sqrt{c+\sin^2 t}$ and
obtain
\bea
I_2\simeq \left\{\begin{array}{rl}\int_{\pi-\theta_2}^{\theta_1}P(t,U_1\psi^{1-
\beta_0})dt+\psi P(\theta_1,U_1\psi^{1-\beta_0})\;,&\beta_0<1,\\
\int_{\pi-\theta_2}^{\theta_1}P(t,U_2)dt+\psi P(\theta_1,U_2)\;,&\beta_0=1,\\
\int_{\pi-\theta_2}^{\theta_1}P(t,U_3\psi^{2(1-\beta_0)})dt+\psi P(\theta_1,
U_3\psi^{2(1-\beta_0)})\;,&\beta_0>1.\end{array}\right.\label{nod_as4}
\eea
Substituting (\ref{nod_as4}) into (\ref{nod_as1}) and making use of asymptotics
(\ref{I2_asympt}) we find
\bea
U\psi^{2-\beta_0}=I_1^{\ast}+\psi\sin\theta_1+s\left\{\begin{array}{rl}
I_1^{\ast}-\frac{U_1}{2}\psi^{1-\beta_0}\ln\left(\tan\frac{\theta_1}{2}\tan
\frac{\theta_2}{2}\right)+\psi P(\theta_1,U_1\psi^{1-\beta_0})\;,&\beta_0<1,\\
\int_{\pi-\theta_2}^{\theta_1}P(t,U_2)dt +\psi P(\theta_1,U_2)\;,&\beta_0=1,\\
\frac1{\sqrt{U_3}}\psi^{\beta_0-1}\int_{\pi-\theta_2}^{\theta_1}\sin^2 t\;dt+
\psi P(\theta_1,U_3\psi^{2(1-\beta_0)})\;,& \beta_0>1,\end{array}\right.
\label{nod_as5}
\eea

Solve equation (\ref{nod_as5}) in two cases. First, if $\theta_1+\theta_2\neq
\pi$, then preserving the leading terms in $\psi$ we get for the ${\sf Nod^+}
\;(\theta_1+\theta_2>\pi)$ solutions,
\bea
U\psi^{2-\beta_0}=2I_1^{\ast},\;\;\beta_0<1;\quad U\psi^{2-\beta_0}=I_1^{\ast}+
\int_{\pi-\theta_2}^{\theta_1}P(t,U_2)dt,\;\;\beta_0=1;\quad
U\psi^{2-\beta_0}=I_1^{\ast},\;\;\beta_0>1,\nonumber
\eea
which yields $U=I_1^{\ast}$, $\beta_0=2$. In the case of the ${\sf Nod^-}$
nodoid and $\theta_1+\theta_2<\pi$ we get
\bea
U\psi^{2-\beta_0}=\left\{\begin{array}{rl}\frac{U_1}{2}\psi^{1-\beta_0}\ln
\left(\tan\frac{\theta_1}{2}\tan\frac{\theta_2}{2}\right)\;,&\beta_0<1\\
I_1^{\ast}-\int_{\pi-\theta_2}^{\theta_1}P(t,U_2)dt\;,&\beta_0=1\\
I_1^{\ast}\;,&\beta_0>1\end{array}\right.,\label{nod_as7}
\eea
satisfied for $U=I_1^{\ast}$, $\beta_0=2$ only. Thus, in the generic setup
$\theta_1+\theta_2\neq\pi$ the both nodoidal menisci have divergent curvature,
\bea
H\simeq -\frac{\cos\theta_1+\cos\theta_2}{\psi^2}\;,\label{nod_as7a}
\eea
In case $\theta_2=\pi/2$ its expression coincides with estimate (\ref{b1})
derived by simple considerations.

Consider a special case $\theta_1+\theta_2=\pi$ and $\theta_1\neq 0,\pi$, for
${\sf Nod^+}$ rewriting (\ref{nod_as5}) in leading terms,
\bea
U\psi^{2-\beta_0}=2\psi\sin\theta_1,\;\beta_0<1;\quad U\psi^{2-\beta_0}=
\frac{2U\sin\theta_1}{2U-\sin\theta_1}\psi,\;\beta_0=1;\quad
U\psi^{2-\beta_0}=\psi\sin\theta_1,\;\beta_0>1\;,\nonumber
\eea
which is satisfied for $U=\frac{3}{2}\sin\theta_1$, $\beta_0=1$, i.e.,
\bea
H=\frac{3\sin\theta_1}{2\psi}\;,\quad\theta_1\neq 0,\pi\;.\label{nod_as8}
\eea
In case ${\sf Nod^-}$ with $\theta_1+\theta_2=\pi$, $\theta_1\neq 0,\pi$, we
have from (\ref{nod_as5}) after substitution of $U_1$, $U_2$ and $U_3$ defined
in (\ref{nod_as2})
\bea
U\psi^{2-\beta_0}\!=\!-2U\psi^{2-\beta_0},\;\beta_0<1;\quad U\psi^{2-\beta_0}\!
=\!2\frac{U-\sin\theta_1}{2U-\sin\theta_1}\psi\sin\theta_1,\;\beta_0=1;\quad
U\psi^{2-\beta_0}\!=\!\psi\sin\theta_1,\;\beta_0>1\nonumber
\eea
The first and third equations cannot be satisfied due to restrictions on
$\beta_0$ and $U\neq 0$. The second equation
\bea
U(2U-\sin\theta_1)=2(U-\sin\theta_1)\sin\theta_1\;,\quad\beta_0=1\;,
\label{nod_as11}
\eea
does not admit real solutions, so the ${\sf Nod^-}$ meniscus is forbidden in
the special case $\theta_1+\theta_2=\pi$.

For $-2<d<0$ a singular point $\psi_*=\arccos(1+d)>0$ does exist and in its
vicinity $\psi-\psi_*=\eta\ll 1$ we obtain $\Psi\simeq\eta\sin\psi_*$.
Choosing the power law of divergence, $H\simeq V\eta^{-\beta_1}$, $\beta_1>0$,
we find $c\simeq 4V^2\eta^{-2\beta_1}\sin^2\psi_*$. Write the leading in
$\eta$ terms of integrals $I_1$ and $I_2$
\bea
I_1=I_1^{**}+\eta\sin t_1^*,\quad I_2=\int_{t_2}^{t_1^*}P(t,4V^2\sin^2\psi_*
\eta^{-2\beta_1})dt+\eta P(\theta_1^*,4V^2\sin^2\psi_*\eta^{-2\beta_1}),
\nonumber
\eea
where $I_1^{**}=-(\cos t_1^*+\cos\theta_2)$ and $t_1^*=\theta_1+\psi_*$, and
substitute them into (\ref{nod_as1}),
\bea
2V\eta^{1-\beta_1}\sin\psi_*\simeq I_1^{**}+\eta\sin t_1^*+\frac{s\eta^{
\beta_1}}{2V\sin\psi_*}\left(\int_{t_2}^{t_1^*}\sin^2 t\;dt+\eta\sin^2t_1^*
\right).\label{nod_as12}
\eea
In general case, $\theta_1+\theta_2+\psi_*\neq \pi$ we have for both nodoids
${\sf Nod^+}\;(\theta_1+\theta_2+\psi_*> \pi)$ and ${\sf Nod^-}\;(\theta_1+
\theta_2+\psi_*<\pi)$,
\bea
H\simeq-\frac{\cos(\theta_1+\psi_*)+\cos\theta_2}{2}\cdot\frac{1}{\psi-\psi_*}
\;.\label{nod_as13}
\eea
In case $\theta_2=\pi/2$ its expression coincides with estimate (\ref{b1a})
derived by simple considerations.

The special case $\theta_1+\theta_2+\psi_*=\pi$ leads to
\bea
2V\eta^{-\beta_1}\sin\psi_*\simeq\sin t_1^*+\frac{s\sin^2t_1^*}{2V\sin\psi_*}
\eta^{\beta_1},\nonumber
\eea
satisfied by $\beta_1=0$ and $H$ does not diverge in vicinity of the critical
value $\psi_*$. This conclusion holds for any other (non power law) divergence
$H\simeq g(\eta)$ when $\theta_1+\theta_2+\psi_*=\pi$,
\bea
2Vg(\eta)\sin\psi_*\simeq\sin t_1^*+\frac{s\sin^2t_1^*}{2V\sin\psi_*}
\frac1{g(\eta)}.\nonumber
\eea

Show that for $\theta_1+\theta_2+\psi_*=\pi$ the ${\sf Nod^-}$ meniscus is
forbidden while the ${\sf Nod^+}$ meniscus is allowed for $\psi>\psi^*$. First
use (\ref{psi_cond_sphere_general}) for $n=0$ to find the value $\phi_0^+$ at
which the sphere ${\sf Sph_0^+}$ is observed:
$$
(1+d)\sin(\theta_1+\phi_0^+)+\sin\phi_0^+\cos\theta_2=\sin\theta_1.
$$
Direct computation shows that $\phi_0^+=\psi_*=\pi-\theta_1-\theta_2$ satisfies
the above equation, so that the sphere ${\sf Sph_0^+}$ exists at $\psi=\psi_*$.
As the meniscus ${\sf Nod^-}$ exists in the range $\psi <\psi_*=\psi_*$ which
is forbidden due to intersection, we conclude that ${\sf Nod^-}$ cannot be
observed in this special case. The same time, the meniscus ${\sf Nod^+}$ is
allowed for $\psi>\phi_0^+=\psi_*$. The value of the ${\sf Nod^+}$ curvature at
$\psi=\psi_*$ can be obtained by noting that it is equal to the ${\sf Sph_0^+}$
curvature that reads $H=\sin(\theta_1+\psi_*)/\sin\psi_*$. The Table 3 (where
$\eta=\psi-\psi_*$ and $\psi_*=\arccos(1+d)$) summarizes the asymptotic behavior
of the ${\sf Nod^{\pm}}$ menisci curvature.
\begin{center}
{\bf Table$\;$3}.\\
\vspace{-.5cm}
$$
\begin{array}{|c|c||c|c|c|c|}\hline
& d&\theta_1+\theta_2+\psi_*<\pi & \theta_1+\theta_2+\psi_*=\pi & \theta_1+
\theta_2+\psi_*>\pi \\\hline\hline
{\sf Nod^-}&=0&-(\cos(\theta_1+\psi_*)+\cos\theta_2)\eta^{-2} &
\mbox{forbidden} & \mbox{forbidden} \\\hline
{\sf Nod^-}&<0&-(\cos(\theta_1+\psi_*)+\cos\theta_2)\eta^{-1} &
\mbox{forbidden} & \mbox{forbidden} \\\hline
{\sf Nod^+}&=0&-& (3/2\sin\theta_1)\eta^{-1} & -(\cos(\theta_1+\psi_*)+\cos
\theta_2)\eta^{-2}\\\hline
{\sf Nod^+}&<0&-& \sin(\theta_1+\psi_*)/\sin\psi_* &
-(\cos(\theta_1+\psi_*)+
\cos\theta_2)\eta^{-1}\\\hline
\end{array}
$$
\label{ta4}
\end{center}
The empty entries in Table 3 indicate that the ${\sf Nod^+}$ meniscus does not
exist in the vicinity of $\psi_*$ contrary to the "forbidden" entry that means
that the corresponding meniscus does not exist in the whole range $\psi_*\le
\psi\le\pi$.
\section{Computation of Elliptic Integrals}\label{appendix3}
\setcounter{equation}{0}
In this appendix we derive formulas for computation of the elliptic integrals
used in the main text.
\subsection{Conjugation of Elliptic Integrals}\label{app31}
Here we prove that
\bea
\int_0^z\frac{dt}{\sqrt{\sin^2 t+c}}=\frac1{\sqrt{c}}\;\overline{\int_0^z
\frac{dt}{\sqrt{1+\frac{\sin^2 t}{c}}}}\;,\label{dd1}
\eea
where ${\overline A}(z)$ stands for complex conjugation of the function $A(z)$.
The case $c>0$ is trivial and the operation ${\overline A}(z)$ can be omitted
there. Consider negative $c$ and rewrite the l.h.s. of (\ref{dd1}) as follows
\bea
R(\nu;0,z)=\int_0^z\frac{dt}{\sqrt{\sin^2 t-\nu^2}}\;,\quad c=-\nu^2\;,
\label{dd2}
\eea
and focus on two cases:
\begin{enumerate}
\item $\sin^2 t-\nu^2\leq 0$, when $0\leq t\leq z$,
\item $\sin^2 t-\nu^2\leq 0$, when $0\leq t\leq z_{\star}$, and $\sin^2 t-
\nu^2\geq 0$, when $z_{\star}\leq t\leq z$.
\end{enumerate}
In the first case the integral in (\ref{dd2}) is purely imaginary,
\bea
R(\nu;0,z)=\int_0^z\frac{dt}{\sqrt{\sin^2 t-\nu^2}}=\frac1{i\nu}\int_0^z
\frac{dt}{\sqrt{1-\nu^{-2}\sin^2 t}},\label{dd3}
\eea
where an integral in the r.h.s. of (\ref{dd3}) is real (positive). Thus,
equality (\ref{dd1}) holds also in this case. In the second case write $R(\nu;
0,z)$ as a sum $R(\nu;0,z)=R(\nu;0,z_{\star})+R(\nu;z_{\star},z)$,
\bea
R(\nu;0,z_{\star})=\int_0^{z_{\star}}\frac{dt}{\sqrt{\sin^2 t-\nu^2}}\quad,\quad
R(\nu;z_{\star},z)=\int_{z_{\star}}^z\frac{dt}{\sqrt{\sin^2 t-\nu^2}}.
\label{dd4}
\eea
The first integral in (\ref{dd4}) for $\sin^2 t-\nu^2\leq 0$, is purely
imaginary and can be calculated using (\ref{dd3})
\bea
R(\nu;0,z_{\star})=\frac1{\pm i\nu}\int_0^{z_{\star}}\frac{dt}{\sqrt{1-\nu^{-2}
\sin^2t}}\;,\quad\sqrt{c}=\pm i\nu\;.\label{dd5}
\eea
Equality (\ref{dd1}) holds for $R(\nu;0,z_{\star})$. The second integral $R(
\nu;z_{\star},z)$, where $\sin^2 t-\nu^2\geq 0$, is positive, so $R(\nu;0,z)$
can be represented as follows,
\bea
\int_0^z\frac{dt}{\sqrt{\sin^2 t-\nu^2}}=\frac1{\pm i\nu}\left(\int_0^{z_{
\star}}\frac{dt}{\sqrt{1-\nu^{-2}\sin^2t}}\pm i\nu\int_{z_{\star}}^z
\frac{dt}{\sqrt{\sin^2t-\nu^2}}\right)\;.\label{dd5a}
\eea
Consider now another integral,
\bea
T(\nu;z_{\star},z)=\frac1{\pm i\nu}\int_0^z\frac{dt}{\sqrt{1-\nu^{-2}\sin^2 t}}
-R(\nu;0,z)=\frac1{\pm i\nu}\int_{z_{\star}}^z\frac{dt}{\sqrt{1-\nu^{-2}
\sin^2t}},\label{dd7}
\eea
which can be rewritten as follows
\bea
T(\nu;z_{\star},z)=-\frac1{\nu}\int_{z_{\star}}^z\frac{dt}{\sqrt{\nu^{-2}
\sin^2t-1}}=-\int_{z_{\star}}^z\frac{dt}{\sqrt{\sin^2t-\nu^2}},\label{dd8}
\eea
which is a negative number. Comparing the latter with (\ref{dd4}) we obtain
$T(\nu;z_{\star},z)=-R(\nu;z_{\star},z)$.

Combining the last equality with (\ref{dd7}) we obtain
\bea
\frac1{\pm i\nu}\int_0^z\frac{dt}{\sqrt{1-\nu^{-2}\sin^2 t}}=\frac1{\pm i\nu}
\left(\int_0^{z_{\star}}\frac{dt}{\sqrt{1-\nu^{-2}\sin^2 t}}\mp i\nu
\int_{z_{\star}}^z\frac{dt}{\sqrt{\sin^2 t-\nu^2}}\right).\label{dd10}
\eea
By comparison (\ref{dd5a}) and (\ref{dd10}) we find finally
\bea
\int_0^z\frac{dt}{\sqrt{\sin^2 t-\nu^2}}=\frac1{\pm i\nu}\;\overline
{\int_0^z\frac{dt}{\sqrt{1-\nu^{-2}\sin^2t}}}\;.\label{dd11}
\eea
Keeping in mind that we have taken $\sqrt{c}=\pm i\nu$ in (\ref{dd5} --
\ref{dd7}) and (\ref{dd10} -- \ref{dd11}), we arrive at (\ref{dd1}).
\subsection{Computation of Elliptic Integrals at Special Limit Values
$t_{\ast}^{\pm}$}\label{app32}
The integrals $\hat I_2=I_2(t_{\ast}^-,t_{\ast}^+)$, $\hat I_3=I_3(t_{\ast}^-,
t_{\ast}^+)$ and $\hat J_2=J_2(t_{\ast}^-,t_{\ast}^+)$ enter numerous formulas
for unduloids ${\sf Und_n^{s}}$ so that it is instructive to find their
explicit expression through the complete elliptic integrals of the first $K$
and second $E$ kind. In derivation we used relations from \cite{wl09b,wl09c}.
We start with the general relations
\be
E(t_{\ast}^++\pi n,k)=k[E(-c)-(c+1)K(-c)]+2nE(k^2), \ \ F(t_{\ast}^++\pi n,k)=
\sqrt{-c}K(-c)+2nK(k^2),\label{Elliptic030102}
\ee
using them with $n=0,-1$ for $t_{\ast}^+,t_{\ast}^-$, respectively. From
definition (\ref{I_2}) of $I_2$ integral we obtain for $\hat I_2$
$$
\hat I_2=2\sqrt{c}[E(k^2)-K(k^2)-k(E(-c)-K(-c))],
$$
where the expression in the square brackets simplifies to $iE(1+1/c)$ leading
to
\be
\hat I_2=2\sqrt{-c}E(1+1/c).\label{I_2_hat}
\ee
We also find
\be
\frac{d\hat I_2}{dc}=-\frac{cE(1+1/c)+K(1+1/c)}{(1+c)\sqrt{-c}}.
\label{dI_2_hat/dc}
\ee
Using (\ref{J2}) it is easy to check by direct computation that
$$
\hat J_2=\frac{1+c}{3}\hat I_2+\frac{E(t_{\ast}^-,k)-E(t_{\ast}^+,k)}{3},
$$
and we find
\be
\hat J_2=\frac{2(1+c)}{3}\sqrt{-c}E(1+1/c)+\frac{2}{3}[i E(-c)-i(1+c)K(-c)-
\sqrt{c}E(-1/c)].\label{J_2_hat}
\ee
Finally, using (\ref{I_3}) we find
$$
\hat I_3=-2[icK(-c)+\sqrt{c}K(-1/c)],
$$
and using \cite{wl09c} we arrive at
\be
\hat I_3=-2\sqrt{-c}K(1+1/c).\label{I_3_hat}
\ee
Collecting the expressions (\ref{I_2_hat},\ref{I_3_hat}) and using the
definition (\ref{K}) we find
\be
\hat K_s=K_{\mp}(t_{\ast}^-,t_{\ast}^+)=4\sqrt{1+c}+2s\sqrt{-c}[2E(1+1/c)-K(
1+1/c)].\label{K_mp_hat}
\ee
Finally, consider integral $\hat I_4=I_4(t_{\ast}^-,t_{\ast}^+)=I_{4c}+I_{4d}$,
which is written as a sum of a constant term $\hat I_{4c}$ and a divergent part
$\hat I_{4d}$. This representation follows from (\ref{I_4}) where the second
term diverges as $c+\sin^2 t_{\ast}^{\pm}=0$ and we find
$$
\hat I_{4d}=\frac{\sin2t_{\ast}^+}{(1+c)\sqrt{c+\sin^2 t_{\ast}^+}}=\frac{2\tan
t_{\ast}^+}{\sqrt{c+\sin^2 t_{\ast}^+}}.
$$
Introducing $c=(4\epsilon^2-1)\sin^2 t_{\ast}^+$ where $\epsilon\to 0$ we
obtain
\be
\hat I_{4d}=\frac{1}{\epsilon\cos t_{\ast}^+}=\frac{1}{\epsilon\sqrt{1+c}}.
\label{I4_hat_divergence}
\ee
Turning to the constant term in $\hat I_4$ we compute the first term in
(\ref{I_4}) using the relations (\ref{Elliptic030102}) and find
\be
\hat I_{4c}=\frac{2}{\sqrt{c}}\left\{K(k^2)-\frac{c}{1+c}[E(k^2)-kE(-c)]
\right\}.\label{I4_hat_const0}
\ee
Using the relation from \cite{wl09c}
$$
E(z)=\sqrt{z}E(1/z)-iE(1-z)+izK(1-z)+(1-z)K(z),
$$
we find for $z=-1/c$
$$
E(-1/c)-\sqrt{-1/c}E(-c)=-iE(1+1/c)-(i/c)K(1+1/c)+\frac{1+c}{c}K(-1/c).
$$
Substituting it in (\ref{I4_hat_const0}) and comparing the result with
(\ref{dI_2_hat/dc}) we find
\be
\hat I_{4}=-2\hat I_2'(c)+\frac{1}{\epsilon\sqrt{1+c}}.\label{I4_hat}
\ee

\end{document}